\documentclass[12pt]{iopart}

\newcommand{\re}{\ref}
\newcommand{\be}{\begin{equation}}
\newcommand{\ee}{\end{equation}}
\newcommand{\la}{\label}
\newcommand{\ber}{\begin{eqnarray}}
\newcommand{\eer}{\end{eqnarray}}

\newcommand{\bra}{\langle}
\newcommand{\ket}{\rangle}
\usepackage{iopams}  
\usepackage{graphicx}
\usepackage{epsfig}
\begin{document}
\maketitle
\eqnobysec

\title[The LIT  method]{The Lorentz Integral Transform  (LIT) method and 
its applications to perturbation induced reactions 
%with few body systems
}

\author{V D Efros$^1$, W Leidemann$^2$, G Orlandini$^2$, N Barnea$^{3,4}$}

\address{$^1$ Russian Research Centre `Kurchatov Institute',  
Kurchatov Square, 1, 123182 Moscow, Russia }

\address{$^2$ Dipartimento di Fisica, Universit\`a di Trento 
and Istituto Nazionale di Fisica Nucleare, Gruppo Collegato di Trento,  
I--38100 Trento Italy}

\address{$^3$ The Racah Institute of Physics, The Hebrew University,   
91904 Jerusalem, Israel}

\address{$^4$ Institute for Nuclear Theory, University of Washington, 
Seattle, WA 98195, USA}

\ead{orlandin@science.unitn.it}

\date{today}
\begin{abstract}

The LIT method has allowed {\it ab initio} calculations of electroweak
cross sections in light nuclear systems.
This review presents a description of the method from both a
general and a more technical point  of view, as well as 
a summary of the results obtained by its application.
The remarkable features of the LIT approach, which make it particularly
efficient in dealing with a general reaction involving continuum states,
are underlined. Emphasis is given on the results obtained for 
electroweak cross sections of few--nucleon systems. Their 
implications for the present understanding of microscopic 
nuclear dynamics are discussed. 
 
\end{abstract}

\tableofcontents

%*********************************************************
%*********************************************************
\section{Introduction}\label{sec:INTRO}
%*********************************************************
%*********************************************************

%*********************************************************
\subsection{Preliminary remarks}\label{sec:PRELIMINARY}
%*********************************************************
A very challenging problem in quantum mechanics is the {\it ab initio}
calculation of the cross section for a perturbation induced 
reaction involving a many--body system.
With {\it ab initio} we intend a calculation that requires 
a Hamiltonian $\hat H$ and the kinematic conditions of the reaction  as only
inputs, and treats all degrees of freedom of the many--body system 
explicitly (microscopic approach). 
In general, if the reaction implies a state belonging to the continuum
spectrum of $\hat H$,  the challenge may become 
enormous since one has to deal with a many--body scattering problem, which 
may lack a viable solution already for a very small 
number of constituents in the system. 
One is then forced to introduce various approximations
that are either based on rather general physical considerations
or need to be validated by experiment. This situation, however, is particularly 
unsatisfactory in cases where experiments cannot be performed, like
e.g. for some nuclear reactions of astrophysical relevance, or where the
object under investigation is just the main input of the calculation,
i.e. the potential $\hat V$ in the Hamiltonian. This situation is typical for nuclear physics
as well as for any `non--fundamental' theory, where one would like to test the 
reliability of the `effective' degrees of freedom in the Hamiltonian and of 
the interaction  $\hat{V}$. In this case the comparison
between theoretical results and experimental data is expected
to give that information, but such a  comparison risks to be inconclusive 
if the quality of the approximation is not under control. 
In both cases described above an accurate {\it ab initio}  calculation may be 
demanded.

The difficulty in calculating a many--body cross section 
involving continuum states
can be understood if one considers that at a given energy
the wave function of the system may have many different components (channels)
corresponding to all its partitions into fragments of various sizes.
Already in a rather small system of four constituents the two--, the three--
and the four--body break--up channels contribute at energies beyond the so--called
four--body break--up threshold. In configuration space the task consists in finding
the solution of the four--body Schr\"odinger equation with the proper
boundary conditions. It is just the implementation of the boundary conditions for 
a continuum wave function which constitutes the main obstacle to the practical solution
of the problem. In fact, the necessary matching of the wave function  to the 
oscillating asymptotic behaviour (sometimes even difficult to be defined unambiguously)  
is not feasible in practice.
In momentum space the situation is as complicated. The proper extension of 
the Lippmann--Schwinger equation to a many--body system has been formulated long ago
with the  Faddeev--Yakubowski equations~\cite{FADDEEV:1961,YAKUBOWSKY:1967}. However,
because of the involved analytical structure of their kernels and the number of 
equations itself, it would be very hard to solve the problem directly with their help,
at energies above the four--fragment break--up threshold, even for a number of constituents as 
small as four.
The great merit of the LIT method is that it allows to avoid all the complications 
of a continuum calculation, reducing the difficulties to those encountered in a 
typical bound--state problem, where the boundary conditions are well defined
and much easier to implement.

%*********************************************************
\subsection{Some historical notes}\label{sec:HISTORY}
%*********************************************************

The LIT method~\cite{EFROS:1994} is the natural extension of an original idea~\cite{EFROS:1985}
to calculate reaction cross sections with the help of integral transforms. 
This kind of approach 
is rather `unconventional'. It starts from the consideration that 
the amount of information contained in the  wave function is  
redundant with respect to the transition matrix elements needed in the cross 
section. Therefore, one can avoid 
the difficult task of solving the Schr\"odinger
equation. Instead one can concentrate directly on the matrix elements. With the help
of theorems based on the closure property of the Hamiltonian eigenstates,
it is proved that these matrix elements (or some combinations of them) can be 
obtained by a calculation of an integral transform with a suitable kernel, 
and its subsequent inversion. The main point is that for some kernels the 
calculation of the transform
requires the solution of a Schr\"odinger--like equation with a source, and that 
its solutions have asymptotic conditions similar to a bound state. In this sense
one can say that the integral transform method reduces the {\it continuum} 
problem to a much less problematic {\it bound--state--like} problem.   

The form of the kernel in the integral transform is crucial. 
The reason is that in order to get 
the quantities of interest the transform needs to be inverted. 
Since it is normally calculated numerically it is affected by 
inaccuracies, and inverting an inaccurate transform is somewhat problematic.
Actually, when the inaccuracies in the
input transform tend to decrease, and a proper regularization is used in the course of inversion, 
the final result approaches the true one for various kernels ~\cite{TIKHONOV:1977}. 
However, the quality of the result of the inversion may vary substantially according to
the form of the kernels, even for inaccuracies of similar size in the transforms.
In particular, when, for a specific kernel, the accuracy of the transform is 
insufficient, the result may be corrupted with oscillations superimposing the true solution. 
%The reason is that an inversion of the transform is always required and it is well known
%that for some kernels the inversion of an integral transform is an 
%{\it ill--posed problem}~\cite{TIKONOV:1977} and may suffer from large instabilities 
%(see~\sref{sec:INV}).

In~\cite{EFROS:1985} the Stieltjes kernel was proposed and its reliability
was tested and discussed in simple model studies. Later, in a test of the 
method on a realistic electromagnetic cross section, calculated also in the conventional
way for the deuteron~\cite{EFROS:1993},
it has been found that the use of the Stieltjes kernel is not satisfactory, since 
it leads to quite inaccurate results. The problem with the  Stieltjes kernel can 
be understood if one notices that its form is not qualitatively
different from that of the Laplace kernel. In fact it is well known that 
the problem of the inversion of a Laplace  
transform is extremely ill posed when the input is numerically noisy and incomplete
~\cite{ACTON:1970}.  Nevertheless the use of Laplace transforms 
is common in various fields of physics, from condensed matter to lattice QCD,
and  elaborated algorithms (e.g. the maximum entropy method~\cite{JAYNES:1978}) 
are sometimes employed for its inversion.

The problems encountered in inverting  the Stieltjes as well as the Laplace 
transform has led to the conclusion that, differently from those two cases,
the `best' kernel should be of a finite range. Its extension should be,  
roughly speaking,  about the range of the quantity to be obtained as the result 
of the inversion. This allows to eliminate  extraneous low--frequency oscillations in the
inversion results, even at a moderate accuracy of the input integral transform, while
high--frequency oscillations are excluded by a regularization (see~\sref{sec:INV}).
%conviction that the `best' kernel should have a bell shaped 
%form in order that the transform  contains as much localized information as possible 
%on the function searched in the inversion, thus minimizing instabilities. 
At the same time, of course, the transform has to be calculable in practice. 
In~\cite{EFROS:1994} it has been found
that the Lorentzian  kernel satisfies both requisites and the 
analogous test, as had been performed in~\cite{EFROS:1993} for the Stieltjes 
kernel, has led to very accurate results. 

The present work constitutes the first comprehensive review article of the LIT method.
Only short summaries of this approach have been published in~\cite{EFROS:1999a,EFROS:2001}. 
Overviews of the LIT method and of its applications have been presented at various conferences
~\cite{LEIDEMANN:2002,LEIDEMANN:2006}

%*************************************************************
\subsection{Descriptive plan of the review}\label{sec:PLAN}
%*************************************************************

In~\sref{sec:LIT_TH} it is shown that the integral transform method 
in general, and the LIT in particular, can be formulated for any perturbation induced
reaction, of inclusive (\sref{sec:INCL}) 
as well exclusive (\sref{sec:EXCL}) character, involving a general system of N interacting particles. 
There it becomes evident that another remarkable feature of the method is 
that the calculation of an inclusive cross section 
%, which in a conventional approach is very laborious, because it requires the 
%preliminary calculation of all the various  exclusive probabilities, 
turns out to be simpler than the calculation of any single exclusive cross section.
Among the exclusive reactions we focus in particular on the two--fragment case (\sref{sec:TWOBODYBR}), 
since up to now the LIT applications to exclusive reactions have  been limited to this kind of reactions.
The application of the LIT method to the case of charge fragments requires some clarification, 
therefore we discuss it in~\sref{sec:COULOMB}. The application of the LIT method to other reactions 
like for example those which are not induced by perturbation (strong reactions)
is discussed in~\sref{sec:OTHER}.

Once the general idea of the method is outlined, in~\sref{sec:LIT_PRACT} we turn to its
practical implementation. In order to apply the method in an efficient way
it is necessary to understand better the role of the parameters of the
Lorentzian kernels. This is illustrated in sections~\ref{sec:WIDTH} and~\ref{sec:CENTROID},
after having defined a few key quantities in~\sref{sec:DISCONT}.
Some particular cases, e.g. energy--dependent transition operators, are discussed 
in~\ref{sec:PARTICULAR}. The inversion of the transform
is discussed in~\sref{sec:INV}, while in sections~\ref{sec:EIGENVALUE} and~\ref{sec:LANCZOS} 
a more technical description 
on the practical implementation of the LIT method is given.

In~\sref{sec:COMMENTS} the LIT approach is compared to other methods that exhibit 
some similarities to the LIT method. 

%Numerical aspects as well as a comparison with similar algorithms used 
%n the literature are discussed.

As already stated above, in all cases the core of the LIT approach resides in 
a non--homogeneous Schr\"odinger--like equation, whose solution has asymptotic 
conditions similar to a bound state, and therefore can be solved with a 
bound--state technique.
In~\sref{sec:RESOLMETH} we describe one of such techniques that has 
been successfully applied to study nuclear, atomic and molecular 
few--body systems
~\cite{FENIN:1972,BALLOT:1980,ROSATI:1992,BARNEA:1999a,KRIVEC:1990,KRIVEC:1998}.
It belongs to the class of methods that make use of expansions on complete 
sets. In this specific case the basis set is given by hyperspherical harmonics (HH) functions. 
The approach  is called the HH method and comes in 
two variations, which differ essentially only by the way the convergence 
to the exact solution is `accelerated'. The first one (known as CHH)  
 makes use of a {\it correlation} function, which
{\it correlates} the pure HH basis in order  to reflect the characteristics 
of the potential. In the second one (known as EIHH) an effective interaction is introduced,
which is  built from the bare potential by means of a similarity transformation.
Details on CHH and EIHH are given in Appendices A and B. The results presented 
in~\sref{sec:RES} have been obtained using both approaches.

In~\sref{sec:APPL} the formalism of the electroweak perturbation induced reactions,
discussed in~\sref{sec:RES}, are described in greater detail.

Section~\ref{sec:NUM} contains four numerical tests of the LIT method.
In sections~\ref{sec:INVDEUT} and \ref{sec:SMOOTH} two interesting test cases 
for the reliability of the inversion procedure, as well as of the whole LIT approach, 
are presented. 
In particular in~\sref{sec:SMOOTH} the comparison between a response 
obtained via the LIT method and the corresponding LCZR of~\sref{sec:DIAG} is discussed.
In~\sref{sec:ALPHA} a test on the inversion algorithm is performed on $^4$He and
finally, in~\sref{sec:EIGENV_LANCZOS} a comparison
between the eigenvalue method described in~\sref{sec:EIGENVALUE} and the use of the Lanczos 
algorithm to calculate the LIT (\sref{sec:LANCZOS}) is presented.

Section~\ref{sec:RES} presents a selection of the results of the various LIT applications.
Up to now the method has been tested and largely applied to reactions 
induced by a perturbation (the electroweak probe) on a strong interacting 
system (the nucleus). There is a reason for this choice, which we explain in the following.

In~\sref{sec:SUMMARY} a brief summary of the review is given.

As already pointed out above, the strength of an 
{\it ab initio} method like the LIT resides in allowing to study the effective degrees
of freedom in the Hamiltonian and the reliability of the potential $\hat V$. Nowadays this is 
a much debated topics in nuclear physics and the LIT method represents a unique tool
to clarify some important issues. In particular it can address the question of 
the nature of the nuclear force. Questions like whether the
force is exclusively of two--body nature or three-- or more body forces are required
to describe the nuclear phenomenology can be addressed also in reactions with more than three nucleons.
 At the same time one is able
to discriminate among potentials derived within purely phenomenological or 
semi--phenomenological (boson exchange) or effective field theory approaches. 
But more than that and differently from the purely strong reactions, electroweak 
reactions with nuclei can provide additional information on the relevant degrees 
of freedom in nuclear physics. In fact 
the real or virtual photons, as well as neutrinos, probe not only the explicit 
degrees of freedom in the Hamiltonian (protons and neutrons), but also the implicit 
ones that generate the exchange (charge or weak) currents.
In the examples reported here particular emphasis on these aspects is given.

%In~\sref{sec:SUMMARY} we summarize the main contributions of the LIT method
%to the better understanding of nuclear dynamics. The future prospects 
%for an extension of its applications to heavier nuclear systems as well as to
%other fields of physics are also discussed.

%*********************************************************
%********************************************************* 
\section{The LIT method: general theory}\label{sec:LIT_TH}
%*********************************************************
%*********************************************************

%************************************************
\subsection{Inclusive processes}\label{sec:INCL}
%************************************************

In inclusive processes the quantities of interest have the following structure
\begin{equation}
r(E)=
\sum\!\!\!\!\!\!\!\int\,d\gamma
\langle Q|\Psi_\gamma\rangle\langle\Psi_\gamma|Q'\rangle
\delta(E_\gamma-E), \label{rr}
\end{equation}
where  $|\Psi_\gamma\rangle$ are solutions to the dynamic equation  
\be
(\hat{H}-E_\gamma)|\Psi_\gamma\rangle=0,
\ee
and $\hat{H}$ is the Hamiltonian of the system. 
The set $|\Psi_\gamma\rangle$  is assumed to be complete and orthonormal, 
\be  
\sum\!\!\!\!\!\!\!\int\,d\gamma|\Psi_\gamma\rangle\langle\Psi_\gamma|=1.\la{cl}
\ee
The integration and summation here and in~(\re{rr}) go 
over all discrete states and continuum spectrum states in the set. 
%The discrete spectrum contribution to (\re{rr}) may be written as ($\gamma\to n$,
%$n$ integer number) 
%\begin{equation}
%\sum_n r_n(E)\equiv\sum_n\langle Q|\Psi_n\rangle\langle\Psi_n|Q'\rangle
%\delta(E_n-E)).
%\end{equation}
We suppose that the norms $\langle Q|Q\rangle$ and $\langle Q'|Q'\rangle$ are finite. 

In the case of perturbation--induced reactions 
one has $|Q\rangle=\hat{O}|\Psi_0\rangle$, $|Q'\rangle=\hat{O}'|\Psi_0\rangle$, where 
$|\Psi_0\rangle$ is the initial state in a 
reaction (generally the ground state of the system undergoing the perturbation),
and $\hat{O}$, $\hat{O}'$ are transition operators. Then one has
\begin{equation}
r(E)=
\sum\!\!\!\!\!\!\!\int\,d\gamma
\langle\Psi_0|\hat{O}^\dag|\Psi_\gamma\rangle
\langle\Psi_\gamma|\hat{O}'|\Psi_0\rangle
\delta(E_\gamma-E). \label{re}
\end{equation}
Here  
$|\Psi_\gamma\rangle$ is a set of final states. If $\hat{O}=\hat{O}'$, the quantity (\re{re}) 
may represent a response function or, in general, a contribution to the response of the system
to a perturbative probe transferring energy $E$ to it.

When the energy $E$ and the number of particles in a system increase
the direct calculation of the quantity $r(E)$
becomes prohibitive (only in case that $E$ becomes larger than the interaction
energy, perturbation theory can be used). The difficulty is related to the fact that in these cases
a great number of continuum spectrum
states $|\Psi_\gamma\rangle$ contribute to $r(E)$ and the
 structure of these  states is very complicated. 

The approach to  overcome this difficulty that is presented here
can be considered as a generalization of the sum rule approach,
since the use of the closure property of the Hamiltonian eigenstates plays a fundamental role.
Consider for example a simple sum rule for the quantity (\re{re}), based on the closure
property (\re{cl}) i.e.
\be    
\sum\!\!\!\!\!\!\!\int\,r(E)dE =\langle Q|Q'\rangle.\la{sum}
\ee
The  calculation of this quantity is much easier than a 
direct calculation of $r(E)$ itself,
since it requires the knowledge of $|Q\rangle$ and $|Q'\rangle$ only. This can be obtained 
with bound--state methods since we have supposed that $|Q\rangle$ and $|Q'\rangle$ have finite norms.
However, this sum rule contains only a limited information on $r(E)$. In order to get
much more information about it we consider instead 
an integral transform
\be
\Phi(\sigma)=\sum\!\!\!\!\!\!\!\int\, K(\sigma,E)\,r(E)\,dE\la{phi}
\ee
with a smooth kernel $K$ (specified below). This yields
\begin{eqnarray}
\Phi(\sigma)&=&\sum\!\!\!\!\!\!\!\int\,d\gamma
\langle Q|\Psi_\gamma\rangle K(\sigma,E_\gamma)
\langle\Psi_\gamma|Q'\rangle\nonumber\\
&=&\sum\!\!\!\!\!\!\!\int\,d\gamma
\langle Q|\hat{K}(\sigma,\hat{H})|\Psi_\gamma\rangle 
\langle\Psi_\gamma|Q'\rangle.\la{r}
\end{eqnarray}
Using the closure property (\re{cl}) one obtains
\be
\Phi(\sigma)=\langle Q|\hat{K}(\sigma,\hat{H})|Q'\rangle.\la{s}
\ee
Therefore equation (\re{s}) may be viewed as a generalized sum rule depending on 
a continuous parameter $\sigma$. 
With a proper choice of the kernel $K$ 
the right--hand side of~(\re{s}) may be calculated using {\em bound--state} type
methods. Once $\Phi(\sigma)$ is available (\re{phi}) may be solved to obtain $r(E)$
via an inversion of the transform. 

Our choice of the kernel $K(\sigma,E)$ is such 
that both the calculation of $\Phi(\sigma)$ and the inversion of (\re{phi}) are feasible. 
We choose~\cite{EFROS:1994}
\be
K(\sigma,E)=\frac{1}{(E-\sigma^*)(E-\sigma)}.\la{k}
\ee
Notice that the energy parameters $\sigma$ that we consider are complex.
For convenience we define them as
\be
\sigma=E_0+\sigma_R+i\sigma_I\,,\la{not}
\ee
where $E_0$ is the ground--state energy, and $\sigma_I\neq0$, so that
$K(\sigma,E)$ is actually a Lorentzian function centered on $E_0+\sigma_R$,
having  $\sigma_I$ as a halfwidth
\begin{equation}
K(\sigma_R,\sigma_I,E)=\frac{1}{(E-E_0-\sigma_R)^2+\sigma_I^2}\,.\la{kRI}
\end{equation}
Then the integral transform (\re{phi}) becomes
\begin{equation}
L(\sigma_R,\sigma_I)=\sum\!\!\!\!\!\!\!\int\,dE\,
\frac{r(E)}{(E-E_0-\sigma_R)^2+\sigma_I^2}\,.\la{lorentz}
\end{equation}
Here and in the following the integral transform $\Phi(\sigma)$ with a Lorentz kernel 
is denoted by $L(\sigma_R,\sigma_I)$.
Using the definition (\re{re}) it is easy to show that the quantity (\re{lorentz}) may be  represented as 
\be
L(\sigma_R,\sigma_I)=\langle\tilde{\Psi}|\tilde{\Psi}'\rangle\,,\la{ov}
\ee
where the `LIT functions' $\tilde{\Psi}$ and $\tilde{\Psi}'$ are given by
\begin{eqnarray}
|\tilde{\Psi}\rangle&=&\left(\hat{H}-E_0-\sigma_R-i\sigma_I\right)^{-1}\hat{O}|\Psi_0\rangle\,, \la{psi1}\\
|\tilde{\Psi}'\rangle&=&\left(\hat{H}-E_0-\sigma_R-i\sigma_I\right)^{-1}\hat{O}'|\Psi_0\rangle\,.\la{psi2}
\end{eqnarray}
These functions are solutions to the inhomogeneous equations
\begin{eqnarray}
\left(\hat{H}-E_0-\sigma_R-i\sigma_I\right)|\tilde{\Psi}\rangle&=&\hat{O}|\Psi_0\rangle\,, \la{eq1}\\
\left(\hat{H}-E_0-\sigma_R-i\sigma_I\right)|\tilde{\Psi}'\rangle&=&\hat{O}'|\Psi_0\rangle\,. \la{eq2}
\end{eqnarray}
Let us  suppose that  $\hat{O'}=\hat{O}$ ($\hat{O}=\hat{O}'$). In this case  $L(\sigma)$
equals to $\langle\tilde{\Psi}|\tilde{\Psi}\rangle$ ($\langle\tilde{\Psi}'|\tilde{\Psi}'\rangle$). 
Since for $\sigma_I\ne0$
the integral in (\re{lorentz}) does exist, the norm of $|\tilde{\Psi}\rangle$ ($|\tilde{\Psi}'\rangle$)
is finite. This implies
that $|\tilde{\Psi}\rangle$ and $|\tilde{\Psi}'\rangle$ are  {\em localized} functions.
%If $\hat{O}\ne\hat{O}'$  then $|\tilde{\Psi}\rangle$ 
%and $|\tilde{\Psi}'\rangle$ are localized under the condition that $|Q\rangle=\hat{O}|\Psi_0\rangle$
%and $|Q'\rangle=\hat{O}'|\Psi_0\rangle$ are themselves localized (which is actually the case in most 
%physical applications).
Consequently,~(\re{eq1}) and~(\re{eq2}) can be solved with bound--state type methods.
Similar to the problem 
of calculating a bound state it is sufficient to impose the only condition that the 
solutions  of~(\ref{eq1}) and~(\ref{eq2}) are localized. This means that
in contrast to continuum spectrum problems, in order 
to construct a solution, it is not necessary here to reproduce a complicated large 
distance asymptotic behaviour in the coordinate representation or singularity structure 
in the momentum representation. This is a very substantial simplification.\footnote{In the 
case of Faddeev--Yakubowsky equations one also needs not impose boundary conditions to
get a solution. However, this is achieved due to an involved structure of these equations
that seriously complicates calculations already for $A=4$. In contrast, we use a
simpler Schr\"odinger operator.}

%, having to deal only with
%simple large distance asymptotic behavior in the coordinate representation or singularity
%structure in the momentum representation. 

Obviously, localized solutions to~(\re{eq1}) and~(\re{eq2})
are unique. Once $L(\sigma)$ is calculated  $r(E)$ is obtained by inversion of 
the integral transform with a Lorentzian kernel (\re{lorentz}) (`Lorentz integral transform'). 
The inversion of the LIT will be discussed in~\sref{sec:INV}.

Before ending this section we should mention that 
if the Hamiltonian is rotationally invariant it is useful to expand 
the states $\hat{O}|\Psi_0\rangle$ and  $\hat{O}'|\Psi_0\rangle$
over states possessing given values $J$ and $M$ of the
angular momentum and its projection. Then the whole calculation may be done    
in separate subspaces of states belonging 
to given $J$ and $M$. Furthermore, the calculations are $M$ independent.
An explicit example is discussed in~\sref{sec:MULT}.

%************************************************
\subsection{Exclusive processes}\label{sec:EXCL}
%************************************************

Here we illustrate the LIT method for perturbation induced exclusive processes. 
Some comments on non--perturbative processes will be listed as well, at 
the end of this section.

The relevant matrix element that one has to calculate in the case of an exclusive
perturbation induced reaction is 
\be
M_{fi}(E)=\langle\Psi^-_f(E)|\hat{O}|\Psi_0\rangle\,, \la{mfi}
\ee
where  $\hat{O}$ is a perturbation operator that causes the transition,
$|\Psi_0\rangle$ is again the localized initial state, while $|\Psi^-_f(E)\rangle$ 
represents the continuum
final state with energy $E$. The calculation proceeds as follows~\cite{EFROS:1985}.
We set~\cite{GOLDBERGER:1964}
\be
|\Psi_f^-(E)\rangle={\cal A}|\phi_f(E)\rangle+(E-\hat{H}-i\eta)^{-1}|\bar{\phi}_f(E)\rangle\,,
\la{psim}
\ee
where ${\cal A}|\phi_f(E)\rangle$ is the `channel' function, $\cal A$ being the antisymmetrizer
\footnote{Note that here any expression containing $x \pm i\eta$ is understood to be 
evaluated at $\eta > 0$ with the limit $\eta \to 0$ then taken.}.
In general this will be 
the properly antisymmetrized product of fragment bound states 
times their {\it free} relative motion sub--states of given momenta (we temporarily
neglect the long--range Coulomb interaction between fragments in the final state; the Coulomb case is
discussed in ~\sref{sec:COULOMB}).
To deal with states of well defined rotational quantum numbers one can write ${\cal A}|\phi_f(E)\rangle$
as properly antisymmetrized product of fragment states and free relative motion
sub--states of given orbital angular momentum quantum numbers, coupled to proper total angular momentum.
%For antisymmetric states the linear combination of $\phi_f^{NS}$ is provided by the
%antisymmetrization operator i.e. $\phi_f={\cal A}\phi_f^{NS}$, where~\cite{GOLDBERGER:1964}
%\be 
%{\cal A}=N^{-1/2}\sum_{i=1}^{N}\epsilon_i\hat{P}_i.
%\ee
%Here the permutation $\hat{P}_i$ interchanges particles between fragments and 
%$\epsilon_i$ are the parities of $\hat{P}_i$. 
The states $|\bar{\phi}_f(E)\rangle$ in (\re{psim}) are
\be
|\bar{\phi}_f\rangle\equiv(\hat{H}-E){\cal A}|\phi_f\rangle\equiv{\cal A}\hat{V}_f|\phi_f\rangle\,, \la{pb}
\ee
where $\hat{V}_f$ is the interaction between particles belonging to different fragments in the 
non--antisymmetrized state $|\phi_f\rangle$

Using (\re{psim}), we rewrite the reaction amplitude (\re{mfi}) as
\be
M_{fi}(E)=\langle{\cal A}\phi_f(E)|\hat{O}|\Psi_0\rangle+
\langle\bar{\phi}_f(E)|(E-\hat{H}+i\eta)^{-1}\hat{O}|\Psi_0
\rangle\,.\la{mm1}
\ee
The first term in (\re{mm1}) is  called the Born term and can be 
computed directly. The second term takes into account the final
state interaction and represents the difficult part of the problem.
However, if we  represent it as a sum over the discrete spectrum $E_n$
and integral over the continuous energies $E'$, i.e. in the form
\begin{eqnarray}
\langle\bar{\phi}_f(E)|(E-\hat{H}&+&i\eta)^{-1}\hat{O}|\Psi_0
\rangle=\sum_n (E-E_n)^{-1}F_{fi}(E,E_n) \nonumber\\
&+&\int_{E_{th}}^\infty
(E-E'+i\eta)^{-1}F_{fi}(E,E')dE',\label{ssst2}
\end{eqnarray}
where the form factor $F_{fi}$ is defined as
\begin{equation}
 F_{fi}(E,E')=\sum\!\!\!\!\!\!\!\int d\gamma\langle\bar{\phi}_f(E)|\Psi_\gamma\rangle
\langle\Psi_\gamma|\hat{O}|\Psi_0\rangle\delta(E_\gamma-E')\,,
\la{ff}
\end{equation}
we realize that (\ref{ff}) has the same formal structure as (\re{rr}), with $E\rightarrow E'$. 
Therefore the form factor
(\re{ff}) can be calculated at a given $E$ value 
as a solution to the equation
\be
\Phi_{fi}(E,\sigma)=\sum\!\!\!\!\!\!\!\int\, K(\sigma,E')F_{fi}(E,E')dE'.\la{inteq}
\ee
This equation is similar to (\re{phi}). Therefore $F_{fi}(E,E')$ can be obtained
calculating first $\Phi_{fi}(E,\sigma)$ from~(\ref{s}) and then inverting the transform.

Also in this case we adopt the Lorentz kernel. However, 
it is useful to express it as 
\be
K(\sigma,E')=\frac{1}{2i\sigma_I}\left(\frac{1}{E'-\sigma}-
\frac{1}{E'-\sigma^*}\right)\,,\la{kp}
\ee
where $\sigma=\sigma_R+i\sigma_I$. Then~\cite{EFROS:1999}:
\be
L_{fi}(E,\sigma)
=(2i\sigma_I)^{-1}
\langle\bar{\phi}_f(E)(|\tilde\Psi_1\rangle -|\tilde\Psi_2\rangle),\la{texc}
\ee
where
\begin{eqnarray} 
|\tilde\Psi_1\rangle&=&(\hat{H}-\sigma_R -i\sigma_I)^{-1}\hat{O}|\Psi_0\rangle, \la{ppsi1}\\
|\tilde\Psi_2\rangle&=&(\hat{H}-\sigma_R +i\sigma_I)^{-1}\hat{O}|\Psi_0\rangle.\la{ppsi2}
\end{eqnarray}
Therefore $|\tilde\Psi_1\rangle$ and $|\tilde\Psi_2\rangle$ are the solutions of the two inhomogeneous equations
\begin{eqnarray}
\left(\hat{H}-\sigma_R -i\sigma_I\right){|\tilde\Psi_1\rangle}&=&\hat{O}|\Psi_0\rangle\,,\la{Eqs1}\\ 
\left(\hat{H}-\sigma_R +i\sigma_I\right){|\tilde\Psi_2\rangle}&=&\hat{O}|\Psi_0\rangle. \la{Eqs2}
\end{eqnarray}
Equation~(\ref{texc}) is convenient for several reasons.  The solution of~(\ref{Eqs1}) or~(\ref{Eqs2})  
involves only $\sigma$ and is independent of the energy $E$. In other words, to obtain $M_{fi}(E)$ 
one needs to solve these equations for a set of $\sigma$ values 
irrespective to $E$. In fact only one of the equations is to be solved since 
$|\tilde\Psi_1\rangle$ and $|\tilde\Psi_2\rangle$ are related to each other 
in a simple way. Furthermore, the dynamic calculation to be performed is independent 
of the final state channel $f$. 
%Therefore transitions to different final states may
%be obtained as quadratures from the same dynamic calculation. Thus to obtain the transition 
%amplitude one needs the form factor $F_{fi}$ entering (\ref{ssst2}).
%It is obtained via inverting the Lorentz transform (\ref{texc}) i.e. solving the integral equation
%of  (\ref{inteq}) form.  The singular integral in (\ref{ssst2})  is calculated as 
%for two reasons. The first one is that $|\tilde\Psi_1\rangle$ and  $|\tilde\Psi_2\rangle$
%are related to each other in a simple way, so that only one
%dynamical calculation is required.  The second one is that the solution of (\ref{Eqs1}) or (\ref{Eqs2})
%involves only $\sigma$ and is independent of the energy $E$.  
%In other words, to obtain the spectrum 
%$M_{fi}(E)$ 
%one needs to solve one of those two equations for a set of $\sigma$ values, 
%irrespective to $E$. Then $M_{fi}(E)$ is obtained in a simple way: 

Inserting  $|\tilde\Psi_1\rangle$ and $|\tilde\Psi_2\rangle$
into (\ref{texc}) one obtains $L_{fi}(E,\sigma)$.
Then the Lorentz transform (\ref{texc}) is inverted, i.e. the integral equation~(\ref{inteq}) 
is solved for $F_{fi}(E,E')$, and 
the singular integral in~(\ref{ssst2}) is calculated as
\begin{eqnarray}\label{pv}
\int_{E_{th}}^\infty(E-E'&+&i \eta)^{-1}F_{fi}(E,E')dE'= 
-i\pi F_{fi}(E,E)\nonumber\\ &+& P\int_{E_{th}}^\infty(E-E')^{-1}F_{fi}(E,E')dE'\,.
\end{eqnarray}
After having calculated the non--trivial term 
$\langle\bar{\phi}_f(E)|(E-\hat{H}+i\eta)^{-1}\hat{O}|\Psi_0\rangle$ 
from~(\ref{ssst2}) it is added in (\ref{mm1}) to the Born term to obtain 
the matrix element $M_{fi}(E)$. \footnote{
It is convenient to calculate the principal value integral entering here as
a sum of ordinary integrals, 
$P\int_{E_{th}}^\infty\frac{f(E')}{E-E'}dE'=\int_{E_{th}}^{E-\Delta}\frac{f(E')}{E-E'}dE'
-\int_{E-\Delta}^{E+\Delta}\frac{f(E')-f(E)}{E'-E}dE'+
\int_{E+\Delta}^\infty\frac{f(E')}{E-E'}dE'$, where $\Delta$ is arbitrary, provided that 
$E_{th}<E-\Delta$.}

%***********************************************************************
\subsubsection{Two--fragment break--up reactions.}\label{sec:TWOBODYBR}
%***********************************************************************

In case of two--body break--up reactions the calculation can also proceed in a different
way. In fact this alternative method has been used in the various exclusive LIT applications
discussed in~\sref{sec:RES}.

The procedure is the same as described by~(\ref{mfi})-(\ref{inteq}), but the Lorentz kernel
is taken in the form~(\ref{k}) instead of~(\ref{kp}). Then one has 
\begin{equation}
L_{fi}(E,\sigma)=\langle\tilde\Psi_a|\tilde\Psi_b\rangle\,,\la{ab}
\end{equation}
with 
\begin{eqnarray} 
|\tilde\Psi_a\rangle&=&(\hat{H}-\sigma)^{-1}\hat{O}|\Psi_0\rangle, \la{ppsia}\\
|\tilde\Psi_b\rangle&=&(\hat{H}-\sigma)^{-1}|\bar\phi_f\rangle.\la{ppsib}
\end{eqnarray}
Obviously $\tilde\Psi_{a,b}$ are the solutions of the two inhomogeneous equations
\begin{eqnarray}
\left(\hat{H}-\sigma_R-i\sigma_I\right){|\tilde\Psi_a\rangle}&=&\hat{O}|\Psi_0\rangle\,,\la{Eqsa}\\ 
\left(\hat{H}-\sigma_R-i\sigma_I\right){|\tilde\Psi_b\rangle}&=&|\bar\phi_f\rangle\,. \la{Eqsb}
\end{eqnarray}

In~\cite{LAPIANA:2000} the general and the two--fragment break--up methods have been used
for calculating the $d(e,e'p)n$ cross section. There were only small differences in the results
(about 1\%). 
%From the numerical point of view, however, the LIT of (\ref{ab}) is probably less 
%affected by numerical errors than the LIT of (\ref{texc}), since it avoids to perform 
%the difference $|\tilde\Psi_1-\tilde\Psi_2\rangle$. On the other hand one has to solve
%(\ref{Eqs2}) for every energy $E$ in order to obtain $M_{fi}(E)$.

%**************************************************************************************
\subsubsection{Case of fragments interacting via Coulomb potential.}\label{sec:COULOMB}
%**************************************************************************************

In order to take into account the long--range Coulomb interaction between the fragments
one may proceed as follows. 

Let us denote with $\hat{U}_f^{C}$ the 
average Coulomb potential between fragments, i.e. the Coulomb potential between the 
charges of the fragments concentrated in their centres of mass. 
Let ${\cal A}|\varphi_f^{(-)}\rangle$ be the continuum state given by the properly 
antisymmetrized product of the bound states of fragments times the 
`plane--plus--ingoing wave' sub--state $|\psi^-\rangle$ of their relative motion in this 
average Coulomb potential. 
The latter is the wave function satisfying
\begin{equation}
(\hat{T}_{rel}+\hat{U}^C_f-E)|\psi^-\rangle=0\,,
\end{equation}
where $\hat{T}_{rel}$ is the kinetic energy of this relative motion. (${\cal A}|\varphi_f^{(-)}\rangle$ 
may as well be the corresponding angular--momentum coupled state of this type).
In analogy with~(\ref{pb}), let us denote with $|\bar{\varphi}_f^{(-)}\rangle$ the properly 
antisymmetrized state
\begin{equation}
|\bar{\varphi}_f^{(-)}\rangle\equiv(\hat{H}-E){\cal A}|\varphi_f^{(-)}\rangle\equiv
{\cal A}(\hat{V}_f-\hat{U}^C_f)|\varphi_f^{(-)}\rangle\,.
\end{equation}
Then one may write (see equation (108) in chapter 5 of~\cite{GOLDBERGER:1964})
\be
|\Psi_f^-(E)\rangle={\cal A}|\varphi_f^{(-)}(E)\rangle+(E-\hat{H}-i\eta)^{-1}|\bar{\varphi}_f^{(-)}(E)\rangle\,.
\ee
Because of this relationship one may use the same procedure as in sections~\ref{sec:EXCL} or 
~\ref{sec:TWOBODYBR} with the replacements $|\phi_f\rangle\rightarrow|\varphi_f^{(-)}\rangle$ and
$|\bar{\phi}_f\rangle\rightarrow|\bar{\varphi}_f^{(-)}\rangle$. However, one has to consider that, 
different from the case with no Coulomb interaction, 
the corresponding sub--states $|\psi^-\rangle$ are no longer plane waves. This problem is easily overcome
when the Coulomb interaction acts only between one pair of fragments since in this case
$|\psi^-\rangle$ is known exactly.

%**************************************************************************************************
\subsection{Comments on the application of the method to other kinds of processes}\label{sec:OTHER}
%**************************************************************************************************

With the integral transform approach one can also obtain the amplitudes of 
non--perturbative reactions of the general type, without calculating continuum states. 
In the non--perturbative case the procedure is 
rather similar to that described above for perturbation--induced transitions~\cite{EFROS:1985}. 
In fact in this case one has to calculate a matrix element like in~(\ref{mfi}), where,
however, $\hat{O}|\Psi_0\rangle$ is replaced by
$|\bar{\phi}_i(E)\rangle$, i.e. the initial state of the reaction, defined in a similar way as the 
state $|\bar{\phi}_f(E)\rangle$ in~(\ref{pb}). The state $|\bar{\phi}_i(E)\rangle$ normally corresponds to two 
colliding fragments. The key point is therefore again to calculate the form factor
of (\ref{ff}) type with the above mentioned replacement.
%which instead of $\hat{O}\Psi_0$ includes the state $\bar{\phi}_i(E)$ that 
%pertains to the initial state of the reaction 
%and is defined similar to the state $\bar{\phi}_f(E)$ above.  
The transformation with the Lorentz kernel may be employed to calculate this form factor. 
No such calculations have yet been performed. Since in this case 
the source term $|\bar{\phi}_i(E)\rangle$ in the equations corresponding to (\ref{ppsi1}) and  
(\ref{ppsi2}) depends on $E$, the corresponding dynamic calculation has to be done 
separately for each $E$ value. 
%The form factor  discussed here
%(as well as that in (\re{ff})) may be calculated with the procedure 
%considered in~\sref{sec:DIAG}, employing diagonalization of the Hamiltonian 
%in a finite subspace of localized functions \footnote{Other procedures to calculate reactions 
%employing this diagonalization have also been considered
%~\cite{KUKULIN:2005}.}.

Semi--inclusive processes, like e.g. $(e,e'N)$ reactions, are sometimes discussed 
in the framework of the so--called spectral function approximation~\cite{FRULLANI:1984}. 
%In this case there is no need in application of the above approach. 
Although the corresponding cross section is not an inclusive 
one, but only semi--inclusive, it may be calculated with the procedure of the type 
presented in~\sref{sec:INCL}. 
The calculation of the spectral function in the $A$--particle case is reduced to the 
corresponding calculation of a quantity of (\re{rr}) type
in the $(A-1)$--particle subsystem, with the state $|Q\rangle$ depending on the momentum of the 
knocked--out nucleon~\cite{EFROS:1998}. 

A non--perturbative amplitude of inelastic scattering of a fast projectile on a target 
in the framework of the Glauber approximation~\cite{GLAUBER:1955} is of the form $\langle\Psi_f|Q\rangle$,
where $|\Psi_f\rangle$ is the continuum final state of the target and $|Q\rangle$ includes its initial
state and the profile function playing the role of the transition operator. This amplitude
has the same structure as the perturbative amplitude of (\re{mfi}) and therefore can be calculated 
in a similar way.  No such calculations are known. 

%*********************************************************************
%********************************************************************* 
\section{The LIT method: practical implementation}\label{sec:LIT_PRACT}
%***********************************************************************
%*********************************************************************

%*****************************************************************
\subsection{Discrete and continuum spectra}\label{sec:DISCONT}
%****************************************************************

Separating in $r(E)$, defined in (\ref{rr}),  the discrete and continuum spectrum contributions, 
one may write
%and remembering the definition of $e=E-E_0$, the Lorentz transform (\re{lorentz}) reads as 
\begin{eqnarray}
L(\sigma_R,\sigma_I) & = &
\sum_n\frac{r_{n}}{(\sigma_R-e_{n})^2+\sigma_I^2}+
\int_{e_{th}}^\infty de\frac{r(e)}{(\sigma_R-e)^2+\sigma_I^2}\,\nonumber\\
&\equiv &L_D(\sigma_R,\sigma_I) + L_C(\sigma_R,\sigma_I)
\la{tra}
\end{eqnarray}
where we have introduced the excitation energy $e=E-E_0$. In particular
$e_n$ are the discrete excitation energies $e_n=E_n -E_0$ and $r_n$ represent the 
corresponding contributions to the function $r$
\begin{equation}
r_n=\langle Q|\Psi_n\rangle\langle\Psi_n|Q'\rangle\,.\label{dc}
\end{equation}
The second term in  (\re{tra}) is related to the continuum part of the spectrum with 
$e_{th}$ representing the continuum threshold energy ($e_n<e_{th}$) and
\begin{equation}
r(e)=
\int_{e_{th}+E_0}^\infty \,dE_\gamma\,\langle Q|\Psi_\gamma\rangle\langle\Psi_\gamma|Q'\rangle
\,\delta(E_\gamma-E)\,.
\la{co}
\end{equation} 

The aim of the LIT method is to obtain both $r_n$ and $r(e)$ solving (\re{tra}) with 
$L(\sigma_R,\sigma_I)$ as an input. It is expedient to use 
$L$ calculated at a fixed $\sigma_I$ in a range of $\sigma_R$ values. 
One may also use $\sigma_I=\sigma_I(\sigma_R)$.
The first task is to get $L_D(\sigma_R,\sigma_I)$ i.e.
the discrete contributions to (\re{tra}). 
One way to get the contributions from the discrete levels is to calculate these levels explicitly to
obtain  $|\Psi_n\rangle$ and $e_n$ and evaluate directly the overlaps~(\ref{dc}) and their 
contributions to the response function.
%, and then use them to obtain the overlaps in~(\ref{dc}), and therefore $r_n$. 
Alternatively one can extract these contributions by calculating the transform
$L=\langle\tilde{\Psi}|\tilde{\Psi}'\rangle$ 
at a very small $\sigma_I$ value
for a range of $\sigma_R$ values between zero and  $e_{th}$. For fixed $\sigma_I$, in the vicinity 
of a level $e_n$, the transform $L$ is dominated 
by the corresponding single term,
\begin{equation}
 L_D(\sigma_R,\sigma_I)\simeq\frac{r_{n}}{(\sigma_R-e_{n})^2+\sigma_I^2}.\label{discrete}
\end{equation}
This $\sigma_R$ dependence allows extracting $e_n$ and $r_n$.
The procedure is convenient when
one deals with few well separated levels. 
%(If $L$ is obtained using complete diagonalization of the Hamiltonian as below in (\re{phin}), 
%this gives discrete contributions to (\re{tra}) directly). 
The accuracy in determining $e_n$ and $r_n$
may affect the continuous contribution $r(e)$ to be obtained subsequently.

After calculating $L(\sigma_R,\sigma_I)$ 
from (\re{ov}) and subtracting the discrete contributions from it,
the following integral equation of the first kind with the Lorentz 
kernel is to be solved to obtain $r(e)$.
\begin{equation}
L_C(\sigma_R,\sigma_I)=
\int_{e_{th}}^\infty de\frac{r(e)}{(\sigma_R-e)^2+\sigma_I^2}.\la{ie}
\end{equation}
To simplify the notation in the following we will omit the $\sigma_I$ dependence 
and use $L(\sigma_R)$ to indicate its continuum part only, i.e. $L_C(\sigma_R)$,
unless specified explicitly.

%************************************************************************
\subsection{Role of the width of the Lorentzian kernel}\label{sec:WIDTH}
%************************************************************************

Let us comment on the choice of the $\sigma_I$ value. The  resulting response $r(e)$
from an `exact' calculation of $L$ and subsequent inversion 
should be independent of this choice.
However, the rate of convergence in the calculation
may depend on $\sigma_I$. 
Let us show that the value of $\sigma_I$ determines the relative error in $r(e)$
manifesting in components with high frequencies. The LIT's of such high frequency components
become very small because of 
the averaging out in the integral. Denoting with $r_\nu\exp(i\nu e)$ 
the component of $r(e)$ with frequency $\nu$
and with $L_\nu$ its transform, one can show
that if $|\nu|\sigma_I$ is less or about unity one has 
$(\sigma_I/\pi)|L_\nu|\simeq|r_\nu|$ while if $|\nu|\sigma_I\gg1$
one has
\begin{equation}
 \frac{\sigma_I}{\pi}|L_\nu|\simeq\frac{1}{\pi|\nu|\sigma_I}\frac{1}
{1+(\sigma_R-e)^2/\sigma_I^2}|r_\nu|.\la{rel}
\end{equation}
When one solves (\re{ie}) numerically one may adopt that 
the errors  $\delta L$ in the transform and $\delta r$ in the  response 
are related to each other in accordance with 
\be
\delta L(\sigma_R)=
\int_{e_{th}}^\infty de\frac{\delta r(e)}{(\sigma_R-e)^2+\sigma_I^2}.\la{ieer}
\ee
Let us denote  $\delta L_\nu$ the component of $\delta L$ that corresponds to 
the component $\delta r_\nu\exp(i\nu e)$ of $\delta r$. Furthermore,
let us use the notation $\delta L_0$, $L_0$,  $\delta r_0$ and  $r_0$
for the components $\delta L_\nu$, $L_\nu$, $\delta r_\nu$ and  $r_\nu$
with a low frequency. When $|\nu|\sigma_I\gg 1$,  $|L_\nu|$ substantially decreases 
according to (\re{rel}) and it may occur that $|\delta L_\nu/\delta L_0|\gg|L_\nu/L_0|$.
This gives
$|\delta r_\nu/r_\nu|\gg|\delta r_0/r_0|.$
Therefore it is desirable that for all the frequencies $\nu$ substantial to 
reproduce the main structure of $r(e)$
the  quantity $|\nu|\sigma_I$ would not be large. To this aim, in order to solve
the integral equation 
(\re{ie})  a regularization is applied, which
suppresses high frequencies (see~\sref{sec:INV}). When the mentioned condition
is fulfilled this ensures approximate stability of $r(e)$ thus obtained and its closeness
to the true solution.

From the above point of view smaller $\sigma_I$ values are preferable. On the other hand, 
when $\sigma_I$ decreases one approaches the scattering regime in (\re{eq1}) and  (\re{eq2}),
which makes 
it harder to obtain $L$ with bound--state type methods.
For example, suppose that those equations are solved in a subspace of localized functions 
of dimension $N$ (see~\sref{sec:EIGENVALUE}). If we denote $e_n^N=E_n^N-E_0^N$ where $E_n^N$ 
are the eigenvalues of the Hamiltonian matrix in this
subspace then the approximate total transform $L_N$ (including both discrete and continuum
contributions) has the structure
\be
L_N=\sum_{n=1}^N
\frac{\gamma_{n}^N}{(\sigma_R-e_{n}^N)^2+\sigma_I^2}.\la{phin}
\ee 
Here the terms with lower $e_n^N$ values represent the discrete contributions in (\re{tra}) 
while other terms
correspond to a discretization of the continuum. When $N$ increases the density of the latter 
type $e_n^N$ values increases
as well providing a considerable number of them 
within a range $\sigma_R-\sigma_I\le e_n^N\le\sigma_R+\sigma_I$. 
When this occurs overlaps of peaks from separate contributions 
to (\re{phin}) lead to a smooth $L_N$ 
approaching the true $L$ as $N$ tends to infinity. However, the smaller is $\sigma_I$
the higher should be the $N$ value to represent the true $L$ with a given accuracy. Therefore 
performing a calculation with too small a value of $\sigma_I$ is not expedient. 

%***********************************************************************************
\subsection{Role of the centroid  of the Lorentzian kernel}\label{sec:CENTROID}
%***********************************************************************************

Let us comment on the range of $\sigma_R$ values for which (\re{ie}) is to be solved. 
If one adopts that the spectrum $r(e)$ to be obtained extends over the range 
$e_{th}\le e\le e_{max}$ then it is reasonable to employ $L(\sigma_R)$ approximately in the range 
$e_{th}-\sigma_I\le \sigma_R\le e_{max}+\sigma_I$. At such conditions complete information
on $r(e)$ contained in $L(\sigma_R)$ is used.

Suppose that 
 $r(e')$ does not change considerably in the range $e-\sigma_I\le e'\le e+\sigma_I$ and that the $r$ values
in this range are not much smaller than  the values beyond this range. Then performing the integration
in (\re{ie})  one may  approximately take $r$ out of the integral at the point $e=\sigma_R$.
If $e>e_{th}+\sigma_I$ this gives
\be
r(e)\simeq\frac{\sigma_I}{\pi}L(e).\la{del}
\ee
Thus the input transform itself looks similar to the output response. 

At large $|\sigma_R|$ the asymptotics
\be
L(\sigma_R)\simeq \sigma_R^{-2}\sum\!\!\!\!\!\!\!\int\,r(e)de=
\sigma_R^{-2}\langle Q|Q'\rangle.\la{asym}
\ee
can be used to calculate the input transform (see the sum rule~(\re{sum})). However, one 
may prove that the asymptotics (\re{asym}) is valid when and only when $e^2r(e)$ tends
to zero as $e$ tends to infinity. Moreover one can notice that 
at large $|\sigma_R|$ the regime (\re{asym}) occurs when
$\sigma_R^{-2}\langle Q|Q'\rangle$ is much higher then $\pi r(\sigma_R)/\sigma_I$ while
the regime (\re{del}) takes place in the opposite case.

%***********************************************************************
\subsection{Remarks on particular cases}{\label{sec:PARTICULAR}
%***********************************************************************

Equation (\re{re}) corresponds to a non--relativistic calculation. It is implied that 
the center of mass subspace is
separated out and states entering the response are defined in the relative motion subspace. 
The operators $\hat O$ and $\hat O'$ are `internal' operators acting in the latter subspace. 
The corresponding matrix elements are taken between internal initial and 
final states with given momenta, with the momentum conservation delta function omitted. 
The energy $E$ pertains to the relative motion subspace.
In some cases, in order to take into account relativistic corrections one 
needs to consider
responses defined in such a way, but 
with the energy conservation delta function that corresponds to relativistic
kinematics. Up to some factors such responses may be expressed in the same form 
as the non--relativistic responses (\re{re}) 
so that the present considerations are applicable (see~\cite{EFROS:2005}).   
 
Performing closure to get (\re{s}) we have assumed that the transition 
operators $\hat O$ and $\hat O'$ do not include dependence on the final--state energy $E_\gamma$. 
But in some cases, like e.g. in the electrodisintegration problem 
(see~\sref{sec:ELECTRON}) these operators are of 
the form $\sum_\alpha G_\alpha \hat{O}_\alpha \la{de}$, where $G_\alpha$ are form factors 
that depend on $E_\gamma$. 
In that case the response functions in the cross section may be written as
\begin{equation}\label{part}
R=\sum_\alpha G_\alpha ^2 r_{\alpha\alpha} +\sum_{\alpha<\beta}G_\alpha 
G_\beta (r_{\alpha\beta}+r_{\alpha\beta}^*),\la{GG}
\end{equation}
where $r_{\alpha\alpha}$ are functions of the form (\re{re}) with $\hat{O}=\hat{O}'=\hat{O}_\alpha$, and
$r_{\alpha\beta}$ are again functions of that form  with $\hat{O}=\hat{O}_\alpha$, $\hat{O}'=\hat{O}_\beta$. 
All the $r_{\alpha\beta}$ are 
independent of $E_\gamma$ and are calculated as described above. Extra factors
depending on $E_\gamma$
may also arise when the continuity equation is used to express some parts of the
electric multipole operators of the current in terms of the density multipole operators 
\cite{BACCA:2007}.
In this case one may act in the same way.

The operators $\hat{O}$ and $\hat{O}'$ may include a more complicated dependence on $E_\gamma$.
Then
one may proceed as follows~\cite{REISS:2003}. One replaces the operators $\hat{O}(E_\gamma)$ and $\hat{O}'(E_\gamma)$
with those $\hat{O}(\bar{E})$ and $\hat{O}'(\bar{E})$ depending on a fixed energy $\bar{E}$ rather than 
$E_\gamma$. The response $r(E)$ is replaced with the arising subsidiary response $r'(E,\bar{E})$, 
and the latter is calculated as 
described above. If $r(E)$ in some $E$ range is of interest then $r'(E,\bar{E})$
is to be found for a set of $\bar{E}$ values lying in the same range. After that $r(E)$ can
be obtained via interpolation as $r'(E,\bar{E}=E)$.

%*********************************************************
\subsection{Inversion of the LIT}\label{sec:INV}
%*********************************************************

As already pointed out, the main advantage of the LIT method to study reactions
is that one avoids to solve the many--body scattering problem. One solves instead,
with bound--state methods, equations of the form (\ref{eq1}). The knowledge of those solutions 
leads to  the LIT  of the function $r$ of interest. A crucial part of the method 
is then the inversion of this integral transform. In~\sref{sec:DISCONT} we have seen how to 
obtain the 
discrete contributions to $r$, so one is left with the most challenging part represented by
the solution of~(\ref{ie}). The inversion of this integral transform
has to be made with care, since it
is unstable with respect to high frequency oscillations  as explained
in~\sref{sec:WIDTH}.  
%Let us consider a well defined $r(\nu)$ to which we add a
%high-frequency term $\Delta r^\Omega (\nu)$. The latter leads
%to an additional $\Delta L^\Omega (\sigma_R,\sigma_I)$ in the transform. For any
%amplitude of the oscillation $\Delta L^\Omega$ decreases with increasing $\Omega$.
%This means that for some vale of $\Omega$  $\Delta L^\Omega$ may be 
% smaller than the size of the errors in the
%calculation. Therefore in this case  $\Delta_\Omega R$ cannot be discriminated. By reducing
%the error in the calculation one can push the frequency of the undiscriminated $\Delta_\Omega
%R$ to higher and higher values. However, even if the excluded frequency
%range is physically not relevant, 
%one cannot simply find a solution
%of the response by application of the inverse operator, since the unphysical
%oscillations cannot be separated from the solution. 
%they can never be excluded. 
In this sense the LIT inversion problem
belongs to the class of so--called ill--posed problems. 

In~\cite{TIKHONOV:1977} the mathematical aspects
of such problems are studied. It is also described how 
solutions of ill--posed problems can be obtained adopting  
a regularization scheme. Therefore, following ~\cite{TIKHONOV:1977}
we have implemented a specific regularization in inverting the Lorentz transforms
obtained in all the calculations reported in this review. This has led to very safe inversion
results. Alternative inversion methods are discussed in~\cite{ANDREASI:2005}. They
can be advantageous in case of response functions with more complex structures.
Up to now, however, such complex structures have not been encountered
in actual LIT applications, since
the various considered $r(e)$ have normally (i) a
rather simple structure, where essentially only a single peak of $r(e)$ has to be
resolved, or (ii) a more complicated structure, which however can be
subdivided into a sum of simply structured responses, where the various LITs can be 
inverted separately. The second case (ii) has already been encountered in~\cite{EFROS:1994} and 
is also discussed  in~\sref{sec:MULT}.
% where the inclusive longitudinal deuteron electron response $R(e)$ was calculated at
%a constant momentum transfer. 
%There it was necessary to separate out the Coulomb
%monopole and quadrupole transitions,
%which lead to a shoulder of the corresponding $R(e)$ at the break--up threshold,
%while the rest of the response shows the typical quasi--elastic peak structure.

The present `standard' LIT inversion method consists in the following ansatz for
the response function
\begin{equation}
r(e') = \sum_{n=1}^{N_{\it max}} c_n \chi_n(e',\alpha_i) \,
\label{sumr}
\end{equation}
with $e'=e-e_{th}$, where $e_{th}$ is the threshold energy for the break--up
into the continuum. The $\chi_n$ are given functions with nonlinear parameters $\alpha_i$.
A basis set frequently used  for LIT inversions is
\begin{equation}
\label{bset}
\chi_n(\epsilon,\alpha_i) = \epsilon^{\alpha_1} \exp(- {\frac {\alpha_2 \epsilon} {n}}) \,.
\end{equation}
In addition also possible information on narrow levels
could be incorporated easily into the set $\chi_n$.
Substituting such an expansion into the right hand side of~(\ref{ie}) (here too we omit
to write the $\sigma_I$ dependence) one obtains
\begin{equation}
L(\sigma_R) =
\sum_{n=1}^{N_{\it max}} c_n \tilde\chi_n(\sigma_R,\alpha_i) \,,
\label{sumphi}
\end{equation}
where
\begin{equation}
\tilde\chi_n(\sigma_R,\alpha_i) =
\int_0^\infty de' {\frac {\chi_n(e',\alpha_i)} {(e'-\sigma_R)^2 + \sigma_I^2}}
\,\,.
\end{equation}
For given $\alpha_i$ the linear parameters $c_n$ are determined from a least--square best fit of
$L(\sigma_R)$ of equation~(\ref{sumphi}) to the calculated
$L(\sigma_R)$ of equation~(\ref{ov}) for a number of $\sigma_R$ points
much larger than $N_{\it max}$. 

If one uses the basis set~(\ref{bset}), in performing the
least--square fit of $L(\sigma_R)$ one should vary the nonlinear parameter 
$\alpha_2$ over a rather large range. If no Coulomb interaction acts between fragments,
with the proper $\alpha_1$ the factor $\epsilon^{\alpha_1}$ in~(\ref{bset}) 
reproduces the energy dependence at threshold of a response function, or a form factor.
This correct $\alpha_1$ value can be obtained from the case when the short--range interaction between 
fragments is absent. Therefore, to have good quality results in the threshold region one may adopt 
this $\alpha_1$ value. 
In the usual case when the lowest open channel is that of two charged fragments the factor $\epsilon^{\alpha_1}$ 
in~(\ref{bset}) may also be replaced with the corresponding Gamow factor to take into account the 
Coulomb interaction. Alternatively, one may include the parameter $\alpha_1$ in the 
least--square fit. However, it should be noted that the least--square procedure 
may not be sensitive enough to a contribution from the threshold region to ensure the proper behaviour.

For every value of $N_{\it max}$ the overall best fit is 
selected and then the procedure is repeated for $N'_{\it max}=N_{\it max}+1$ till
a stability of the inverted response is obtained and taken as inversion
result. A further increase of $N_{\it max}$ will eventually reach a point, where the
inversion becomes unstable leading typically to random oscillations. The
reason is that $L(\sigma_R)$ of equation~(\ref{sumphi}) is not determined
precisely enough so that a randomly oscillating $r(e)$ leads to a better
fit than the true response. If the accuracy in the determination of $L(\sigma_R)$
from the dynamic equation is increased then one may include more basis functions in the expansion
(\ref{sumphi}).
%If one takes the basis set~(\ref{bset}) one should vary
%the nonlinear parameter $\alpha_2$ over a rather large range, while one can
%determine $\alpha_1$ from the possibly known threshold behavior of $r$ or 
%vary it within a reasonable range.
%For every value of $N_{\it max}$ the overall best fit is 
%selected and then the procedure is repeated for $N_{\it max}=N_{\it max}+1$ till
%a stability of the inverted response is obtained and taken as inversion
%result. A further increase of $N_{\it max}$ will eventually reach a point, where the
%inversion becomes unstable leading typically to random oscillations. The
%reason is that $L(\sigma_R)$ of equation~(\ref{sumphi}) is not determined
%precisely enough so that a randomly oscillating $r(e)$ leads to a better
%fit than the true response. 

It is evident that the number of functions $N_{max}$
plays the role of a regularization parameter and has to be chosen within the
above mentioned stability region. Normally such a stability region is reached
without greater problems. However, in principle it can happen that one does not
find a stable result. In such cases one can either try to improve the precision of the
calculated $L(\sigma_R)$ or use different basis sets, or enlarge
the flexibility of the basis set~(\ref{bset}) by taking e.g. $\{\chi^\beta(e,\alpha_i), \beta_{\it min}
\le \beta \le \beta_{\it max} \}$, 
\begin{equation} 
\label{bset2}
\chi_1^\beta(e,\alpha_i) = e^{\alpha_1} \exp(- \alpha_2 e),
\,\,\,\,\,\,\chi_n^\beta(e,\alpha_i) = e^{\alpha_1} \exp(- {\frac {\alpha_2 e}
{n \beta}}), \,\,\,\,\,\, n\geq 2
\end{equation}
with suitable values for $\beta_{\it min}$ and $\beta_{\it max}$.

In case that the inversion leads to a
response that exhibits an unexpected structure it is useful to decrease the
parameter $\sigma_I$ in order to have a better resolution in the transform.

We mention that according to what is said above any basis set may be used 
only up to a certain value of $N_{max}$. 
Information which can be parametrized only via $\chi_n$ with $n>N_{max}$ is lost anyway.
Therefore it is important to work with a basis set, where a relatively small $N_{max}$
leads to a high--quality estimate of $r(e)$. Varying the
nonlinear parameters $\alpha_i$ of $\chi_n$ (and also $\beta$ in case of $\chi_n^\beta$)
is equivalent to using many different basis sets. Usually
it is no problem to find a proper set.

The reliability of the inversion method outlined above has already been
proved for various cases~\cite{EFROS:1994,LAPIANA:2000,GOLAK:2002}. A discussion about two additional 
interesting cases can be found in~\sref{sec:NUM}.

%**********************************************************************
\subsection{Calculation of the LIT via the eigenvalue method}
\label{sec:EIGENVALUE}
%***********************************************************************

As we have seen in~\sref{sec:LIT_TH} the LIT method can be used to
reformulate a scattering
problem as a Schr\"odinger--like equation with source terms
which depend on the kind of reaction under consideration. These equations, 
which
we will call the `LIT equations` are essentially the same for any reaction, 
differing by the source 
term i.e. by their right hand side (see~(\ref{eq1}), (\ref{eq2}) and (\ref{Eqsb})). 
In all cases the asymptotic boundary conditions are bound--state--like.
Consequently the solutions of these equations can be found with similar
methods as for the
bound--state wave functions. Therefore the problem is much simpler than solving
the Schr\"odinger equation for the continuum spectrum.
%Bound state solutions of the Schr\"odinger equation can be found in 
%different ways. One can search for a direct numerical solution of the differential
%equations in coordinate space or of the integro--differential equations in 
%momentum space. For example, in the three--body case this corresponds to Faddeev 
%calculations (for the solution of the LIT equation with the Faddeev approach 
%see~\cite{MARTINELLI:1995}).Alternatively,  one can use expansions  of the LIT functions
%on some basis set of localized functions (bound-state-like boundary condition). 
%These expansion methods become more and more  advantageous with respect
%to the former one, with increasing particle number. For this reason, and because it has been used
%for the vast majority of the LIT applications we discuss it in the following in greater detail.
Quantum mechanical bound--state problems may be solved in 
various ways. One can search for a direct solution of the dynamic equations 
(differential, integro--differential or integral) either in the coordinate 
or momentum representation. Alternatively, one can employ expansions over some 
basis set of localized functions. These expansion methods become more and more 
advantageous than the above mentioned ones as the number of particles increases. 
Methods of all these types may be employed to calculate the localized LIT functions
as well. In the following we discuss in greater detail the expansion methods to 
find the LIT functions. This is done both for the just mentioned reason and because they 
have been used for the vast majority of the LIT applications. (For a solution of the 
three--body LIT equation with the direct Faddeev--type approach see~\cite{MARTINELLI:1995}).

The truncation of the basis set converts the bound--state Schr\"odinger equation into a
matrix eigenvalue problem and the LIT equations into a set of linear equations.
These equations can be solved with various iteration methods
and also Gauss type non--iteration ones. Such strategies have the drawback that
one should solve these equations many times,
%Thus in principle one can solve the LIT equation through matrix
%inversion. Although feasible this strategy is not recommended since
%one should solve the LIT equation many times, 
as many as the number of
$\sigma_R$ values that one needs for a proper inversion of the transform.
In this section and in the following one we present two better 
strategies for calculating
$L(\sigma)$. The first strategy, which we call the eigenvalue method and 
which
we discuss here, involves the full diagonalization of the Hamiltonian 
matrix and expresses
the LIT through its eigenvalues. This method is instructive from a theoretical point 
of view.
%The spectral analysis provided by it leads to important conclusions
%about the way the continuum is treated in the LIT method. It also points 
%into
%its power and limitations.

Regardless of the reaction under consideration and the process
that one wants to study the LIT method requires the calculation of the overlap
(see~(\ref{ov})--(\ref{psi2}), or~(\ref{texc})--(\ref{ppsi2}), 
or~(\ref{ab})--(\ref{ppsib}))
\begin{eqnarray}\label{L_ssp}
L(\sigma_R,\sigma_I)&=&\langle\tilde\Psi|\tilde\Psi'\rangle\nonumber\\
&=&\bra Q | \frac{1}{(\hat{H}-E_0-\sigma_R+i\sigma_I )}
           \frac{1}{(\hat{H}-E_0-\sigma_R-i\sigma_I )}
| Q'\ket\,,
\end{eqnarray}
where  $|Q'\ket$ and $|Q\ket$ contain the information
about the kind of reaction one is considering.
%The bound--state--like boundary conditions implemented in the basis set imply that
%$H$ has only a discrete spectrum. Only eigenstates $|\nu\rangle$ of $H$ with eigenvalues
%$\epsilon_{\nu}<0$ are real physical states. All other states are non physical 
%discrete eigenstates that result from the imposed boundary conditions.
%%Denote by $\{|\nu\ket\,\epsilon_{\nu}\}$ the set of eigenvalues and
%%eigenvectors of $H$. 
%The eigenstates $|\nu\ket$ are a complete set of basis functions in the truncated space
%and can be used for expanding the LIT states,
%\begin{eqnarray} \label{psi_nu}
 %   |\widetilde{\psi} \ket &=&
 %      \sum_{\nu \neq Phys}\frac{1}{\epsilon_{\nu}-E_0-\sigma_R+i\sigma_I}
 %      \bra \nu | Q \ket |\nu \ket\\
%    |\widetilde{\psi}' \ket &=&
%       \sum_{\nu \neq Phys}\frac{1}{\epsilon_{\nu}-E_0-\sigma_R+i\sigma_I}
%       \bra \nu | Q ' \ket |\nu \ket \;.
%\end{eqnarray}
%%As the LIT only considers transition into the continuum 
%Since we consider here only the continuum contribution $L_C$ the sums 
%in~(\ref{psi_nu})
%are restricted to the {\it non--physical} states, while the physical ones
%(bound states) are excluded. Substituting~(\ref{psi_nu}) into
%(\ref{L_ssp}) yields the following expression for the LIT,
Seeking $|\tilde{\Psi}\rangle$ and $|\tilde{\Psi}'\rangle$ as expansions over $N$ localized basis states, 
it is convenient to 
choose as basis states $N$ linear combinations of states that diagonalize the Hamiltonian matrix. 
We denote these combinations $|\varphi_\nu^N\rangle$ and the eigenvalues $\epsilon_\nu^N$. 
The index $N$ is to remind that they  both depend on $N$. 
If the continuum starts at $E = E_{th}$ then at sufficiently high $N$ the states $|\varphi_N\rangle$ having 
$\epsilon_\nu^N< E_{th}$ will represent approximately the bound states. The other states  will 
gradually fill in the continuum as $N$ increases. The expansions of our localized  LIT functions 
read as 
\begin{equation}\label{nu1}
|\tilde{\Psi}\rangle =\sum_\nu^N\frac{\langle\varphi_\nu^N|Q\rangle}
{\epsilon_\nu^N-E_0-\sigma_R-i\sigma_I}|\varphi_\nu^N\rangle\,,
\end{equation}
\begin{equation}\label{nu2}
|\tilde{\Psi}'\rangle=\sum_\nu\frac{\langle\varphi_\nu^N|Q'\rangle}
{\epsilon_\nu^N-E_0-\sigma_R-i\sigma_I}|\varphi_\nu^N\rangle\,.
\end{equation}	
Substituting  (\ref{nu1}) and (\ref{nu2}) into (\ref{L_ssp}) 
yields the following expression for the LIT,
\begin{equation}\label{L_epsnu}
L(\sigma_R,\sigma_I)=\sum_{\nu} \frac{\bra Q| \varphi_\nu^N \ket \bra \varphi_\nu^N | Q' \ket}
                    {(\epsilon_{\nu}^N-E_0-\sigma_R)^2+\sigma_I^2}  \;,
\end{equation}
and for $|Q '\ket=|Q \ket$
\begin{equation}\label{L_epsnu_diag}
L(\sigma_R,\sigma_I )=\sum_{\nu} \frac{ |\bra \varphi_\nu^N | Q \ket|^2}
                     {(\epsilon_{\nu}^N-E_0-\sigma_R)^2+\sigma_I^2}\;.
\end{equation}
From~(\ref{L_epsnu}) and (\ref{L_epsnu_diag}) it is clear that
$L(\sigma_R,\sigma_I)$ is a sum of Lorentzians. The inversion of the LIT contributions
from the states with $\epsilon_\nu^N < E_{th}$ gives the discrete part of a response function, 
whereas the inversion of the rest gives its continuum part. The spacing between 
the corresponding eigenvalues with $\epsilon_\nu^N > E_{th}$ depends on $N$ and in a given energy 
region the density of these eigenvalues increases with $N$. 
(Since the extension of the basis states grows with $N$, this resembles the increase 
of the density of states in a box, when its size increases.) 
For the reliability of the inversion procedure one needs to reach the regime when one has 
a sufficient number of levels $\epsilon_\nu^N$ within the $\sigma_I$ extensions of the Lorentzians,
as will be  illustrated in~\sref{sec:ALPHA}.
%with fixed width $\sigma_I$ centered around the discrete continuum eigenvalues of $H$. The spacing
%between the eigenvalues $\epsilon_{\nu}$ depends on the boundary conditions. In particular
%the density of the states increase with the size of the `box'. The choice of a proper density of 
%states is relevant for the reliability of the inversion procedure, as will be illustrated in 
%\sref{sec:ALPHA}.

%\begin{figure}[h]
%\centerline{
%\rotatebox{0}{
%\resizebox{!}{!}
%{\epsfig{file=lit_sigmaI_4He_MT13_photo_totINV.eps}}}}
%\vskip 1cm
%\caption{\label{Fig1:sigmaI_effect}
%         The effect of $\sigma_I$ on the LIT.}
%\end{figure}

%************************************************************************
\subsection{Calculation of the LIT via the Lanczos algorithm}
\label{sec:LANCZOS}
%************************************************************************

In this section a second strategy to obtain the LIT is described.
This utilizes the Lanczos algorithm ~\cite{LANCZOS:1950,GOLUB:1983} to express the LIT as a continuous fraction. 
%using making use  explain why Lanczos approach is the best way for 
%calculating the LIT. It is formulated in terms of matrix vector
%multiplications and is therefore very effective when dealing with sparse
%matrices. Moreover 
It turns out that for obtaining an accurate LIT only a
relatively small number of Lanczos steps are needed.
As the number of particles in the system under consideration increases the
number of basis states grows up very rapidly  and the Lanczos approach seems at present 
the only viable method to calculate the LIT.

The motivation to use the Lanczos approach comes from the observation that from the computational 
point of view the calculation of the LIT is much more 
complicated and demanding than finding the ground state wave function of 
an $A$--particle system.
In fact, to  obtain the ground--state wave function one needs only to find the lowest
eigenvector of the Hamiltonian matrix. On the contrary, as it is clear from~(\ref{L_epsnu}) 
and (\ref{L_epsnu_diag}), 
the complete spectra of $\hat H$ over a wide energy range should be known
to calculate the LIT. Therefore it is no surprise that the
computational time and the memory needed to calculate $L(\sigma)$ have been the
limiting factors in extending the LIT method for systems with more than four
particles.
It turns out that these obstacles can be overcome if 
the LIT method is reformulated using the Lanczos algorithm.
%This algorithm has enormous advantages in extracting the 
%extreme eigenvalues and the related eigenvectors of large symmetric matrices as well
%as in solving large linear equation problems. Utilizing its power to the
%calculation of the LIT has represented  a major step towards the application of the  
%LIT method to larger systems.
Following reference~\cite{MARCHISIO:2003}, in this section it is shown
how this can be done. 

To this end we assume that the source states $|Q \ket$ and $|Q '\ket$ are
real and 
rewrite the LIT in the following form
\begin{equation} \label{lorelan1}
 L(\sigma)=-\frac{1}{\sigma_I} \mbox{Im} 
\left\{ \bra Q | \frac{1}{\sigma_R+i\sigma_I+E_0-
\hat{H}} | Q'\ket \right\}\:\mbox{.}
\end{equation}
A similar relation connects the response function $r(e)$ to 
the Green's function
\begin{equation} \label{lorelan2}
  r(e)=-\frac{1}{\pi} \mbox{Im} \left\{\lim_{\eta\to 0} 
  G(e + i \eta + E_0)\right\}
\:;\hspace{6mm} 
  G(z)=\bra Q | \frac{1}{z-\hat{H}} | Q' \ket \:,
\end{equation}  
provided that $z = e + i \eta$ is replaced by $\sigma_R + i \sigma_I$.
This is not surprising since the properly
normalized Lorentzian kernel
is one of the representations of the  $\delta$--function and $\sigma_I/\pi\,L(\sigma_R)  
\rightarrow r(\sigma_R)$ for $\sigma_I\rightarrow 0$. In condensed 
matter calculations~\cite{DAGOTTO:1994,HALLBERG:1995,KUEHNER:1995} the Lanczos
algorithm  has
been applied to the calculation of the Green function with a small value of 
$\eta$, and its imaginary part has been interpreted as $r(e)$
directly. This can be done if the spectrum is discrete (or discretized) and
$\eta$ is sufficiently small. In our case we have a genuine continuum
problem and we want to avoid any discretization, therefore we calculate
$L(\sigma_R)$ in the same way, i.e. with finite $\sigma_I$ using the 
Lanczos algorithm, but then we antitransform $L(\sigma_R)$ in order to 
obtain $r(e)$.

This section is divided into two parts. First 
%we briefly present the Lanczos algorithm. Then 
we discuss the application of the Lanczos method for 
a symmetric LIT, i.e. for the $|Q '\ket=|Q \ket$ case. This case is somewhat
simpler then the general case discussed in the second part and can be used
as a starting point for it.

%****************************************************************
\subsubsection{The Lanczos algorithm.}\label{sec:LANCZOSALG}
%****************************************************************

The Lanczos algorithm~\cite{LANCZOS:1950,GOLUB:1983} is a technique to solve
large, sparse, hermitian eigenvalue problems. The method involves partial {\it
tridiagonalizations} of the matrix $\hat H$ without generating, however,
intermediate full submatrices. Equally important, is that the extreme
eigenvalues of $\hat H$ tend to emerge long before the tridiagonalization is
complete. 
In order apply the Lanczos algorithm for
calculating the ground state energy 
of an Hamiltonian matrix $\hat H$ one begins by choosing a starter
vector $|\phi_0 \ket$ (also called the Lanczos {\it pivot}), which must have a non--zero
overlap with the ground state $|\psi_0 \ket$. One then proceeds
building the Lanczos orthonormal basis $\{|\phi_i 
\ket \mbox{,}i=0,\ldots \mbox{,} n-1\}$ by applying the 
Lanczos algorithm recursively    
\begin{equation} \label{lana}
  |w_{i+1}\ket =\hat{H}|\phi_i\ket - 
   a_i|\phi_i\ket -b_i|\phi_{i-1}\ket \:,
\end{equation}
where the Lanczos coefficients $a_i$ and $b_i$ are defined as
\begin{equation}
a_i=\bra \phi_i |\hat{H}|\phi_i \ket;\,\,\,
b_i=\parallel |w_i \ket \parallel;\,\,\,
b_0=0;\,,
\end{equation}
and
\begin{equation}
|\phi_i\ket=\frac{|w_i\ket}{b_i};\,\,\,
|\phi_{-1}\ket =0;\,\,\,
\bra \phi_i |\phi_j \ket =\delta_{ij}\mbox{.}
\end{equation}
%The Lanczos coefficients $a_i$ and $b_i$ are defined as
%\begin{equation}
%   a_i=\bra \phi_i |\hat{H}|\phi_i \ket \:\mbox{,} \hspace{3mm} 
%   b_i=\parallel b_i|\phi_i \ket \parallel \:\mbox{,} \hspace{3mm}
%  b_0=0 \:\mbox{.} 
 % \end{equation}
If one applies $M$ Lanczos steps to an $M\times M$ matrix,
then the original 
matrix is reduced into a tridiagonal form. As mentioned above, the power of
the Lanczos algorithm lays in the fact that even after $ m << M$ steps the
eigenvalues  
of the $m \times m $ tridiagonal Lanczos matrix are good approximations of the
extreme eigenvalues of $\hat H$. These eigenvalues converge rapidly with
increasing  number of Lanczos steps. 
Working with finite precision, it is necessary to 
recall that at each step there are 
round--off errors so that after a certain number of steps
the Lanczos vectors loose their orthogonality. In order to 
obtain accurate results this loss has to remain small, otherwise
one has to apply an additional reorthogonalization procedure.

%******************************************************************************
\subsubsection{The symmetric case $|Q '\ket=|Q \ket$.}\label{sec:LANCZOS_SYM}
%******************************************************************************

The application of the Lanczos algorithm to the symmetric LIT,
\begin{equation} \label{lorelan1sym}
 L(\sigma)=-\frac{1}{\sigma_I} \mbox{Im} 
\left\{ \bra Q | \frac{1}{\sigma_R+i\sigma_I+E_0-
\hat{H}} | Q \ket \right\}\;,
\end{equation}
starts by choosing the normalized source vector
\begin{equation} \label{start}
  |\phi_0\ket = \frac{|Q \ket}{\sqrt{\bra Q| Q \ket}},\la{pivot}
\end{equation}
as the {\it pivot} for the Lanczos basis
and setting $z=E_0+\sigma_R+i\sigma_I$. With the help of these definitions
one can rewrite $L(\sigma)$ as 
\begin{equation} \label{loreins}
  L(\sigma)= - \frac{1}{\sigma_I} \bra Q | Q \ket  
            \mbox{Im} \left\{\bra \phi_0 |
                      \frac{1}{z-\hat{H}}|\phi_0\ket\right\} \;.
\end{equation}
Therefore the LIT depends on the matrix element
\begin{equation}
  x_{00}= \bra \phi_0 |\frac{1}{z-\hat{H}}|\phi_0\ket \:.
\end{equation}
This matrix element can be calculated applying Cramer's rule to the solution of the 
linear system~\cite{DAGOTTO:1994,FULDE:1991}
\begin{equation}
\label{linsys0}
  \sum_n (z-\hat{H})_{mn}x_{n0}=\delta_{m0}\,,
\end{equation}
(where $x_{n0}= \bra \phi_n |\frac{1}{z-\hat{H}}|\phi_0\ket$) which arises from the identity 
\begin{equation}
  (z-\hat{H})(z-\hat{H})^{-1}=I\,,
\end{equation}
on the Lanczos basis \{$|\phi_i\ket \mbox{;} \,\,i=0\mbox{,}\ldots\mbox{,} n-1\}$. 
Using Cramer's rule one gets
\begin{equation}
 x_{00}=\frac{det (M_{00})}{det(z-\hat H)}\,,
\end{equation}
where 
\begin{equation}\label{m00}
M_{00}=\left(\begin{array}{cccc}
1      & -b_{1}    & 0      &\cdots  \\
0      & z-a_{1}   & -b_{2} &\cdots   \\
0      & -b_{2}    & z-a_{2}&\cdots   \\
\vdots &\vdots      &\vdots   & \ddots
\end{array}\right)\:\mbox{,}
\end{equation}
and the $a_n$ and $b_n$ are the Lanczos coefficients.
Defining $D_i$ as the matrix obtained by removing the first $i$ rows and
$i$ columns from $(z-\hat H)$, one  sees that $D_0=(z-\hat H)$,
\begin{equation}
  det(M_{00})=det(D_1)
\end{equation}
and
\begin{equation}\label{detD0}
  det(D_0)=(z-a_0) det(D_1) - b_1^2 det(D_2)\;.
\end{equation}
Thus one has
\begin{equation}\label{x00step1}
 x_{00}=\frac{1}{z-a_0-b_1^2 \frac{det(D_2)}{det(D_1)}}
\end{equation}
The recurrence relation~(\ref{detD0}), is valid for any submatrix $D_i$.
Substituting the appropriate expression for $det(D_1)$ one gets
\begin{equation}\label{x00step2}
 x_{00}=\frac{1}{z-a_0-\frac{b_1^2}{z-a_1-b_2^2 \frac{det(D_3)}{det(D_2)}}}
\end{equation}
In this way one is able to write $x_{00}$ as a continued fraction containing 
the Lanczos coefficients $a_i$ and $b_i$, 
\begin{equation} \label{conti}
  x_{00}=\frac{1}{z-a_{0}-\frac{b^{2}_{1}}{z-a_{1}-
                  \frac{b^{2}_{2}}{z-a_{2}-b^{2}_{3}\ldots}}} \:,
\end{equation}
and thus also the LIT becomes a function of the Lanczos coefficients
\begin{equation} \label{lorelan3}
L(\sigma )=-\frac{1}{\sigma_{I}}\mbox{Im}\left \{ \frac
          {\bra Q | Q \ket
           }{z-a_{0}-\frac{b^{2}_{1}}{z-a_{1}-\frac{b^{2}_{2}}{
          z-a_{2}-b^{2}_{3}\ldots}}}\right \} \:.
\end{equation}
This result illustrates how, using the Lanczos method, 
$L(\sigma )$ is determined 
without solving the LIT equations, i.e. without inverting 
or diagonalizing the Hamiltonian matrix.
%******************************************************************
\subsubsection{The general case.}\label{sec:LANCZOS_GEN}
%******************************************************************

Now we turn to the application of the Lanczos method to the general LIT 
case, i.e. when $|Q \ket\neq|Q '\ket$. One can reformulate the problem in two different ways.

\noindent{\it Reformulation I}

The LIT 
\begin{equation} 
 L(\sigma)=-\frac{1}{\sigma_I} \mbox{Im} 
\left\{ \bra Q | \frac{1}{z-\hat{H}} | Q' \ket \right\}\;,
\end{equation}
is rewritten as a sum of four symmetric terms through the relation
\begin{eqnarray}
\label{offd}
 \lefteqn{\bra Q |\frac{1}{z-\hat{H}}| Q' \ket =} \cr
    & & \frac{1}{4}(\bra Q |+ \bra Q'  |)\frac{1}{z-\hat{H}}(| Q \ket+  |Q' \ket) - 
        \frac{1}{4}(\bra Q |- \bra Q'  |)\frac{1}{z-\hat{H}}(| Q \ket-  |Q' \ket) \cr
    &-& \frac{i}{4}(\bra Q |+i\bra Q'  |)\frac{1}{z-\hat{H}}(| Q \ket+i |Q' \ket) \cr
    &+& \frac{i}{4}(\bra Q |-i\bra Q'  |)\frac{1}{z-\hat{H}}(| Q \ket-i |Q' \ket)  \:.
\end{eqnarray}
In this form one can apply the results of the previous  section. 

\noindent{\it Reformulation II}

Although quite elegant the previous reformulation requires 
four separate LIT calculations.
%may be somewhat problematic from a practical point of view. 
%In exclusive reactions where one would like to study the cross section as a
%function of the relative momentum between the outgoing fragments
%one should repeat the calculation many times.
%Each time one should calculate four new sets of Lanczos
%coefficients associated with the state $|Q '\ket$
%=  \hat{\mathcal{V}}^{\alpha}|\phi_{\alpha}\ket $.
%This has to be compared with a single inversion of the Hamiltonian matrix.
In~\cite{MARCHISIO:2003} a powerful alternative reformulation has been derived,
which allows the calculation of all the non--diagonal 
LITs, just with one set of Lanczos coefficients.
Writing the LIT in the form (see (\ref{kp}))
\begin{equation} \label{loreaga}
  L(z)
      =-\frac{1}{2i\sigma_I} \bra Q |
       \left(\frac{1}{z-\hat{H}}-\frac{1}{z^*-\hat{H}}\right) 
       | Q' \ket  \:\mbox{,}
\end{equation}
choosing as previously the {\it pivot}~(\ref{start})
%\begin{equation} \label{pivot}
%  |\phi_0\ket = \frac{|Q \ket}{\sqrt{\bra Q| Q \ket}},
%\end{equation}
and assuming that after $n$ steps the Lanczos vectors approximately 
form a complete basis, i.e. 
$\sum_{i=0}^{n-1}|\phi_i\ket\bra\phi_i|\simeq 1$,
one gets
\begin{eqnarray} \label{lorefast}
L(\sigma) &=& \frac{i}{2\sigma_I} \sum_{i=0}^{n-1} 
          \bra Q  |\phi_i \ket 
          \left[ \bra \phi_i|\frac{1}{z  -\hat{H}} | Q '\ket
                -\bra \phi_i|\frac{1}{z^*-\hat{H}} | Q '\ket \right] 
\cr
%          &=& -\frac{\sqrt{\bra Q | Q \ket}}{\sigma_I} 
%          \sum_{i=0}^{n-1} \bra Q' | \phi_i \ket 
%          \mbox{Im}\left\{\bra \phi_i|\frac{1}{z-\hat{H}}|\phi_0\ket\right\}
%\\
          &=&-\frac{\sqrt{\bra Q' | Q' \ket}}{\sigma_I} \sum_{i=0}^{n-1} 
          \bra Q |\phi_i \ket\mbox{Im}\{x_{i0}\} \:,
\end{eqnarray}
where the {\it pivot} $|\phi_0 \ket$ is defined as in~(\ref{pivot}) with $|Q\ket\rightarrow |Q'\ket$
and the matrix elements
\begin{equation}
   x_{i0} = \bra \phi_i|\frac{1}{z-\hat{H}}|\phi_0\ket \:\mbox{,}
\end{equation}
can be written as continued fractions of 
the Lanczos coefficients in a similar way as in~(\ref{conti}). One can notice
that $\mbox{Im}\{x_{i0}\}= \mbox{Im}\{x_{0i}\}$, therefore
in the following we describe a simple algorithm for a recursive calculation
of all the $x_{0i}$ starting from $x_{00}$. 

\Eref{linsys0} can be generalized to 
\begin{equation} 
\label{linsyst01}
\sum_n(z-\hat{H})_{mn} x_{ni} = \delta_{mi} \:\mbox{.}
\end{equation} 
From this one can obtain all the $x_{0i}$ matrix elements
for $n=0$ and $m=i$.
The matrix element $x_{01}$ is obtained solving the linear system
\begin{equation}
\label{linsys01}
\left( \begin{array}{cccc}
z-a_{0}  & -b_{1}  & 0       & \cdots\\
 -b_{1}  & z-a_{1} & -b_{2}  & \cdots \\
  0      & -b_{2}  & z-a_{2} & \cdots \\
\vdots    &\vdots    & \vdots   & \ddots
\end{array}\right) \left( \begin{array}{c}
x_{01}\\
x_{11}\\
x_{21}\\
\vdots
\end{array}\right) =\left( \begin{array}{c}
0\\
1\\
0\\
\vdots
\end{array}\right) \:\mbox{.}
\end{equation}
Using Cramer's rule one has
\begin{equation}
\label{cram01}
x_{01}=\frac{det(M_{01})}{det(z-\hat{H})}\:\mbox{,}
\end{equation}
where $M_{01}$ has the following form
\begin{equation}
\label{m01}
M_{01}=\left(\begin{array}{cccc}
0     & -b_{1}    & 0      &\cdots  \\
1     & z-a_{1}   & -b_{2} &\cdots   \\
0     & -b_{2}    & z-a_{2}&\cdots   \\
\vdots &\vdots      &\vdots   & \ddots
\end{array}\right)\:\mbox{.}
\end{equation}
Since
\begin{eqnarray}
\label{determinants} 
det(M_{01})&=&b_1 det( D_2) \:;\\
det(z-\hat{H})&=&(z-a_{0})det(D_{1})-b^{2}_{1}det(D_{2})\:\mbox{,}
\end{eqnarray}
one has
\begin{equation}
\label{x01sta}
x_{01} = \frac{b_1 det (D_2)}{(z-a_0)det (D_1)-b_1^2 det (D_2)}\:\mbox{.}
\end{equation}
In~\sref{sec:LANCZOS_SYM} it has been shown that for evaluating $x_{00}$, 
one has to rewrite it in terms of the ratio between $det (D_2)$ and $det (D_1)$.
In an analogous way, $x_{01}$ must be written in terms 
of the ratio $\frac{det (D_3)}{det(D_2)}$.
To this aim it is useful to rearrange the denominator of~(\ref{x01sta})
in such a way that the ratio between 
$det (D_3)$ and $det (D_2)$ appears in the formula. This is achieved 
by using the following equation
\begin{equation}
\label{detd1} 
det (D_1) =(z-a_1) det (D_2) -b_2^2 det (D_3) \:\mbox{.}
\end{equation}
In fact one has 
\begin{eqnarray}
\label{x01fin}
 x_{01} &=& \frac{b_1}{(z-a_0)(z-a_1)-b_1^2-(z-a_0)b_2^2
      \frac{det(D_3)}{det D_2}} \cr
%\cr
%        &=& \frac{1}{(z-a_0)\frac{(z-a_1)}{b_1}-b_1-\frac{(z-a_0)}{b_1}b_2^2
%\frac{det(D_3)}{det(D_2)}} \\
 &=& \frac{1}{(z-a_0)\frac{(z-a_1)}{b_1}-b_1-\frac{z-a_0}{b_1} \frac{b^{2}_{2}}
{z-a_{2}-\frac{b^{2}_{3}}{z-a_{3}-\frac{b^{2}_{4}}{z-a_4-b_5^2...}}}} \:\mbox{.}
\end{eqnarray} 
Proceeding in an analogous way and using in addition the  relation 
\begin{equation}
\label{folid}
det(D_2) =(z-a_2)det(D_3)-b_3^2 det(D_4) 
\end{equation}
 one obtains
\begin{eqnarray}
\label{x02fin}
 x_{02} &=& \frac{b_1 b_2 det(D_3)}{(z-a_0)det(D_1)-b_1^2 det(D_2)} \cr
        &=& \left[\frac{(z-a_2)}{b_2}\left(\frac{(z-a_1)(z-a_0)}{b_1}-b_1\right)-
\frac{(z-a_0)}{b_1}b_2\right.\cr
\nonumber &-& \left. \frac{1}{b_2}\left(\frac{(z-a_1)(z-a_0)}{b_1}-b_1\right)b_3^2
\frac{det(D_4)}{det(D_3)} \right ]^{-1} \\
\nonumber &=&\left [ \frac{(z-a_2)}{b_2}\left(\frac{(z-a_1)(z-a_0)}{b_1}-b_1\right)-
\frac{(z-a_0)}{b_1}b_2\right.\\
&- & \left. \frac{1}{b_2}\left(\frac{(z-a_1)(z-a_0)}{b_1}-b_1\right) \frac{b^{2}_{3}}
{z-a_{3}-\frac{b^{2}_{4}}{z-a_4-\frac{b_5^2}{z-a_5-b_6^2...}}}\right ]^{-1} \:\mbox{.}
\end{eqnarray}
At this point it is clear that defining
\begin{equation}
   g(\nu)=-\frac{b_{\nu}^2}{z-a_{\nu}
          -\frac{b_{\nu+1}^2}{z-a_{\nu+1}
          -\frac{b_{\nu+2}^2}{z-a_{\nu+2}
          -\frac{b_{\nu+3}^2}{\cdots}}}}
\end{equation}
one has
\begin{equation} \label{xoo}
  x_{00}=\frac{1}{z-a_0+g(1)}\,;\,\,\,\,\,\,
x_{01}=\frac{1}{(z-a_1)\lambda_{00}-b_1+\lambda_{00} g(2)} \:,
\end{equation}
%and
%\begin{equation} \label{xo1}
% x_{01}=\frac{1}{(z-a_1)\lambda_{00}-b_1+\lambda_{00} g(2)} \:,
%\end{equation}
where $\lambda_{00}={z-a_0}/{b_1}$
%\begin{equation} \label{lambdao}
%  \lambda_{00}=\frac{z-a_0}{b_1}
%\end{equation}
is obtained from $x_{00}$.
Now a new parameter 
\begin{equation}
\label{lambda1}
\lambda_{01}=\frac{(z-a_1)\lambda_{00}-b_1}{b_2}  
\end{equation}
can be obtained from $x_{01}$, leading to the evaluation of
\begin{equation}
\label{xo2}
x_{02}=\frac{1}{(z-a_2)\lambda_{01}-b_2\lambda_{00}+\lambda_{01} g(3)} \:,
\end{equation}
which generates the parameter $\lambda_{02}$
\begin{equation}
\label{lambda22}
\lambda_{02}=\frac{(z-a_2)\lambda_{01}-b_2\lambda_{00}}{b_3} 
\end{equation}
and so on. Therefore one has
\begin{equation}
\label{xon}
x_{0i}=\frac{1}{(z-a_i)\lambda_{0\,i-1}-b_i\lambda_{0\,i-2}+\lambda_{0\,i-1}g(i+1)} 
\end{equation}
with
\begin{equation}
\label{lambdan}
\lambda_{0i}=\frac{(z-a_i)\lambda_{0\,i-1}-b_i\lambda_{0\,i-2}}{b_{i+1}} \:\mbox{,}
\end{equation}
where $\lambda_{0\,-1}=1$.
%%%%%%%%%%%%%%%%%%%%%%%%%%%%%%%%%%%%%%%%%%%%%%%%%%%%%%%%%%%%%%%%%%%%%%%%
%results

%************************************************************
%************************************************************
\section{Comments on related approaches}\label{sec:COMMENTS}
%************************************************************
%************************************************************

%**********************************************************************************
\subsection{Integral transform methods with other kernels}\label{sec:OTHERKERNELS}
%**********************************************************************************

A microscopic study of few--body responses with the integral transform approach
has been suggested in~\cite{EFROS:1980,EFROS:1985}. In~\cite{EFROS:1980} the Stieltjes kernel 
\be
K(\sigma,E)=\frac{1}{E+\sigma}\la{st}
\ee
has been employed to this purpose. 
The values of $\sigma$ are real here and $-\sigma$ does not belong to the spectrum.
The corresponding transform 
\be
\Phi(\sigma)=\langle\Psi_0|\hat{O}^\dag(\hat{H}+\sigma)^{-1}\hat{O}|\Psi_0\rangle,\la{tst}
\ee
can be obtained with standard bound--state type methods. 
The inversion has not been considered. Instead it has been suggested to compare
$\Phi(\sigma)$ with the corresponding transform of the experimental $r(E)$.
Such a procedure is inferior to the direct comparison between the experimental and theoretical
$r(E)$. In fact, when one passes from responses to transforms interesting physical effects, that
can be visible only in certain limited ranges of $E$, can be obscured in the transform. In fact 
they can be spread out on a larger $\sigma_R$ range by the integral with such a kernel. 
Moreover, experimental data in the whole range of $E$ are required to perform the comparison 
and they may be missing.
Sometimes the lack of data is compensated by theoretical results obtained within some model assumption. 
However, it may be difficult then to judge how appropriate these assumptions are, given the above mentioned 
spreading of that information by the integral. 

Besides the Stieltjes kernel, also the kernel 
\be
K(\sigma,E)=\frac{1}{(E+\sigma)^2},\la{st1}
\ee
has been considered in~\cite{EFROS:1985}.
In addition it has been shown how to calculate with this approach
not only exclusive perturbation--induced reactions, but also general
strong--interaction induced reactions. The necessity of an inversion 
of the transforms, in order to compare directly the responses with the experimental data, 
has also been underlined, since it is a necessary element for such applications.
To this aim accurate inversions of the transforms with the kernels (\re{st}), (\re{st1})
have been performed in some model problems. However, it has been realized that
a considerably higher accuracy in the input transforms than in the output responses was required.
This is just the important difference with the case of the Lorentz kernel. 

In~\cite{EFROS:1993}, the realistic problem of the deuteron electrodisintegration has been 
addressed using the Stieltjes kernel~(\re{st}). The transforms (\re{tst}) have been calculated 
using standard routines for solving ordinary differential equations. 
It has been found that these transforms do not ensure sufficiently stable inversion.
The conclusion is that the Lorentz transform is preferable even if the calculation of the
Stieltjes transform may be somewhat easier. 

The Laplace kernel
\be
K(\sigma,E)=\exp(-E/\sigma)\la{kl}
\ee
leads to the transform (\re{s})
\be
\Phi({\sigma})=\langle Q|\exp(-\hat{H}/\sigma)|Q'\rangle\la{pl}
\ee
of $r(E)$ defined in~(\re{rr}). Monte Carlo techniques for calculating quantities of the form 
(\re{pl}) have been studied  
starting from~\cite{THIRUMALAI:1983} in connection with the determination of thermodynamic Green 
functions.\footnote{In~\cite{CARLSON:1998} the authors refer to~\cite{BAYM:1961}  
inappropriately in relation to the present approach.  
Indeed, in~\cite{BAYM:1961} methods to calculate any many--body quantities like (\re{pl})
are not discussed. This paper is devoted exclusively 
to a procedure for reconstructing the Fourier or Laplace transform of the 
thermodynamic Green functions from some discrete Fourier
coefficients. This procedure is based on the periodicity--in--time  boundary condition 
of the thermodynamic Green functions at finite temperatures (the period is $(kT)^{-1}$) 
and has no relevance to the present zero temperature problems.}

The problem of the inversion of the Monte Carlo Laplace transforms of temperature dependent
condensed matter response functions has been addressed in~\cite{SILVER:1990,GUBERNATIS:1991}.
The input transforms can be obtained with only a limited accuracy
and the inversion is problematic. In~\cite{CARLSON:1992} the transform (\re{pl}) of a few--body ($\alpha$--particle)
response has been calculated using a known Green function Monte Carlo type technique. 
In~\cite{CARLSON:1992} an attempt has been made
to perform a comparison with experiment 
at the response level, inverting the Laplace transform 
of the response. However, information sufficient to judge on the quality of the inversion
is not listed there. In subsequent papers the authors shifted to comparisons between theory
and experiment at the level of the Laplace transform
In~\cite{CARLSON:2002} the Laplace transforms (\re{pl}) of electrodisintegration responses of a number 
of nuclei have been  compared with the transforms (\re{phi})  with the kernel 
(\re{kl}) of experimental data.

%\footnote{In~\cite{CARLSON:1992} an attempt has been  done 
%to perform a comparison with experiment 
%at the response level, inverting the Laplace transform 
%of the response. However, information sufficient to judge on the quality of the inversion
%is not listed there. In subsequent papers the authors shifted to comparisons between theory
%and experiment at the level of the Laplace transform}
%&&&
%The inversion was done via an approximation of the response 
%with a linear combination of three known functions. 
%Their parameters were fitted  to the transform. This attempt
%is not justified since there is no guarantee that such a procedure with a different number
%of functions would lead to the same results.  Inversion stability with respect 
%to the number of basis functions representing a response was not established
%in~\cite{CARLSON:1992}.} 

%******************************************************
\subsection{The method of moments}\label{sec:MOMENTS}
%******************************************************

The method of moments (MM) has been  used to calculate responses in atomic physics 
(see e.g.~\cite{LANGHOFF:1980,REINHARDT:1980}).
As an input one uses the moments 
\be
M_n=\sum\!\!\!\!\!\!\!\int\,(E+\bar{E})^{-n}r(E)dE.\la{mom}
\ee 
They are calculated with the help of closure as explained in~\sref{sec:INCL}. 
The procedure is a recursive one.
Comparing the MM with the LIT approach one finds a few similarities and one noticeable difference.
The former consist in that also (\ref{mom}) can be considered as a kind of mapping 
depending on two parameters: in the LIT case $\sigma_R$ and $\sigma_I$, in
the MM case $n$ and $\bar E$. Moreover, in both cases one of them is kept constant 
($\sigma_I$ or $\bar E$) and varies the other one ($\sigma_R$ or $n$). 
In this way one deals  essentially with a 
mapping of $E$ into $\sigma_R$ or $n$. The important difference  consists in the fact that 
the continuous variable $E$ is mapped in our case to a parameter which
can be  varied continuously ($\sigma_R$) while in the MM case $n$ can only
be discrete.

While positive results have been obtained with the MM  approach, it has some
limitations. The response
is reconstructed from the moments using the classical machinery. First of all one has to be sure
that all used moments exist. Then the machinery is rather
complicated and it includes solving nonlinear problems. Another shortcoming 
is related to the loss of accuracy in higher moments because of the necessity to calculate 
them recursively, while  a rather high accuracy in its moments is required
to reconstruct the detailed behaviour of $r(E)$. 

Another interesting approach to calculate electron scattering response funnctions
has been proposed in~\cite{ROSENFELDER:1980}. There a few `cumulant expansions'
have been suggested, that could restrict the number of moments necessary for
reconstracting the energy spectrum. To our knowledge 
such an approach has not been investigated further. 

%*****************************************************************************************
\subsection{Response function from transform in the $\sigma_I\to 0$ limit}\label{sec:ETA}
%*****************************************************************************************

Using in (\re{rr}) the  Lorentzian representation of the delta function 
\be
\delta(E_\gamma-E)=\,
\frac{1}{\pi}\,{\rm lim}_{\eta\to\infty}\,{\rm Im}\,\frac{1}{E_\gamma-E-i\eta}=
\,{\rm lim}_{\eta\to\infty}\,\frac{\eta}{\pi}\,\frac{1}{(E_\gamma-E)^2+\eta^2}
\ee
one realizes that $r(E)$ may be calculated as
follows
\be
r(E)={\rm lim}_{\sigma_I\rightarrow0}r_{\sigma_I}(E),\qquad r_{\sigma_I}(E)=\frac{\sigma_I}{\pi}
\langle\tilde{\Psi}_{\sigma_I}|\tilde{\Psi'}_{\sigma_I}\rangle,\la{eeta}
\ee
where 
\begin{eqnarray}
|{\tilde \Psi}_{\sigma_I}\rangle&=&(\hat{H}-E-i{\sigma_I})^{-1}|Q\rangle,\la{eta1}\\
|{\tilde \Psi}'_{\sigma_I}\rangle&=&(\hat{H}-E-i{\sigma_I})^{-1}|Q'\rangle,\la{eta2}
\end{eqnarray}
%This simply means that (\ref{eta1}) and (\ref{eta2}) coincide with (\ref{psi1})
%and (\ref{psi2}) if $E\to E_0+\sigma_R$. 
Therefore $r_{\sigma_I}(E)\equiv ({\sigma_I}/\pi) L(E,\sigma_I)$.

However, unfortunately the limit $\sigma_I=0$ cannot be realized in a simple way
via an extrapolation of the $\sigma_I$ dependent LIT. If, for example, the 
calculation of $r_{\sigma_I}(E)$ is performed solving~(\ref{eta1}) and~(\ref{eta2}) 
in a subspace of localized functions of a fixed dimension $N$ for many ${\sigma_I}$,
then the extrapolation of its results to ${\sigma_I}=0$
would give a model response of the form
\be
r(E)=\sum_{n=1}^Nr_n\delta(E_n-E).
\ee
This limiting response in a given finite subspace is quite different from
the smooth true response that corresponds to transitions to continuum. 

As already  discussed in ~\sref{sec:WIDTH} the smaller is ${\sigma_I}$ 
the more difficult is to obtain $|{\tilde \Psi}_{\sigma_I}\rangle$ and $|{\tilde \Psi}'_{\sigma_I}\rangle$ and hence 
the response $r_{\sigma_I}(E)$, with the help
of `simple' bound--state type methods.
Therefore we believe that this way of proceeding is more troublesome than 
the LIT method, which relies on the inversion of the transform.
The latter gives an opportunity to perform a dynamic calculation with a 
higher ${\sigma_I}$  value, which is certainly an easier task.

In~\cite{ISHIKAWA:1994} a procedure has been given to calculate the $A=3$ response functions in the framework 
of the Faddeev description of dynamics. The equation of~(\re{eta1}) form 
at ${\sigma_I}\rightarrow 0$ has been transformed to inhomogeneous integral equations of Faddeev type.
Different from the case of  ${\sigma_I}$  finite, the solutions of such equations are not localized.
No inversion is required here.
As a result the summation over contributions to a response from separate final states 
has been performed in a closed form, i.e. using the closure property, like in the LIT case. 

Here we should also mention calculations, made in the same spirit for $A=2-4$, of the non--Born part 
of the T--matrix at complex energy with subsequent analytic continuation~\cite{KAMADA:2003,UZU:2003}.

%*********************************************************************************
\subsection{Methods with Hamiltonian matrix diagonalization}\label{sec:DIAG}
%**********************************************************************************

The response $r(E)$ may also be obtained in the following way~\cite{EFROS:1999} (compare
also~\sref{sec:EIGENVALUE}). 
%The approach based on Eqs. (\re{phi}),
%(\re{s}) is applicable in cases when calculation of the transform (\re{s})
%is feasible. It is feasible
%anyway when the Hamiltonian matrix is diagonalized.
Consider a complete set of localized functions, and let
$|\varphi_\nu^N\rangle$ and $\epsilon_\nu^N$ be eigenstates and eigenvalues of the Hamiltonian matrix in a subspace
that spans the first $N$ states in the set. Then from~(\re{r}) one gets 
\be
\Phi_N(\sigma)=\sum_\nu^N\langle Q|\varphi_\nu^N\rangle K(\sigma,\epsilon_\nu^N)\langle\varphi_\nu^N|Q'\rangle.
\la{phin1}
\ee
It is easy to show that if either $\langle Q|Q\rangle$ or $\langle Q'|Q'\rangle$  
is finite and  the kernel is uniformly bounded ($K(\sigma,E)<C$) then 
$\Phi_N\rightarrow\Phi(\sigma)$
uniformly in $\sigma$ as $N\rightarrow\infty$. 

While  the transform $\Phi_N$ tends to the true transform 
$\Phi$ when $N$ increases,
the corresponding $r_N(E) $ are quite different from each other at any $N$. While the
real $r(E)$ is continuous and smooth,
the  $r_N$ corresponding to (\re{phin1}) consists of transitions to quasi--levels 
(the same states as in (\ref{nu1})--(\ref{L_epsnu_diag})),
\be
r_N(E)=\sum_\nu^N\langle Q|\varphi_\nu^N\rangle\langle\varphi_\nu^N|Q'\rangle\delta(\epsilon_\nu^N-E).
\la{rn}
\ee
The transition amplitudes entering $r_N$ are rather chaotic,
and as $N\rightarrow\infty$ the $r_N$ do not tend to any limit in the normal sense.

To get a reasonable approximation to the true  $r(E)$ from the approximate transform
(\re{phin1}) one should solve 
\be
\Phi_N(\sigma)=\sum\!\!\!\!\!\!\!\int\, K(\sigma,E)r(E)dE,\la{phi1} 
\ee 
imposing a regularization condition. This ensures a
smooth output. In this way one comes to a continuous smooth response $r_N^s(E)$
that does tend to $r(E)$ as $N\rightarrow\infty$. When $N$ increases the difference
between the response $r_N^s(E)$ and the response~(\re{rn}) consists
in components with higher and higher frequencies.

In accordance with the discussion above it is expedient to apply this procedure with
`smoothing--type' kernels, like the Lorentz kernel~(\re{kRI}) 
or the Gaussian kernel
$K(\sigma_R,\sigma_I,E)=\exp[-(E-\sigma_R)^2/\sigma_I^2]$ (for analogy with the Lorentz case
the centroid and width of the Gaussian are indicated by $\sigma_R$ and $\sigma_I$, respectively). 
%Let $\Delta E$ be the range of $E$ values for which
%$K(\sigma_R,\sigma_I,E)$ is not small at a given $\sigma_R$ value.
One may expect that $\Phi_N(\sigma)$ reaches stability at $N$ values that provide
sufficiently many levels 
$\epsilon_\nu^N$ in an energy range corresponding to the widths of the kernels. 

In some shell model studies (see e.g.~\cite{HAXTON:2005}) the following
procedure to calculate response functions has been  adopted. The Hamiltonian has been diagonalized 
in a large model space with the Lanczos algorithm and the model response
(\re{rn}) has been obtained. Then this model response has been smoothed using the Lorentz or the Gaussian
leading to the so--called {\it Lanczos response} (LCZR)
smoothed function, and the result of (\re{phin1}) form has been taken as the final response. 
However, such a quantity can reproduce the true response only up to the width of a smoothing
function. On the other hand, the smaller is
this width the larger dimension $N$ of the model space is required to ensure 
convergence with respect to $N$ of such a smoothed response. 

In the light of our experience with the  LIT method we recommend a better procedure: not to adopt 
the quantity (\re{phin1}) as the final response $r(E)$ but to extract the latter  
as a regularized solution to (\re{phi1}). This would allow using the `smoothing functions',
or kernels, with larger widths and, correspondingly, employing 
shell model spaces of lower dimensions $N$. Different from the above procedure, 
the resolution of the spectrum obtained in this way is better then the width of a smoothing function.  
This occurs because of the improvement in the results when one passes 
from a transform to a response. 

%Speaking of the Lorentz transform (\re{ov}) one may note that  
%diagonalization of the Hamiltonian is not always the optimal way to calculate the transforms.
A numerical example which shows the comparison between the LCZR and that 
obtained by inversion of the LIT is given in~\sref{sec:SMOOTH}.

%**************************************************************************************
\section{Solution of the relevant equations: the hyperspherical harmonics approach} 
\label{sec:RESOLMETH}
%***************************************************************************************

In~\sref{sec:LIT_TH} it has been explained that the LIT method demands a good bound--state resolution  
technique that allows an accurate solution, not only for the ground state wave function 
of the target system, but also for the LIT functions (see~(\ref{eq1}),~(\ref{eq2}) and~(\ref{Eqs2})). 
As was already stated in~\sref{sec:EIGENVALUE}, methods of expansion over basis sets turn out to be 
advantageous with increasing number of particles. 
In this section we shortly summarize one of them, namely the hyperspherical harmonics 
expansion method, since it has been used in most of the applications presented in this review
(for LIT application to the No Core Shell Model (NCSM) approach see~\cite{STETCU:2007}). 

We first introduce the hyperspherical formalism and define the HH basis on which the expansions are made. 
The two practical ways to accelerate the convergence rate of the HH expansion are described 
in the Appendix A and B. In particular, in~\ref{sec:CHH} we  
describe the correlated HH expansion method (CHH), and in~\ref{sec:EIHH}
the method which makes use of an effective interaction defined in a subspace of the HH basis
(EIHH). Most of the results presented in this
review have been obtained using one or the other of these two approaches.

In the HH formalism, which has been successfully applied to study nuclear, atomic 
and molecular few--body systems
~\cite{FENIN:1972,BALLOT:1980,ROSATI:1992,BARNEA:1999a,KRIVEC:1990,KRIVEC:1998},
the Jacobi coordinates are replaced by a single length
coordinate, the hyperradius, and a set of $(3A-4)$ hyperangles.
The HH are the A--body generalization of the two--body spherical harmonics,
and likewise depend only on the hyperangular (angular) coordinates in the
hyperspherical (spherical) decomposition of the A--body (two--body) system.
In general, the wave function can be expanded in a series consisting of
products of HH basis functions and hyperradial basis functions.
This decomposition makes the HH expansion rather flexible in describing the
A--body wave function and allows an accurate description of the wave functions 
at large distances.

%.in order to account for the repulsion between two nucleons at short
%distances nuclear potentials usually include a hard core. Consequently the
%Schroedinger wave function tends to zero at
%short inter--particle distances. In order to reproduce this short distance
%behavior with the
%HH expansion, or in fact with any other method that include polynomial
%expansion, one needs many basis states. Therefore, in such cases, the
%convergence rate of the HH basis turns out to be rather slow.

%*******************************************************************
\subsection{The hyperspherical coordinates}\label{sec:HHCOORD}
%******************************************************************

To introduce the hyperspherical coordinates we start from the set composed by the 
centre of mass coordinate
$\bi{R}=\frac{1}{A}\sum_{i=1}^A \bi{r}_i$
and the normalized reversed order $(A-1)$ Jacobi coordinates (A represents the total number of particles)
\begin{eqnarray} \label{Jacobi}
  \bfeta_1 & = & \sqrt{\frac{A-1}{A}}\Big[\bi{r}_1
                - \frac{1}{A-1}(\bi{r}_2 + \bi{r}_3 + \cdots
                + \bi{r}_{A} )\Big]  \nonumber \\
  \bfeta_2 & = & \sqrt{\frac{A-2}{A-1}}\Big[\bi{r}_{2}
               - \frac{1}{A-2}(\bi{r}_3 + \bi{r}_4 + \cdots
               + \bi{r}_{A} )\Big]  \nonumber \\
  &\ldots&  \nonumber \\
  \bfeta_{A-1} & = & \sqrt{\frac{1}{2}}\Big( \bi{r}_{A-1}
                        - \bi{r}_{A} \Big)\,,
\end{eqnarray}
which describe the positions of the $j$--th particle relative 
to the centre of mass of particles $j+1$ to $A$.
Each Jacobi coordinate $\bfeta_j$ consists of a radial
coordinate $\eta_j$ and a pair of angular coordinates
$\hat{\bfeta}_j \equiv (\theta_j,\, \phi_j)$.

The $(A-1)$ Jacobi  coordinates are then transformed into the hyperangular coordinates
$\alpha_2,\ldots,\alpha_{A-1}$ through the relations
\begin{equation} \label{hyper_n}
  \sin \alpha_n =  \eta_n / \rho_{n} \; ,
\end{equation}
where
\begin{equation}
\label{rho_n}
\rho_{n}^2 = \rho_{n-1}^2 + \eta_n^2 = \sum_{j=1}^n \eta_j^2 \; ,
\end{equation}
with $\rho_0^2=0$.
For $n=A-1$ one has
\begin{equation} \label{hyprho}
  \rho^2 \equiv \rho_{A-1}^2 =
  \frac{1}{A}\sum_{i<j}^A (\bi{r}_i-\bi{r}_{j})^2 \; .
\end{equation}
From this relation it is evident that the hyperradial coordinate $\rho$ is symmetric with respect to
permutations of the single particle coordinates.
After the transformation described above the $3(A-1)$ internal coordinates 
for the $A$--particle system consist of
one hyperradial coordinate $\rho \equiv \rho_{A-1}$, and  $(3A-4)$
hyperangular coordinates
$\hat\Omega \equiv \{\hat{\eta}_1,\, \hat{\eta}_2,\, \cdots,\,
\hat{\eta}_{A-1}, \alpha_2,\, \alpha_3,\, \cdots,\, \alpha_{A-1}\} $.

%These coordinates depend on the set of Jacobi
%coordinates specified in equation (\ref{Jacobi}).

By using hyperspherical coordinates one can write the Laplace operator for
$n$ Jacobi coordinates ($n=1 \ldots N$),
as a sum of two terms
\begin{equation} \label{Laplacen}
  \Delta_{n} = \frac{1}{\rho_{n}^{3n-1}}\frac{\partial}{\partial \rho_{n}}
  {\rho_{n}^{3n-1}} \frac{\partial}{\partial \rho_{n}}
   - \frac{1}{\rho_{n}^2} \hat{K}_n^2 \; .
\end{equation}
The  hyperspherical, or grand angular momentum
operator $\hat{K}_n^2$ of the $n$ Jacobi coordinates can be expressed in
terms of $\hat{K}_{n-1}^2$ and $\hat{\ell}_n^2$ as follows~\cite{EFROS:1972}
\begin{eqnarray} \label{Kn}
 \hat{K}_n^2 &=& - \frac{\partial ^2}{\partial \alpha_n^2} +
 \frac{3n-6 - (3n-2) \cos (2\alpha_n)} {\sin(2\alpha_n)}
 \frac{\partial}{\partial \alpha_n} \nonumber \\
& & +\frac{1}{\cos^2 \alpha_n}\hat{K}_{n-1}^2
+ \frac{1}{\sin^2 \alpha_n} \hat{\ell}_n^2 \; ,
\end{eqnarray}
where we define $\hat {K}_1^2 \equiv \hat{\ell}_1^2$.
The angular momentum operator associated with these $n$ coordinates
is $\hat{\bi L}_n =  \hat{\bi L}_{n-1} + \hat{\boldsymbol{\ell}}_n$
($\hat\bi{L}_0=0$).
The operators $\hat{K}_{n}^2$, $\hat{\ell}_{n}^2$, $\hat{K}_{n-1}^2$,
$\hat{L}_n^2$ and $\hat{L}_n^z$
commute with each other.

%*******************************************************************
\subsection{The hyperspherical harmonics basis}\label{sec:HHBASIS}
%******************************************************************

The hyperspherical harmonics functions ${\cal Y}_{[K_n]}$ are the
eigenfunctions of the hyperangular
operator~(\ref{Kn}). The explicit expression for the HH functions of the first $n$
Jacobi coordinates is given by~\cite{FABRE:1983}
\begin{eqnarray} \label{HH}
 {\cal Y}_{[K_n]} & = & \left[ \sum_{ m_1, \ldots ,m_n }
 \bra \ell_1 m_1  \ell_2 m_2 | L_2 M_2 \ket
 \bra L_{2} M_{2} \ell_3 m_3 | L_3 M_3 \ket \right.\times \ldots \nonumber \\
 & & \; \; \; \times
 \left.\bra L_{n-1} M_{n-1} \ell_n m_n | L_n M_n \ket
 \prod_{j=1}^{n} Y_{\ell_j, \, m_j} (\hat{\eta}_j) \right] 
 \nonumber \\ & &
\times\left[ \prod_{j=2}^{n} 
{\cal N}_{\mu_j}^{\ell_j+\frac{1}{2}, K_{j-1}+ \frac{3j-5}{2}}
(\sin \alpha_j )^{\ell_j} (\cos \alpha_j )^{K_{j-1}}\right.\nonumber \\
& & \left. P_{\mu_j}^{( \ell_j + \frac{1}{2},K_{j-1} + \frac{3j-5}{2})}(\cos (2\alpha_j)) 
 \right] \; ,
\end{eqnarray}
where $Y_{\ell, \, m}$ are the spherical harmonics functions, $P_{\mu}^{(a,b)}$
are the Jacobi polynomials~\cite{ABRAMOWITZ:1965} }and ${\cal N}_{\mu}^{a,b}$ are
normalization constants given by~\cite{EFROS:1972}:
\begin{equation} \label{norN}
  {\cal N}_{\mu}^{a,b}=\left[\frac
                       {2(2\mu+a+b+1)\mu!\Gamma(\mu+a+b+1)}
                       {\Gamma(\mu+a+1)\Gamma(\mu+b+1)}
                       \right]^{\frac{1}{2}}\;.
\end{equation}
The symbol $[K_n]$ stands for the set of quantum numbers $\ell_1,...,\ell_n$,
$L_2,...,L_n$, $\mu_2,...,\mu_n$ and $M_n$.
The quantum numbers $K_j$ are given by
\begin{equation} \label{K_{A-1}}
K_j = 2\mu_j + K_{j-1}+ \ell_j \,\,\,\, ; \,\,\,\,\mu_1 \equiv 0 \; ,
\end{equation}
and the $\mu_j$ are non--negative integers.
By construction, $\rho_n^{K_n} {\cal Y}_{[K_n]}$ is a harmonic polynomial
of degree $K_n$. The HH function ${\cal Y}_{[K_n]}$ is an eigenfunction of
$\hat{K}_n^2$ with eigenvalues $K_n(K_n + 3n - 2)$.

In view of equations~(\ref{rho_n}) and (\ref{hyprho}) it is evident that the harmonic oscillator
(HO) Hamiltonian may be written in the form

\begin{eqnarray}
\frac{\hbar\omega}{2}\sum_{j=1}^N\left(-\Delta_{{\bar\eta}_j}+{\bar\eta}_j^2\right)\equiv \frac{\hbar\omega}{2}
\left(-\Delta_{A-1}+{\bar\rho}^2\right)\nonumber\\
=\frac{\hbar\omega}{2}\left(-\frac{\partial^2}{\partial{\bar\rho}^2}-\frac{3N-1}{\bar\rho}\frac{\partial}{\partial{\bar\rho}}
+\frac{{\hat K}^2}{{\bar\rho}^2}+{\bar\rho}^2\right),
\end{eqnarray}
where ${\bar\eta}_j=\bfeta_j/r_0$, ${\bar\rho}=\rho/r_0$ where $r_0=(\hbar/(m\omega))^{1/2}$ is the
oscillator length. It has the eigenvectors
%\begin{eqnarray}\label{HHO}
%\sum_{j=1}^N\left(-\Delta_{\eta_j}+\omega^2 \eta_j^2\right)&\equiv& -\Delta_N +\omega^2 \rho= \nonumber\\
%&=&
%\left
%(-{\partial^2\over\partial\rho^2}-{(3N-1)\over \rho}{\partial\over\partial \rho}
%+{\hat K^2\over \rho^2}+\omega^2\rho\right) \; ,
%\end{eqnarray}
%has the eigenvectors 
\begin{equation}\label{psiHO}
\Psi_{HO}=R_{n_\rho}(\rho)\,{\cal Y}_{[K]}
\end{equation}
with eigenvalues
\begin{equation}\label{eigenHO}
E_n=\hbar\omega\left({3(A-1)\over2}+n\right)=\hbar\omega\left(
{3(A-1)\over2}+2n_\rho+K\right)\,.
\end{equation}
Therefore the HH $K$--quantum number can be associated with the quanta
of excitations of the HO wave function, assuming no hyperradial excitation.

When applying the HH method to the solution of the A--body problem one is
facing two problems:

(i) in general the HH basis states possess no special symmetry under particle
permutations. For the three-- and four--body problem, this is a minor problem since a direct
(anti) symmetrization of the wave function is still feasible. For larger 
systems, direct symmetrization becomes impractical and there is a 
need for a more sophisticated symmetrization method~\cite{NOVOSELSKY:1994,BARNEA:1997,BARNEA:1998};

(ii) for strongly repulsive potentials the
convergence rate of the HH basis turns out to be rather slow,
therefore two methods have been suggested in order to `accelerate' it. 
The first one (CHH) makes use of  {\it correlation} functions, which
{\it correlate} the pure HH basis in order  to reflect the characteristics 
of the potential. In the second one (EIHH)
the convergence is accelerated by an effective interaction, built 
from the bare potential by means of a similarity transformation.
The description of these two methods can be found in Appendices A and B.

\section{Applications}\label{sec:APPL}
The LIT method has been applied to various perturbation induced reactions with nuclei.
For the strong interacting nucleons the perturbation is represented by the electroweak probe
(photons, electrons, neutrinos) and the cross sections contain the responses of the systems to
the perturbation. In this section the formalisms for such reactions are outlined, putting in evidence
how the cross sections are related to the response functions, which can be calculated with the LIT 
method. It is also discussed which are the relevant operators $\hat O$ and $\hat O'$ to be used for 
each reaction.

%************************************************************ 
\subsection{Electromagnetic reactions}\label{sec:EM}
%***********************************************************

The principal aim in studying of electromagnetic interactions with nuclei is to investigate
the nuclear dynamics and to clarify the role of effective nuclear and subnuclear degrees of freedom.
These are traditionally described by nucleons, mesons, and isobars 
and lead to two-- and three--nucleon forces. In principle the effective degrees of freedom and
the corresponding forces should be related to the more fundamental quark--gluon dynamics of QCD.

A great part of our present knowledge about the internal structure of nuclei has been obtained
from electromagnetic reactions. The electromagnetic interaction has various advantages. It is
a fundamental theory and it is weak enough, so that in most cases lowest--order perturbation
theory is sufficient, allowing a simple interpretation of cross sections and other observables
in terms of charge and current density matrix elements.

In particular the study of the photoabsorption spectra has proved
to be fundamental for the comprehension of the dynamics of a quantum system.
Photoabsorption experiments on nuclear targets  have been performed extensively
since the dawn of nuclear physics until the beginning of the eighties.  Later, however,
this activity
has slowed down somewhat. One of the possible reasons for this may have been  the lack
of theoretical progress, which made the comparison theory--experiment rather
inconclusive. In relation to this, it is interesting to quote here
a sentence from~\cite{FAUL:1981} that expresses this concern. The title of the paper is 
`{\it Photodisintegration of 
$^3$H and $^3$He}'. In the conclusions the authors write: `{\it 
More theoretical work will be needed before the experimental results for the
photodisintegration of three--body nuclei can test quantitatively the charge symmetry of 
the nuclear interaction. 
The present theoretical models are still rudimentary in that they do not include simultaneously
a consistent treatment of the final state interactions...}' and a list follows of other
shortcomings in the calculations available at that time. The authors conclude that `{\it As a 
consequence of these problems the agreement between experimental and theoretical results is, in
the mean, not notably good.}'
The  progress achieved in recent years in the theory of few--body systems, and in particular 
the possibility opened by the LIT method of performing consistent treatments of initial and 
final states, also for systems with $A>3$, has allowed a sort of revival of the field,
with particular emphasis on the lightest nuclei, and has
stimulated new experimental investigations with modern techniques.  In particular,
both inclusive and exclusive cross sections have been addressed.

%%%%%%%%%%%%%%%%%%%%%%%%%%%%%%%%%%%%%%%%%%%%%%%%%%%%%%%%%%%%%%%%%%%%%%%%%%%%%%%%%%%%%%%%%%%%%%%%%%%%%%%%%
%The  progress achieved in the theory of few--body systems in recent years and in particular 
%the possibility opened by the LIT ve been addressed.
%In the following the formalisms for the crossmethod of performing consistent tratments of initial and 
%final states also for systems with $A>3$,
%has allowed a sort of revival of the field, with particular emphasis on the lightest nuclei,
%stimulating new experimental investigations with modern techniques.  In particular,
%both inclusive and exclusive cross sections have been addressed.
%In the following the formalisms for the cross sections of both kinds of reactions is summarized.
%%%%%%%%%%%%%%%%%%%%%%%%%%%%%%%%%%%%%%%%%%%%%%%%%%%%%%%%%%%%%%%%%%%%%%%%%%%%%%%%%%%%%%%%%%%%%%%%%%%%%%%%%
The information about nuclear dynamics that one can get from electron scattering reactions 
is complementary to that obtained by real photons and much richer. The reason is that,
while the energy of the photon fixes the momentum transferred to the target,
an electron can transfer energy and momentum independently. This allows to study
the dynamics of the system at different wavelengths via the elastic or inelastic
form factors (also called `electron scattering response functions') that appear 
in the cross section. In about fifty years of experimental activity in electron scattering 
a wealth of data on targets all over the nuclear table has been collected at low as well as 
intermediate and high energies and momenta.
Various laboratories are still active in this field, both in the United States 
(Jefferson Laboratory) and in Europe (Mainz, Bonn, Darmstadt). In Darmstadt 
a collider of electrons and beams of nuclei far from stability is planned.
Considering both present and planned activities of these laboratories one can notice
that they  are concentrating on `light' targets. 
A special role play the few--body systems $^2$H and $^3$He, since they also serve as
effective neutron targets.

%********************************************************************
\subsubsection{Electron scattering reactions.}\label{sec:ELECTRON}
%**********************************************************************

%Among the various reasons to privilege such targets there is an important one. 
%In fact the possibility to scatter electrons elastically at various intermediate-high 
% momenta allows to use some nuclei (
%e.g. deuteron and $^3$He) as neutron targets and study the still largely unknown 
%internal structure of the neutron. Of course this is possible provided that the 
%nuclear dynamics is under control from the theoretical point of view, in order 
%that the extraction of the neutron 
%information is reliable. So one can say that electrons, even more than photons, give access 
%to the gray zone of the intersection between nuclear and subnuclear physics.

In the one--photon--exchange approximation
the inclusive electron scattering cross section can be written as~\cite{DEFOREST:1966} 
\begin{eqnarray}
{d^2\sigma\over d\Omega\,d\omega}\ =\ \sigma_M\ \bigg[\ {q_\mu^4\over q^4}\,
R_L(q,\omega)\ +
\ \bigr(\ {q_\mu^2\over 2q^2}+\tan^2{\theta\over 2}\ \bigr)\,R_T(q,\omega)\bigg]\,,
\end{eqnarray}
where $\sigma_M$ is the Mott cross section, $\omega$ is the electron energy loss, $q$ 
the magnitude of the  momentum transferred by the electron, $\theta$  the electron 
scattering angle and $q_\mu^2 = q^2 - \omega^2$. The functions $R_L$ and $R_T$ are the 
electron scattering longitudinal
and transverse response functions, respectively. The function $R_L$ represents the response 
of the nucleus to the electron field via the nuclear charge and is given by
\begin{eqnarray}
R_L(q,\omega)&=& \sum \!\!\!\!~\!\!\!\!\!\!\int\, df\langle \Psi_0|
\hat{\rho}^\dag(\bi{q},\omega)| \Psi_f \rangle\langle \Psi_f|\hat{\rho}(\bi{q},\omega)|
 \Psi_0 \rangle \nonumber\\
& & \times \delta(E_f-E_0+\frac{q^2}{2M_T} -\omega)\,,
\label{rl}
\end{eqnarray}
where $\hat{\rho}(\bi{q},\omega)$ represents the charge density operator and $M_T$ is the mass of 
the target nucleus.

The results presented in~\sref{sec:RES} are obtained with the following 
charge density operator
\begin{equation}
\hat{\rho}(\bi{q},\omega)=\sum_{j=1}^A\hat{\rho}_{j}^{nr}(\bi{q},\omega)+\hat{\rho}_{j}^{rc}
(\bi{q},\omega),\label{rho}
\end{equation}
where 
\begin{equation}
 \hat{\rho}_{j}^{nr}(\bi{q},\omega)=\hat{e}_je^{i\bi{q}\cdot\bi{r}_j},\label{nr}
\end{equation}
\begin{equation}
\hat{\rho}_{j}^{rc}(\bi{q},\omega)=-\frac{q^2}{8m^2}\hat{e}_je^{i\bi{q}\cdot
\bi{r}_j} -i\frac{\hat{e}_j- 2 \hat{\mu}_j}{4m^2}\,{\bsigma}_j\cdot(\bi{q}
\times\bi{p}_j) e^{i\bi{q}\cdot\bi{r}_j},\label{rc}
\end{equation}
\begin{equation}
\hat{e}_j=G_E^p(q_\mu^2)\frac{1+\tau_{z,j}}{2}+G_E^n(q_\mu^2)
\frac{1-\tau_{z,j}}{2}
\equiv\frac{1}{2}\left[G_E^S(q_\mu^2)+G_E^V(q_\mu^2)\tau_{z,j}\right],\label{e}
\end{equation}
\begin{equation}
\hat{\mu}_j=G_M^p(q_\mu^2)\frac{1+\tau_{z,j}}{2}+G_M^n(q_\mu^2)
\frac{1-\tau_{z,j}}{2}\label{m}
\equiv\frac{1}{2}\left[G_M^S(q_\mu^2)+G_M^V(q_\mu^2)\tau_{z,j}\right].\label{mu}
\end{equation}
In (\ref{nr})--(\ref{m})  $\bi{r}$, $\bi{p}$, $\bsigma$, and $\btau$ are the nucleon position, 
momentum, spin and isospin operators, respectively, $m$ is the nucleon mass, 
$G_{E,M}^{p,n}$ are the proton/neutron Sachs form factors and $G_{E,M}^{S,V}$ are the 
isoscalar/isovector nucleon form factors.
The two terms in~(\ref{rc}) proportional to $m^{-2}$ are the Darwin--Foldy (DF) 
and spin--orbit (SO) relativistic corrections to the main operator~(\ref{nr}) (see e.g. 
\cite{DEFOREST:1984,FRIAR:1973}). We refer to the main operator~(\ref{nr}) as the 
non--relativistic one, although the dependence of the nucleon form factors on 
$q_\mu^2$ does not allow a non--relativistic interpretation. 

The operator $\hat{\rho}$ can also be conveniently rewritten in 
terms of the isoscalar and isovector charge nucleon form factors from~(\ref{e}), 
\begin{eqnarray}
\hat{\rho}(\bi{q},\omega)= G_E^S(q_\mu^2)\hat{\rho}_s(\bi{q}) + 
G_E^V(q_\mu^2)\hat{\rho}_v(\bi{q}).\label{rsv}
\end{eqnarray} 

As to the transverse response function, it represents the response of the nucleus to 
the electron field via the current density operator $\hat{\bi{j}}_T$ transverse with respect
to $\hat\bi{q}$
\begin{eqnarray}
R_T(q,\omega)&=& \sum \!\!\!\!~\!\!\!\!\!\!\int\, df\langle \Psi_0| \hat{\bi{j}}_T^\dag
(\bi{q},\omega)| \Psi_f \rangle\langle \Psi_f| \hat{\bi{j}}_T(\bi{q},\omega)|
\Psi_0 \rangle  \nonumber\\
& & \times \delta(E_f-E_0+\frac{q^2}{2M_T} -\omega)\,.
\label{rt}
\end{eqnarray}
The current density in~(\ref{rt}) consists first of all of the transverse components of 
a convection and a spin current 
\begin{equation}
\hat{{\boldsymbol j}}_{(1)}({\boldsymbol q},\omega)=\hat{{\boldsymbol j}}^c_{(1)}({\boldsymbol q},\omega)
+\hat{{\boldsymbol j}}^s_{(1)}({\boldsymbol q},\omega)\,,
\end{equation}
with
\begin{eqnarray}
\hat{{\boldsymbol j}}^c_{(1)}({\boldsymbol q},\omega)&=& \frac{1}{2m}\sum_j \hat e_j\,
\{{\boldsymbol p}_j,e^{i\bi{q}\cdot\bi{r}_j}\}\,,\\ 
\hat{{\boldsymbol j}}^s_{(1)}({\boldsymbol q},\omega)&=& \frac{1}{2m} \sum_j 
\hat \mu_j\,{\boldsymbol \sigma}_j \times {\boldsymbol
q}\,e^{i\bi{q}\cdot\bi{r}_j}\,.
\label{currents}
\end{eqnarray}
The index $(1)$ indicates that the operator is one--body.
In order to satisfy the continuity equation, however, the momentum and/or isospin dependent 
$n$--body interaction $\hat V$ requires
a $n$--body current density operator $\hat{{\boldsymbol j}}_{(n)}$. This is called
the interaction (or meson
exchange) current. As is well known, the continuity equation is
not sufficient to determine 
the $n$--body current uniquely. Therefore in principle one needs a
dynamical model for the nuclear potential which reveals the underlying
interaction mechanism. Such a model can be supplied by the meson exchange
picture of the nucleon--nucleon (NN) interaction. Even for a phenomenological
potential  a consistent two--body current can often be 
constructed~\cite{BUCHMANN:1985,RISKA:1985,ARENHOEVEL:2001}. 

If one wants to apply the LIT method to the calculation of the  electron scattering response functions
$R_{L,T}$ one has to deal with an `unpleasant' feature of the operators. In fact they depend on energy via the 
nucleon form factors. In~\sref{sec:PARTICULAR} it has been shown that the operators can be 
put in the form (\ref{part}) and one can deal with the corresponding energy independent $r_{\alpha\beta}^{L,T}$.
%In order to distinguish the response functions relative to the energy independent part of the 
%charge current operator from those relative to the energy independent part of the current
%density operator, we will label them
%$r^L_{\alpha}$ and $r^T_{\alpha}$, respectively.

%**********************************************************
\subsubsection{Multipole Expansions.}\label{sec:MULT}
%*********************************************************

In order to apply the LIT method it is  useful first to put in evidence the rotational 
quantum numbers $J$ and $M$ in each $\alpha\beta$--component
of the response functions 
\begin{eqnarray}
& & r_{\alpha\beta}^L(q,\omega)=\frac{1}{2J_0+1}\sum_{M_0,J,M}
\sum\!\!\!\!\!\!\!\int\,df \langle\Psi_0(J_0,M_0)|\hat{\rho}_\alpha^\dag({\bf q})|\Psi_f(J,M)\rangle \nonumber\\
& &
\times \langle\Psi_f(J,M)|\hat{\rho}_\beta({\bf q})|\Psi_0(J_0,M_0)\rangle \,\delta\left(E_f-E_0+\frac{q^2}{2 M_T}-\omega\right),\la{long}
\end{eqnarray}
\begin{eqnarray}
& & r_{\alpha\beta}^T(q,\omega)=\frac{1}{2J_0+1}\sum_{M_0,J,M}
\sum\!\!\!\!\!\!\!\int\,df \langle\Psi_0(J_0,M_0)|\hat{{\bf j}}_\alpha^{T\dag}({\bf q})|\Psi_f(J,M)\rangle\nonumber\\ 
& & \times\langle\Psi_f(J,M)|\hat{{\bf j}}_\beta^T({\bf q})|\Psi_0(J_0,M_0)\rangle
\,\delta\left(E_f-E_0+\frac{q^2}{2 M_T}-\omega\right)\,,\la{trans}
\end{eqnarray}
where $J_0$ is the total angular momentum of the ground state. Here $f$ labels asymptotic 
quantum numbers other than $J,M$. The labels $\alpha,\beta$ refer to the isoscalar and isovector 
parts of the operators (see discussion about~(\ref{GG})). Then one can
expand the $\omega$--independent parts of the 
charge and current density operators (to simplify the notation we now omit the subscripts `$\alpha$' 
and `$\beta$') as
\ber
\hat{\rho}(\bi{q})=4\pi\sum_{jm}i^j\hat{\rho}_{jm}(q)Y_{jm}^*({\hat {\bf q}}),\la{rhoexp}\\
\hat{{\bf j}}_T(\bi{q})=4\pi\sum_{\lambda={\rm el,mag}}
\sum_{jm}i^{j-\epsilon}\hat{T}_{jm}^{\lambda}(q){\bf Y}_{jm}^{\lambda *}({\hat {\bf q}})\,.
\la{jexp}
\eer
Here ${\hat {\bf q}}=q^{-1}{\bf q}$, and ${\bf Y}_{jm}^{\lambda }$ are electric and magnetic 
vector spherical harmonics 
\cite{VARSHALOVICH:1988}.
The operators $\hat{\rho}_{jm}$ and $\hat{T}_{jm}^{\lambda}$ are
irreducible tensors of rank $j$. 
In (\re{jexp}) we set $\epsilon=0$ in the electric case and $\epsilon=1$ 
in the magnetic case . With 
this choice the matrix elements of the $\hat{T}_{jm}^{\lambda}$ 
operators between the conventional angular momentum states are real.
In~\ref{sec:APMULT} it is shown that one can write, 
\begin{equation}
r^L(q,\omega)=\frac{4\pi}{2J_0+1}\sum_{Jj}(2J+1)[r^L(q,\omega)]_J^j,\label{rr1}
\end{equation}
\begin{equation}
r^T(q,\omega)=\frac{4\pi}{2J_0+1}\sum_{\lambda={\rm el,mag}}\sum_{Jj}
(2J+1)[r^T(q,\omega)]_J^{j\lambda}\,,\label{rr2}
\end{equation} 
where the partial responses $(r^L)_J^j$ and $(r^T)_J^{j\lambda}$ 
correspond to transitions induced by the multipoles $\hat{\rho}_{jm}$ and $\hat{T}_{jm}^\lambda$ 
to states with a given $J$. There it is also shown that 
these partial responses do not depend on $M$ and $m$.
\footnote{The following notation is often found in the literature and will also be used
in~\sref{sec:NEUTRINO}: the longitudinal multipoles of order $j$ are indicated with $C_j$,
the transverse electric with $E_j$ and the transverse magnetic with $M_j$. As an example:  $(r^L)_J^j\equiv r_{C_j}$, $(r^T)_J^{j\lambda}\equiv r_{E_j}$ if $\lambda=el$  
and $(r^T)_J^{j\lambda}\equiv r_{M_j}$ if $\lambda=mag$. Notice that in neutrino scattering (\sref{sec:NEUTRINO})
also multipoles of longitudinal character appear in the response to the $Z_0$ field via the axial current density.
This does not happen in electron scattering since charge conservation can be applied and such multipoles can be
rewritten in terms of the $C_j$.}

The LIT of~(\ref{rr1}) and (\ref{rr2}) are 
\be
 L^L(q,\sigma_R)=
\frac{4\pi}{2J_0+1}
\sum_{Jj}(2J+1)[ L^L(q,\sigma_R)]_J^{j},
\la{suml}\ee
\be
L^T(q,\sigma_R)=
\frac{4\pi}{2J_0+1}\sum_{\lambda={\rm el,mag}}
\sum_{Jj}(2J+1)[ L^T(q,\sigma_R)]_J^{j\lambda}.
\la{sumt}
\ee
The partial transforms entering the right hand sides of (6.19) and (6.20) are calculated 
in subspaces with given $J,M$ and parity from dynamic equations with sources depending on 
$J$ and $j$. The calculation is $M$ independent. 
To obtain the responses~(\ref{rr1}) and  (\ref{rr2}) from these transforms the integral equations of the
form (\re{tra}) are to be solved. Alternatively,
one may invert the single terms separately and then sum up
the corresponding partial contributions to the responses.

%%%%%%%%%%%%%%% Il resto nell'appendice %%%%%%%%%%%%%%%%%%%%%%%%%%%%%%%%%%%%%%%%%%%%%%%%%

%************************************************************ 
\subsubsection{Photoabsorption reactions.}\label{sec:PHOTON}
%***********************************************************

The inclusive photoabsorption cross section is given by the
following expression:

\begin{equation}\label{sigma}
\sigma_\gamma(\omega)=  \frac{4}{3} \pi^2 \,\alpha\, \omega\, R(\omega)\,,
\end{equation}
where $\alpha$ is the fine structure constant, $\omega$ represents the photon energy 
and $R(\omega)$ is the  response of the nucleus to the photon perturbation in 
unretarded dipole approximation, defined as
\begin{equation}\label{response}
   R(\omega)=\sum\!\!\!\!~\!\!\!\!\!\!\int\,df\,
             |\bra \Psi_f|\hat{\bi{D}}| \Psi_0 \ket |^2 \delta(E_f -\omega-E_0) \, .
\end{equation} 
Here $|\Psi_0\rangle$ and
$E_0$ are the nuclear ground state wave function and energy, $|\Psi_f\rangle$ 
and $E_f$ denote eigenstates and eigenvalues of 
the nuclear Hamiltonian $\hat H$, and 
\begin{equation}\label{operator}
\hat{\bi{D}} =\frac{1}{2}\sum_i^A \textbf{r}_i \tau_i^3\,,
\end{equation}
where $A$ is the number of nucleons, $\tau^3_i$  is the third 
component of the isospin operator and $\bi{r}_i$ the coordinate of the $i$th particle in 
the centre of mass frame.

As mentioned above, equations~(\ref{response})--(\ref{operator}) represent the so--called unretarded 
dipole approximation to the
total cross section. The approximations consist in the following: i) neglect of target recoil;
ii) neglect of all electric multipoles $E_j$ of order higher than $j=1$; iii) after application of 
charge conservation, known as `Siegert's 
theorem', reduction to the first order term in the Taylor expansion of 
the Bessel function of order $j=1$ in $\hat{\rho}_{jm}(q)$ (see~\cite{DEFOREST:1966}); iv) neglect of all
magnetic multipoles $M_j$ in the interaction between the photon and the nuclear currents.
It is clear that approximations i)--iii) are justified for `low'--energy photons.
However, the limit of validity of approximation iv) needs to be investigated, because it is
relevant in relation to the problem of  nuclear currents and of their consistency
with the potential, i.e. the problem, already mentioned in the introduction, of understanding the 
implicit degrees of freedom of nuclear dynamics. This can be understood considering the following:
as already mentioned above in iii) the dipole operator $\hat{\bi{D}}$ is the long wavelength limit of the 
$j=1$ term in $\hat{\rho}_{jm}(q)$. The longitudinal $C_1$ operator
(and not the transverse $E_1$ multipole operator) enters the cross section,
because one has used charge conservation connecting
the current density to the charge density, thus one works with the so--called Siegert's operator
(`Siegert theorem'). Therefore, while  
$\hat{\bi{D}}$ is written in terms of single nucleon coordinates, information
on many--body currents is included. 
%This is due to the application 
%of charge conservation in the derivation of the cross section, known as `Siegert theorem'. 
In fact for energies where approximations ii) and iii) are valid, the knowledge of the 
form of the potential is sufficient to include all consistent 
currents in the dipole cross section, without knowing their formal expressions.
However, Siegert's theorem does not apply to magnetic multipoles therefore if iv) is not
valid one needs models for the many--body currents that can be tested in the comparison with
experiments.
 
Of course also the extension in energy of the 
validity of ii) and iii) has  to be investigated. Some studies have been 
performed in the past on the two-- and three--body systems~\cite{ARENHOEVEL:1991,GOLAK:2002}, 
but the situation for heavier nuclei is still unclear. 
%This makes the study of 
%the photoabsorption cross section particularly interesting and important.

Regarding the exclusive reaction of a photon induced two--body break--up of the nucleus 
into a nucleon and a residual $(A-1)$ system, one can write the  
total exclusive cross section $\sigma_\gamma^{TB}$,
within the approximations i)--iv) mentioned above,  as

\begin{equation}\label{sigmaexcl}
\sigma_\gamma^{TB}(\omega) =
\frac{4}{3} \pi^2 \,\alpha \,\omega \,k \,\mu \,\sum_{M_N,M_{(A-1)}}\int\,\,d\Omega_{k}\,
             |\bra \Psi^-_{N,(A-1)}(E_{N,(A-1)})|\hat{\bi{D}}| \Psi_0 \ket |^2 \,,
\end{equation}
where $\mu$ and $k$ denote the reduced mass and the relative momentum of the fragments, respectively, 
$M_N$ and $M_{(A-1)}$ the total angular momentum projections of the nucleon and the $(A-1)$ system, 
$\Psi^-_{N,(A-1)}$ is the final state continuum wave function of the minus type 
pertaining to the $N,(A-1)$ channel~\cite{GOLDBERGER:1964} and $E_{N,(A-1)}$ is the 
energy of the final state
\begin{equation}\label{ENAminus1}
E_{N,(A-1)}=\omega + E_0=\frac{k^2}{2 \mu} + E_{(A-1)}\,,
\end{equation}
with $E_{(A-1)}$ representing the ground state energy of the fragment $(A-1)$.

It is evident that $R(\omega)$ in~(\ref{response}) has the form~(\ref{re}). Therefore in order to calculate
it, one has to invert the LIT~(\ref{ov}) obtained by solving~(\ref{eq1}) ($\hat O = \hat O' =\hat{\bi{D}}$).
In the exclusive case the matrix element in (\ref{sigmaexcl}) can be calculated using the LIT method 
explained in~\sref{sec:EXCL}, again with $\hat O =\hat{\bi{D}}$.

%****************************************************
\subsection{Neutrino reactions}\label{sec:NEUTRINO}
%*****************************************************

The interest in inelastic neutrino reactions with nuclear targets stems from
the role they play in major questions of contemporary physics. Such
reactions are of central importance in various astrophysical phenomena, such
as supernova explosions and the nucleosynthesis of the elements. 

Core collapse supernovae are widely accepted to be a neutrino driven
explosion of a massive star. When the iron core of a massive star becomes
gravitationally unstable it collapses until short--range nuclear forces halt
the collapse and drive an outgoing shock through the outer layers of the
core and the inner envelope. However, the shock loses energy through
dissociation of iron nuclei and neutrino radiation, and gradually stalls, it
becomes an accretion shock. It is believed, but to date not proved, that the
shock is then revived as neutrinos emitted from the collapsed core (the
proto--neutron star) deposit energy in the collapsing layers to overcome the
infall and eventually reverse the flow to an outgoing shock which explodes
the star. 
%Hydrodynamic simulations of a collapsing star, which are
%restricted to spherical symmetry, fail in reviving the shock~\cite{LI01}.
%Lately it was shown~\cite{BU03} that even in full 2-D calculations the shock
%is not revived. 
In order to revive the shock, the neutrinos must deposit
about $1\%$ of their energy in the matter behind the shock. The
latter, which is assumed to be in thermodynamic equilibrium, is composed
mainly of protons, neutrons, electrons, and light nuclei. In
contrast to the fairly known cross sections of neutrinos with electrons and
nucleons, the inelastic interaction of neutrinos with light nuclei such as
$^{4}\mathrm{He}$, $^{3}\mathrm{He}$ or $^{3}\mathrm{H}$ cannot be
accurately measured, and one must rely on theoretical calculations in order
to evaluate the cross sections.

The neutrinos migrating out of the proto--neutron star are in flavor equilibrium
for most of their migration. The electron--neutrinos remain in equilibrium
with matter for a longer period than their heavy--flavor counterparts, due to
the larger cross sections for scattering of electrons and because of charge
current reactions. Thus the heavy--flavor neutrinos decouple from deeper
within the star, where temperatures are higher. Typical calculations yield
temperatures of $\sim 10$~MeV for $\mu$-- and $\tau$-- neutrinos
\cite{WOOSLEY:1988}, which is approximately twice the temperature of
electron--neutrinos.
Consequently, there is a considerable amount of $\nu _{\mu ,\tau }$ with
energies above $20$~MeV that can dissociate the $^{4}\mathrm{He}$, not
to mention other nuclei, through neutral current reaction. 

The flux of neutrinos emitted in the collapse process is sufficiently
large to initiate nucleosynthesis in the overlaying shells of heavy
elements. Neutral reactions of alpha particle and neutrino in the inner helium
shell are part of reaction sequences leading to the production of the
rare $A=7$ lithium and beryllium isotopes~\cite{WOOSLEY:1990}.
Thus, better understanding of the $\nu$--$\alpha$ reaction
can lead to better prediction for the abundances of these elements. 

In the following we shall limit the discussion to neutral current 
neutrino--nuclei reactions. The extension to charged current reactions is
straightforward, due to isospin symmetry.

In the limit of small momentum transfer (compared to the Z--boson rest
mass) and using the multipole expansion (see~\sref{sec:MULT}, including the footnote), 
the cross section can be written as~\cite{DONNELLY:1976} 
\begin{eqnarray}\label{eq:nu_crs} 
   \frac{d\sigma}{dk_{f}} &=&
   \frac{4G^{2}}{2J_{0}+1}k_{f}^{2}
   \int_{0}^{\pi}\sin{\theta}\,d\theta\,\times
   \cr & \times & 
   \left\{
   \left(
   \sin^{2}\frac{\theta}{2}-\frac{q^{\mu }q_{\mu }}{2q^{2}}\cos^{2}\frac{\theta}{2}
   \right)
   \sum_{J,j\geq 1}
   \left[
   {R}^J_{Mj}(q,\omega)+{R}^J_{Ej}(q,\omega)
   \right]
   \right.
    \cr &\mp&
   \sin {\frac{\theta}{2}}\sqrt{\sin^{2}\frac{\theta }{2}-
   \frac{q^{\mu }q_{\mu }}{2q^{2}}\cos ^{2}\frac{\theta }{2}}
   \sum_{J,j\geq 1}2{R}^J_{{Ej},{Mj}}(q,\omega)\,
   \cr &+&
   \left.
   \cos^2\frac{\theta}{2}\sum_{J,j\geq 0} R^J_{{Cj}-\frac{\omega}{q}{Lj}}(q,\omega)
   \right\} \,,
\end{eqnarray}
where $k_f$ is the final momentum of the neutrino and $\theta$ is the scattering angle.
The rest of the notation is the same as
for the electron scattering cross section (\sref{sec:ELECTRON}).
The minus sign $(-)$ in~(\ref{eq:nu_crs}) stands for neutrino scattering and the plus
$(+)$ for antineutrino. The functions $R_{\hat{O},\hat{O}'}(q,\omega)$
%\begin{eqnarray}
%R_{\hat{O},\hat{O}'}(\omega )=\int d\Psi _{f} 
%{\langle \Psi_{0}\mid \mid \hat{O}\mid \mid \Psi _{f}
%\rangle \langle \Psi _{f}\mid \mid \hat{O}'\mid \mid \Psi
%_{0}\rangle \delta \left(E_{f}-E_{0}-\omega+\frac{q^2}{2 M_T}\right)} 
%\end{eqnarray}
are the response functions with respect to the multipole transition operators $\hat{O}$ 
and $\hat{O}'$ (when $\hat{O}=\hat{O}'$ we use the notation $R_{\hat{O}}=R_{\hat{O},\hat{O}}$).
 At the supernova energy regime the neutrinos carry few tens of~MeV. Therefore, it is
both useful and instructive to apply the long wavelength approximation
and restrict the sum in (\ref{eq:nu_crs}) to the leading multipoles.
In the limit $q \to 0$ they are the
Gamow--Teller $E^A_1$,$L^A_1$ and Fermi $C_0^V$ operators which are $q$--independent
and then the axial vector operators
$E^A_2, L^A_2, M^A_1, L^A_0 $ and the vector $C^V_1, E^V_1, L^V_1$,
which are linear in $q$ .
The main low--energy contribution usually comes from the Gamow--Teller and Fermi operators.
However, for double closed shell nuclei such as the $^{4}\mathrm{He}$ nucleus
these operators are suppressed and the leading contribution is linear in $q$.
The complete current is composed of one--body terms (the impulse approximation)
and many--body terms. 
It turns out that in the long wavelength limit the leading axial two--body
current contribute mainly to the Gamow--Teller term~\cite{PARK:2003}.

%***********************************
\section{Numerical examples}
\label{sec:NUM}
%********************************

In this section we perform various numerical tests, both internal to the LIT method itself,
and regarding the comparison with other approaches. We take as test observable the total 
photoabsorption cross section in
unretarded dipole approximation (see~\sref{sec:PHOTON}) of $^2$H and $^4$He. 
In the first two sections two interesting test cases for the reliability 
of the inversion methods, as well as of the whole LIT approach, are presented. 
In~\sref{sec:SMOOTH} we discuss the comparison between a response 
obtained via the LIT method and the corresponding LCZR of~\sref{sec:DIAG}).
Finally, in~\sref{sec:EIGENV_LANCZOS} we present a comparison
between the eigenvalue method and the use of the Lanczos algorithm 
to calculate the LIT (see sections~\ref{sec:EIGENVALUE} and~\ref{sec:LANCZOS}).

%********************************************************************************
\subsection{A test of the inversion algorithm on the deuteron}\label{sec:INVDEUT}
%********************************************************************************

\begin{figure}[htb]
\centerline{\resizebox*{12cm}{7cm}{\includegraphics*[angle=0]{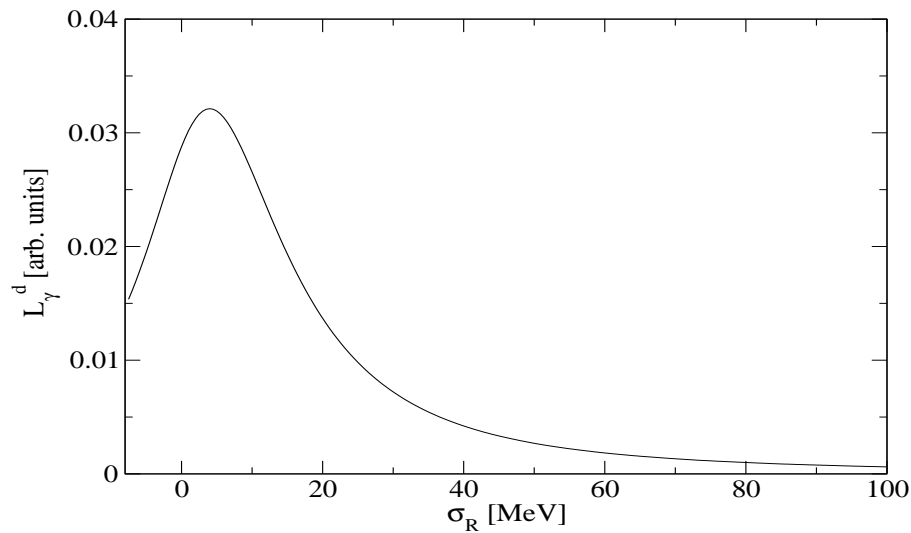}}}
\caption{LIT of the deuteron photoabsorption cross section $L_{\gamma}^d$ of~(\ref{LITdeut}) with
$\sigma_I=10$ MeV.}
\label{I1}
\end{figure}

The deuteron total photoabsorption cross section is calculated
in strict analogy to~\cite{EFROS:1993}. 
Following the notations there, one concentrates on the following LIT:
\begin{equation}\label{LITdeut}
L_\gamma^d(\sigma_R,\sigma_I) = {1\over 3}\sum_{jl}(2 j + 1) \int_0^\infty dr
\,|\phi_{L=1,jl}(\sigma_R,\sigma_I,r)|^2\,.
\end{equation}
Here only electric dipole transitions are taken into account ($L=1$). Then one has to
consider the following combinations for total angular momentum and angular momentum quantum
numbers $j$ and $l$ of the outgoing neutron--proton ($np$) pair:
$(j,l)$ = (0,1), (1,1), (2,1), (2,3). 
The inversion of $L_\gamma^d$ leads to the unretarded dipole response function
(\ref{response}) and the corresponding cross section is given by ~(\ref{sigma}).
The complex $\phi_{L=1,jl}$ are obtained from the 
same radial differential equations
as equations (27) and (28) of~\cite{EFROS:1993}, but with a complex
$\sigma=-\sigma_R+i \sigma_I$ and taking for the excitation operator $\theta$
the dipole operator $\hat{\bi{D}}$ of~(\ref{operator}).
The radial differential equations can be
solved in two different ways: i) direct numerical solution, as normally done for
a two--nucleon problem, and ii) expansion of the solution in a complete set
with a subsequent determination of the expansion coefficients. In this section we use 
method i), while method ii) is taken for the study in~\sref{sec:SMOOTH}.

\begin{figure}[htb]
\centerline{\resizebox*{12cm}{7cm}{\includegraphics*[angle=0]{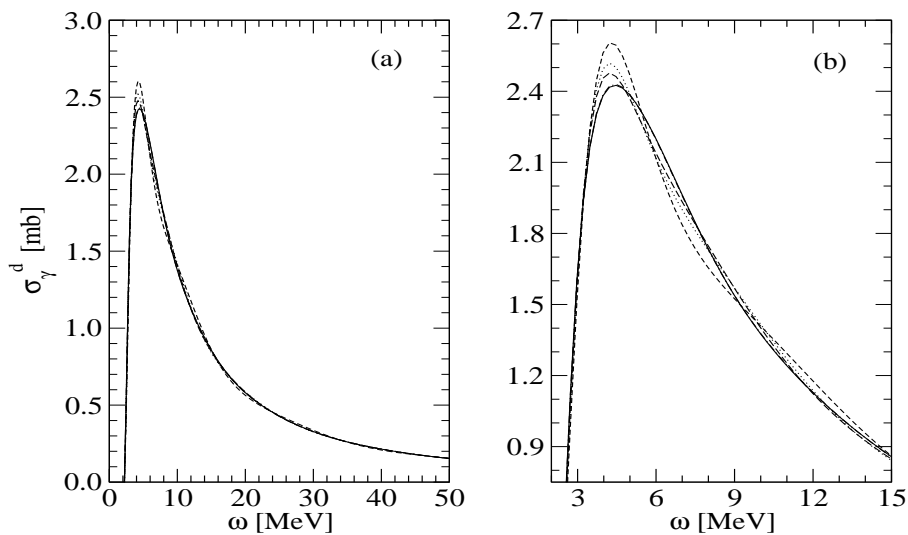}}}
\caption{Total deuteron photoabsorption cross section up to 50 MeV (a) and 15 MeV (b)
from inversion of $L_{\gamma}^d$ with various $N_{max}$-values (see (\ref{sumr}):
10 (short dashed), 15 (dotted), 20 (long dashed), 25 (solid), 26 (dash--dotted).}
\label{I2}
\end{figure}

In~\fref{I1} we show $L_\gamma^d$  calculated with the AV14 NN potential~\cite{WIRINGA:1984}.
One sees that the transform is dominated by a low--energy peak and that further
structures are not present. An inversion of $L_\gamma^d$ with the basis set~(\ref{bset}) 
leads to a rather slow convergence in $N_{max}$ and thus we use the enlarged set~(\ref{bset2})
with $\alpha_1=3/2$, $\beta_{min}=1$ and $\beta_{max}=2$. In the past 
we did not observe a similar slow convergence for other LIT applications.
This is caused by the steep low--energy rise of the cross section combined
with a rather narrow  peak at only about 2~MeV above threshold.
In~\fref{I2} we illustrate the convergence pattern of the cross section as a function of 
$N_{max}$.
One sees that the peak height is steadily reduced with growing $N_{max}$, while 
the cross section is quite stable at $E>10$~MeV. The reduction of the peak height ends with
$N_{max}=25$; $N_{max}=26$ leads to an almost identical result, 
and also for even higher
$N_{max}$ the decrease of the peak height does not continue further.
Therefore we consider the inversion result with $N_{max}=25$ as the final result for 
the cross section.
%One sees that the exact determination of the peak height and
%shape request a rather high number of basis functions. On the other hand the
%final cross section results showsThis is due to
%the already mentioned steep rise of the deuteron response, which almost approaches
%a step function. Nonetheless one can that a stable inversion result is obtained
%(fine tuning of exact result).

\begin{figure}[htb]
\centerline{\resizebox*{12cm}{7cm}{\includegraphics*[angle=0]{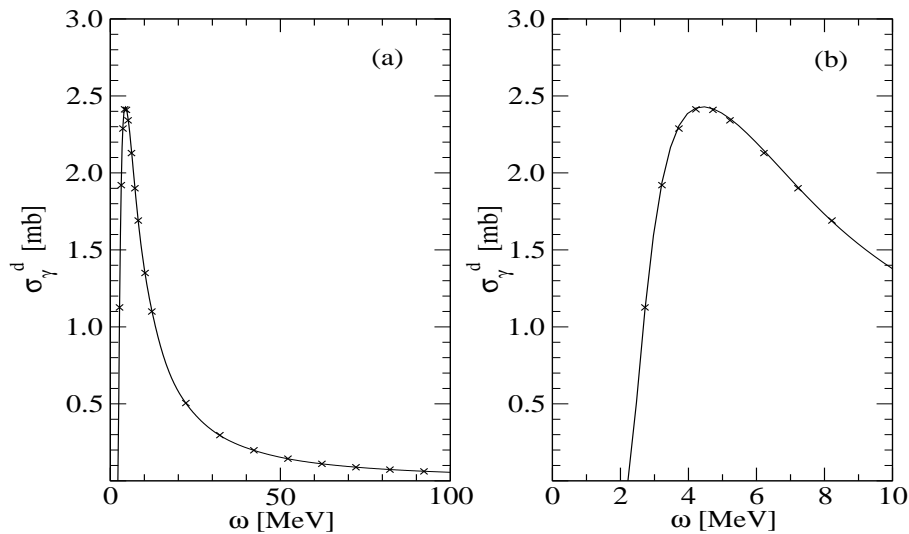}}}
\caption{Total deuteron photoabsorption cross section up to 100 MeV (a) and 10 MeV (b):
LIT result (solid) and from calculation with explicit $np$ final state wave function (x).
}
\label{I3}
\end{figure}

As mentioned one can make a conventional calculation of the 
cross section  by using the proper $np$ scattering states. In~\fref{I3}
we show the result of such a calculation in comparison to that obtained via the LIT 
method, plotted in~\fref{I2}. One observes an excellent
agreement between the two calculations, showing the high precision that can be
obtained with the LIT method.

%*****************************************************************************
\subsection{Lanczos response vs. inverted LIT response}\label{sec:SMOOTH}
%*****************************************************************************

In this section we discuss the comparison  between a response obtained with the LIT method
and the corresponding LCZR calculated as described in~\sref{sec:DIAG}. To this end we reconsider the 
deuteron total photoabsorption cross section (see~\sref{sec:INVDEUT}). However, this time we use
the non--local JISP6 potential~\cite{SHIROKOV:2003}. This potential is parametrized in a harmonic 
oscillator (HO) basis and hence we also expand $\phi_{L=1,jl}$ of~(\ref{LITdeut}) over the same HO basis,
up to a maximal value of the HO quantum number $N_{\it HO}$
and solve for $L_\gamma^d$ with the Lanczos method, as explained in~\sref{sec:LANCZOS}. (Results for such
a calculation with the JISP6 potential are already published in~\cite{BARNEA:2006}).
Here we discuss this case because it allows to  demonstrate in a clear way the difference between  
a LIT inversion result and the LCZR of~\sref{sec:DIAG}. 

\begin{figure}[htb]
\centerline{\resizebox*{12cm}{7cm}{\includegraphics*[angle=0]{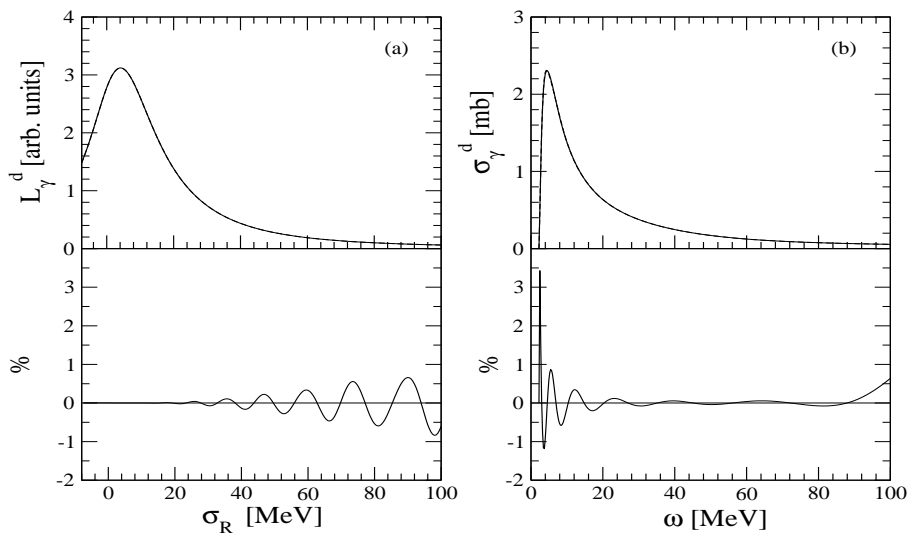}}}
\caption{Deuteron photoabsorption, (a): LIT $L_{\gamma}^d$ of~(\ref{LITdeut}) with
$\sigma_I=10$ MeV and (b): corresponding cross section (b). Upper panels:
absolute results with $N_{HO}=150$ (dashed) and $N_{HO}=2400$ (solid). 
Lower panels: relative difference of results with $N_{HO}=150$ and $N_{HO}=2400$ in \%.
}
\label{I5}
\end{figure}

In~\fref{I5} we show $L_\gamma^d$ and the cross section with the JISP6 potential using
expansions of $\phi_{L=1,jl}$ with $N_{\it HO}=150$ and 2400.
In both cases the maximal number of possible Lanczos steps, i.e. $N_{\it HO}$, has been
carried out. One observes that the increase of $N_{\it HO}$ has no visible
effect on $L_\gamma^d$ (relative differences are below 0.01\% in the whole peak region). 
Thus it is not surprising that also the corresponding cross sections, are practically identical
(inversions of $L_\gamma^d$ are made in a similar
way as discussed for the AV14 case in~\sref{sec:INVDEUT}).

\begin{figure}[htb]
\centerline{\resizebox*{12cm}{7cm}{\includegraphics*[angle=0]{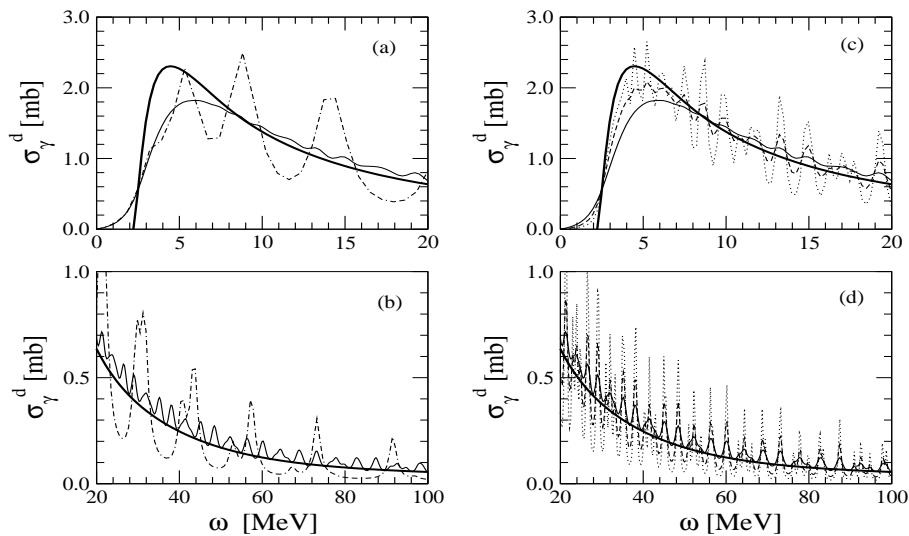}}}
\caption{Deuteron total photoabsorption cross section, LIT result (thick solid)
and various LCZR results:  $\sigma_I=1$ MeV with $N_{HO}=150$ (dash--dotted) and $N_{HO}=2400$ 
(thin solid) in (a) and (b); $N_{HO}=2400$ and $\sigma_I=1$ MeV (thin solid), 
$\sigma_I=0.5$ MeV (dashed) and $\sigma_I=0.25$ MeV (dotted) in (c) and (d).
}
\label{I6}
\end{figure}

It is interesting to compare the obtained cross section with that of
a LCZR approach. The results are illustrated in~\fref{I6},
where for the LCZR a halfwidth $\sigma_I$ of 1~MeV is taken.
With $N_{\it HO}$=150 a very poor result is obtained, exhibiting very
strong unphysical oscillations. With $N_{\it HO}$=2400 the oscillations are considerably 
reduced, but they are still rather pronounced beyond 10~MeV. In addition the cross section
is not reproduced correctly neither in the peak region nor at higher energies.
One also notes a further deficiency: the cross section is non vanishing below
the break--up threshold. If one decreases the halfwidth of the LCZR further
to 0.5 and 0.25~MeV one does not gain much of an improvement, as shown in~\fref{I6} (c),(d)
for the $N_{\it HO}$=2400 case. The oscillations become visible also at lower
energies and their magnitude grows considerably. One would
need to increase $N_{\it HO}$ much further in order to have a realistic result with
a smooth and correct low--energy cross section. 

The oscillatory behaviour of the LCZR is due
to the fact that only a tiny fraction of the Lanczos energy eigenvalues falls
into the energy range of interest, while the vast majority of the about 10$^4$
eigenvalues (there are four different ($j,l$) combinations (see text after~(\ref{LITdeut})
) is located beyond 100~MeV.
On the other hand we want to emphasize that $N_{\it HO}$=150
is fully sufficient to have a converged LIT for $\sigma_I$ of 10~MeV, which after
inversion leads to a practically identical cross section as the $N_{\it HO}$=2400 case
(see~\fref{I5}). This shows that via the LIT approach
the information is processed in a much better way than with the LCZR.
On the other hand for many--body calculations one finds cases in the literature 
(see e.g.~\cite{HAXTON:2005}),
where one can work also successfully with the LCZR. In these
cases one has a huge number of states 
and Hamiltonians (e.g. effective $sd$--shell interactions)
that generate most of the Lanczos eigenvalues in the  range of physical interest.

%**********************************************************************************
\subsection{A test of the LIT method on the $\alpha$--particle}\label{sec:ALPHA}
%**********************************************************************************

\begin{figure}[htb]
\centerline{\resizebox*{12cm}{7cm}{\includegraphics*[angle=0]{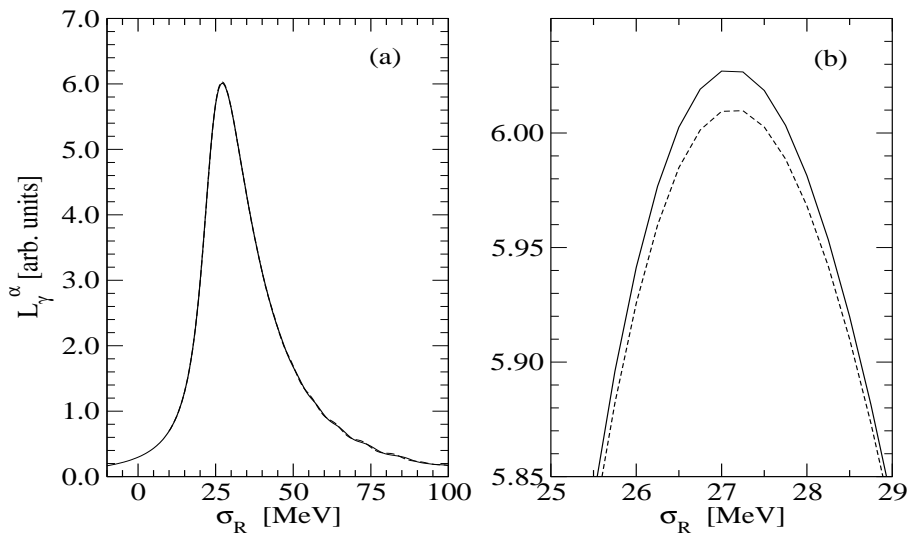}}}
\caption{$^4$He photoabsorption: LIT $L_{\gamma}^{\alpha}$ ($\sigma_I=5$ MeV) up to 100
MeV (a) and a zoom on the peak region (b) for $K_0=8$, $K_{LIT}=9$ (dashed) and for $K_0=14$, $K_{LIT}=15$
(solid).}
\label{I7}
\end{figure}
\begin{figure}[htb]
\centerline{\resizebox*{12cm}{7cm}{\includegraphics*[angle=0]{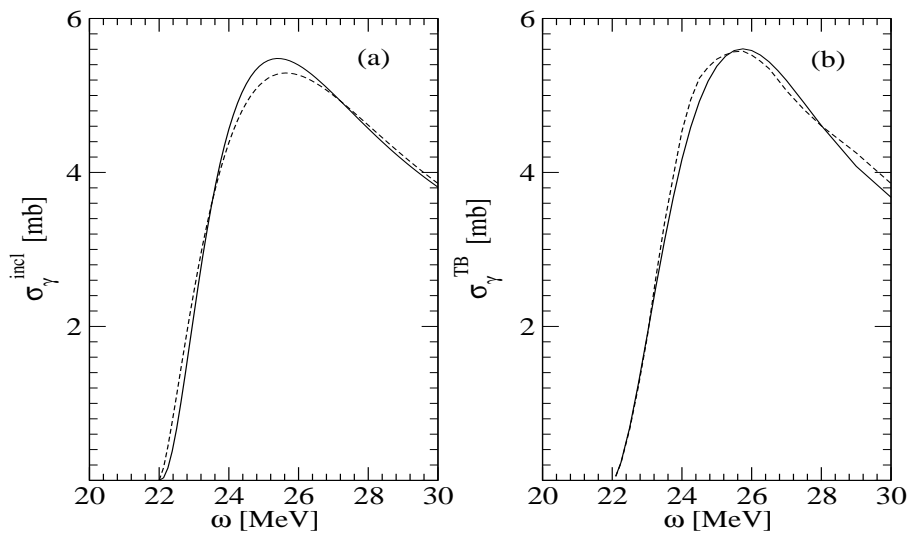}}}
\caption{$^4$He photoabsorption, convergence of HH expansion for $\sigma_{\gamma}^{incl}$
(a) and $\sigma_{\gamma}^{TB}$ (b): $K_0=8$, $K_{LIT}=9$ (dashed) and $K_0=14$, $K_{LIT}=15$
(solid). }
\label{I8}
\end{figure}
In this section we illustrate a test of `internal' consistency for the LIT
method, when applied both to inclusive and exclusive cross sections. To this aim we 
calculate the total photoabsorption cross section of $^4$He in unretarded dipole approximation in
two different ways, namely in the direct way as an 
inclusive reaction (\sref{sec:INCL}) leading to $\sigma_\gamma^{incl}$
and in an indirect way by summing the total cross sections of
the $^4$He$(\gamma,p)^3$H and $^4$He$(\gamma,n)^3$He channels leading 
to $\sigma_\gamma^{TB}$. The latter 
are determined by using the LIT formalism for exclusive reactions 
(see~\sref{sec:TWOBODYBR} and~\ref{sec:COULOMB}).
Both approaches should yield identical results below the three--body break--up threshold,
whereas beyond it  one has 
$\sigma_\gamma^{incl} \ge \sigma_\gamma^{TB}$,
since also other reaction channels are open.
Such a comparison between $\sigma_\gamma^{incl}$ and $\sigma_\gamma^{TB}$ 
has already been made~\cite{QUAGLIONI:2004} for a LIT calculation taking the MT--I/III 
potential~\cite{MALFLIET:1969} as NN interaction. A rather good agreement between $\sigma_\gamma^{incl}$
and $\sigma_\gamma^{TB}$ has been  obtained,
but at very low energies non--negligible differences have been found.
Since the low--energy region is particularly interesting, e.g. for reactions
of astrophysical relevance,
it is important to re--investigate this case.
To this end we choose a very simple NN potential, namely the Volkov model~\cite{VOLKOV:1965},
which is a soft--core central interaction. In addition we neglect the Coulomb force, thus
one has $\sigma_\gamma^{TB} = 2 \,\sigma(^4$He$(\gamma,p)^3$H).
The LIT calculation is carried out using HH expansions via
the EIHH method (see Appendix B). In~\fref{I7} we show  the LIT of the total 
cross section with $\sigma_I=5$~MeV, where
we take the following maximal grand--angular momentum quantum numbers $K$: i) $K_0=8$
(ground state), $K_{LIT}=9$ (LIT function) and ii) $K_0=14$, $K_{LIT}=15$. It is evident that the
difference between the two LITs is very small (about 0.5~\% in the peak region). Nonetheless one encounters
important consequences for the inversion. For the case with $K_{LIT}=9$ one obtains unstable inversions
already for $N_{\rm max} > 7$, while the stability range extends to
$N_{\rm max}=15$ for the $K_{LIT}=15$ case. Similar results are found for the various
inversions which are necessary to determine $\sigma_\gamma^{TB}$.
\begin{figure}[htb]
\centerline{\resizebox*{12cm}{7cm}{\includegraphics*[angle=0]{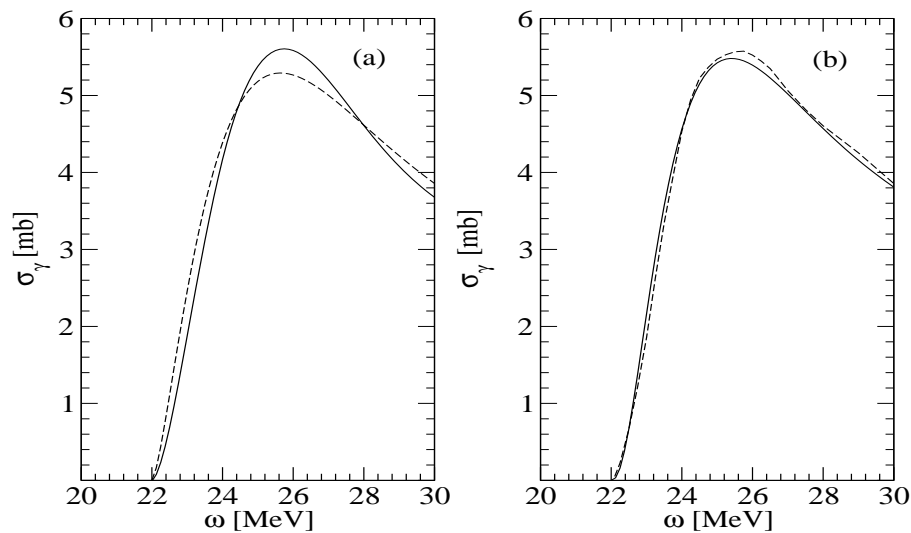}}}
\caption{$^4$He photoabsorption, comparison of $\sigma_{\gamma}^{incl}$ (dashed) and 
$\sigma_{\gamma}^{TB}$ (solid): $K_0=8$, $K_{LIT}=9$ (a) and $K_0=14$, $K_{LIT}=15$ (b).} 
\label{I9}
\end{figure}

In~\fref{I8} we show results for the cross sections $\sigma_\gamma^{incl}$ and 
$\sigma_\gamma^{TB}$ with the two different sets of $K$ values. 
One observes that for $\sigma_\gamma^{incl}(K_0=8,K_{LIT}=9)$ there is a steeper low--energy rise  
and a lower peak cross section than for $\sigma_\gamma^{incl}(K_0=14,K_{LIT}=15)$.
On the contrary for $\sigma_\gamma^{TB}$ one finds a good agreement up to
about 1~MeV, but somewhat different peak shapes. In~\fref{I9}
we compare $\sigma_\gamma^{incl}$ and $\sigma_\gamma^{TB}$ for the
same $K$ value sets. It is evident that with the lower number of retained HH 
one does not have 
consistent results below the three--body break--up threshold of about 28~MeV,
but with a  higher number of them there is a much better picture. 
\begin{figure}[htb]
\centerline{\resizebox*{12cm}{7cm}{\includegraphics*[angle=0]{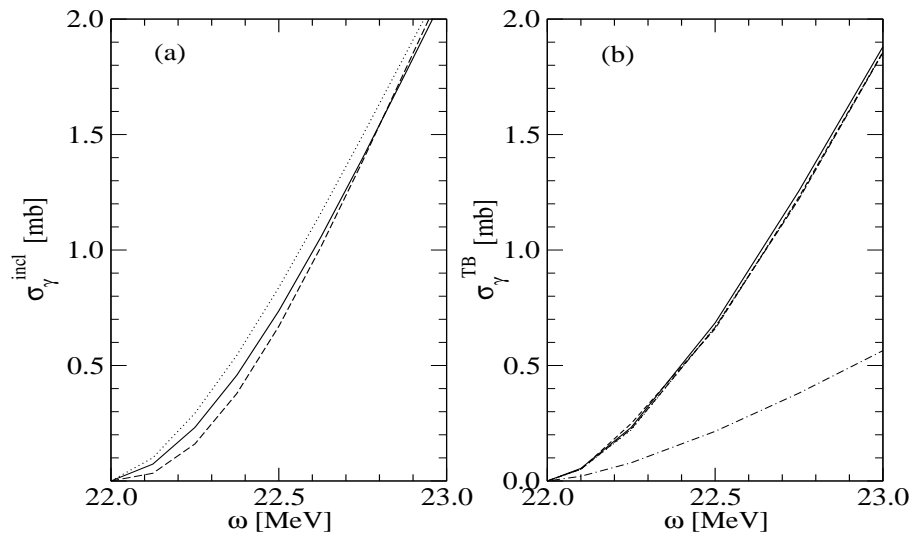}}}
\caption{Low--energy $^4$He total photoabsorption, $\sigma_{\gamma}^{incl}$ (a) and
$\sigma_{\gamma}^{TB}$ (b) with $K_0=14$, $K_{LIT}=15$. Inversion results with
set (3.12) for $\sigma_I=5$ MeV (solid), $\sigma_I=10$ MeV (dotted) and from 
set (3.15) with $\sigma_I=5$ MeV (dashed). In (b) also shown the Born result (dash--dotted).
}
\label{I4}
\end{figure}
In order to judge the quality of the agreement of the results we have also investigated
the uncertainty in the inversions using set~(\ref{bset}) with
two different $\sigma_I$ values ($\sigma_I=5$ and 10 MeV) and using the
inversion set~(\ref{bset2}) with $\sigma_I=5$ MeV. For $\sigma_{\gamma}^{incl}$ with 
$\omega \le 28$ MeV we find a rather energy independent error of about 0.08 mb.
On the contrary the error on $\sigma_{\gamma}^{TB}$ depends on $\omega$. It is close
to zero below 23 MeV, while in the peak region it amounts up to about 0.08 mb.
In~\fref{I4} we illustrate the situation below $\omega = 23$ MeV. It is evident that
one has very stable threshold results for $\sigma_{\gamma}^{TB}$, although
the final state interaction (FSI) contribution is rather large as the comparison 
with the Born result shows. On the other hand 
$\sigma_{\gamma}^{incl}$ exhibits a rather large relative error. The origin
of this rather different behaviour for $\sigma_{\gamma}^{incl}$ and $\sigma_{\gamma}^{TB}$
is due to the rather different LIT methods for inclusive and exclusive reactions.
The main difference lies in the necessity to evaluate  the principle value integral 
in~(\ref{pv}) for the exclusive
calculation. If the pole at $E=E'$ is sufficiently far from the cross section peak,
i.e. for the energies of~\fref{I4}, then small differences in the peak shape do
not matter much in the integral. On the contrary, in case of the inclusive calculation
one obtains one inversion result valid in the whole considered energy range
and small differences in the peak shape propagate to the threshold region
and can lead  to rather large relative errors there. Of course in the present
case one could also try to increase the precision of the calculation by further
increasing $K_0$ and $K_{LIT}$, which then could lead to similarly precise results
as for the deuteron case of~\fref{I3}.

As mentioned above, the low--energy region is of particular interest. Therefore it
is an important finding that $\sigma_{\gamma}^{TB}$ leads to very stable low--energy
results: the convergence in $K$ is rather good (see~\fref{I8}) and the inversion error
is quite small (see~\fref{I4}). 

%*****************************************************************************
\subsection{Eigenvalue versus Lanczos methods} \label{sec:EIGENV_LANCZOS}
%*****************************************************************************

\begin{figure}[htb]
\centerline{\resizebox*{12cm}{12cm}{\includegraphics*[angle=0]{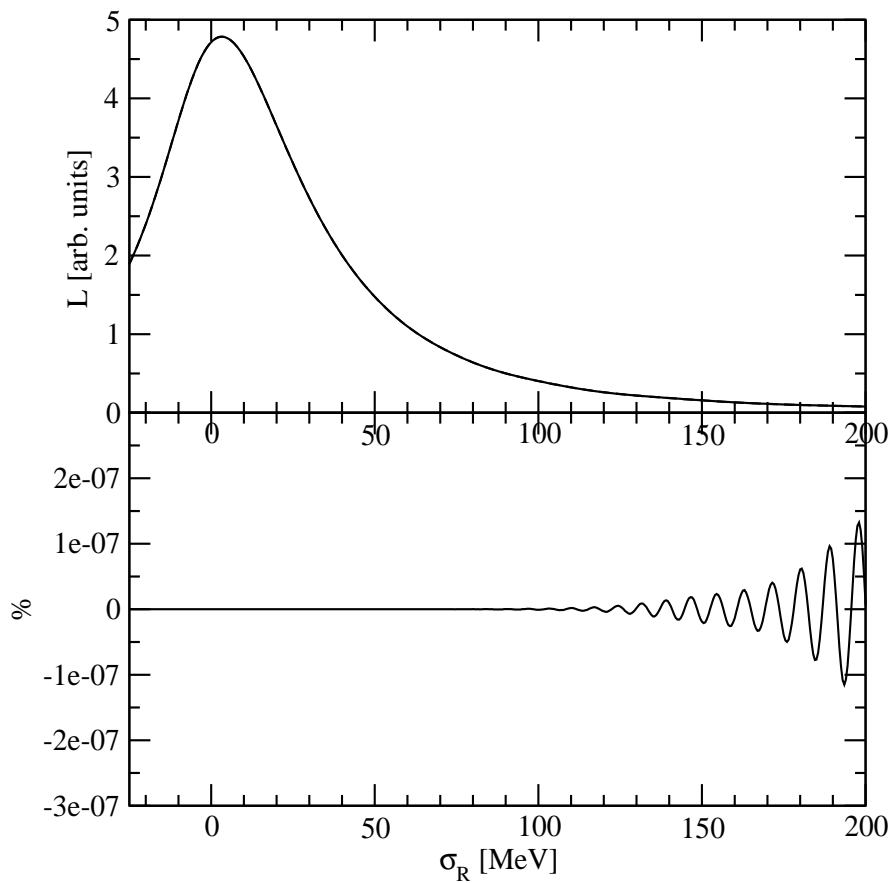}}}
\caption{\label{4He_MT13_photo_LANINV}
         Upper panel: $^4$He photoabsorption inclusive LITs calculated with the
         Hamiltonian inversion and with the Lanczos  algorithm (150 Lanczos coefficients). 
         The two curves are 
         indistinguishable on this scale. In the lower panel the relative difference
         between them is plotted.} 
\end{figure}
In the upper panel of~\fref{4He_MT13_photo_LANINV} 
we compare the inclusive LIT for the photodisintegration of $^4$He (see~\sref{sec:PHOTON})
obtained with the eigenvalue method and with the Lanczos algorithm. 
The results have been obtained expanding the $^4$He ground state $|\Psi_0\rangle$ and 
the LIT functions $|\tilde\Psi\rangle$ using the HH
basis and applying the EIHH method, described in Appendix B, to accelerate the
convergence. For simplicity we have used the 
semirealistic central MT--I/III NN potential~\cite{MALFLIET:1969}.
The ground state has been  expanded on the set of 
four--body (anti)symmetrized hyperspherical harmonics (SYMHH)~\cite{BARNEA:1997} 
characterized by angular momentum $L=0$, spin 
$S=0$, isospin $T=0$ and $T_z=0$, up to grandangular momentum quantum 
number $K_{max}=8$.  Since the dipole operator 
has $j=1$ and no spin degrees of freedom, the LIT function has been expanded on the set of 
four--body SYMHH associated with the quantum numbers $L=1$, $S=0$, 
$T=1$ and $T_z=0$. The dipole operator only allows transition by one unit of $K$, 
therefore it has been sufficient to consider HH states with $K\leq K_{max}=9$. The LIT obtained
with these values of $K_{max}$ have shown a sufficient convergence. 
The number of hyperradial basis functions has been $21$ and the 
parameter $\sigma_I$ has been set equal to 20~MeV. Since one difficulty in the Lanczos 
method is retaining the
orthogonality of the basis vectors, we have checked this point and found that the
overlap between the {\it pivot} and the $1000$'th vector is smaller then
$10^{-7}$. 

In~\fref{4He_MT13_photo_LANINV} one sees that there is an 
excellent agreement between the results obtained in the two ways.
In fact, up to 250~MeV, the maximal difference is lower than $5\cdot 10^{-7}$ 
(see lower panel).
This picture deteriorates a bit when we decrease $\sigma_I$, but even for
$\sigma_I=10$~MeV the difference between the methods is better then $3\cdot
10^{-3}$ for the same $\sigma_R$ range. Thus we can conclude that the 
two methods are essentially equivalent.

In the following we discuss the main advantages of the Lanczos technique,
 which are i) its applicability to huge sparse matrices
and ii) a considerable reduction in
the CPU time.

\noindent i) It seldom happens that one has to deal with a Hamiltonian matrix which is either sparse
or written as an outer product. For such matrices the calculation of the product
$\hat H | \phi \rangle$ goes much faster then $O(n^2)$ ($n$ is the dimension of the matrix), 
and there is a great incentive to use algorithms, like the Lanczos algorithm, that exploits 
this feature. 
In fact, the real power of the Lanczos formulation lies here: it opens up
the way to work with a huge number of basis functions, which direct inversion methods
or full diagonalization methods fail to address. 
\begin{figure}[htb]
\centerline{\resizebox*{12cm}{12cm}{\includegraphics*[angle=0]{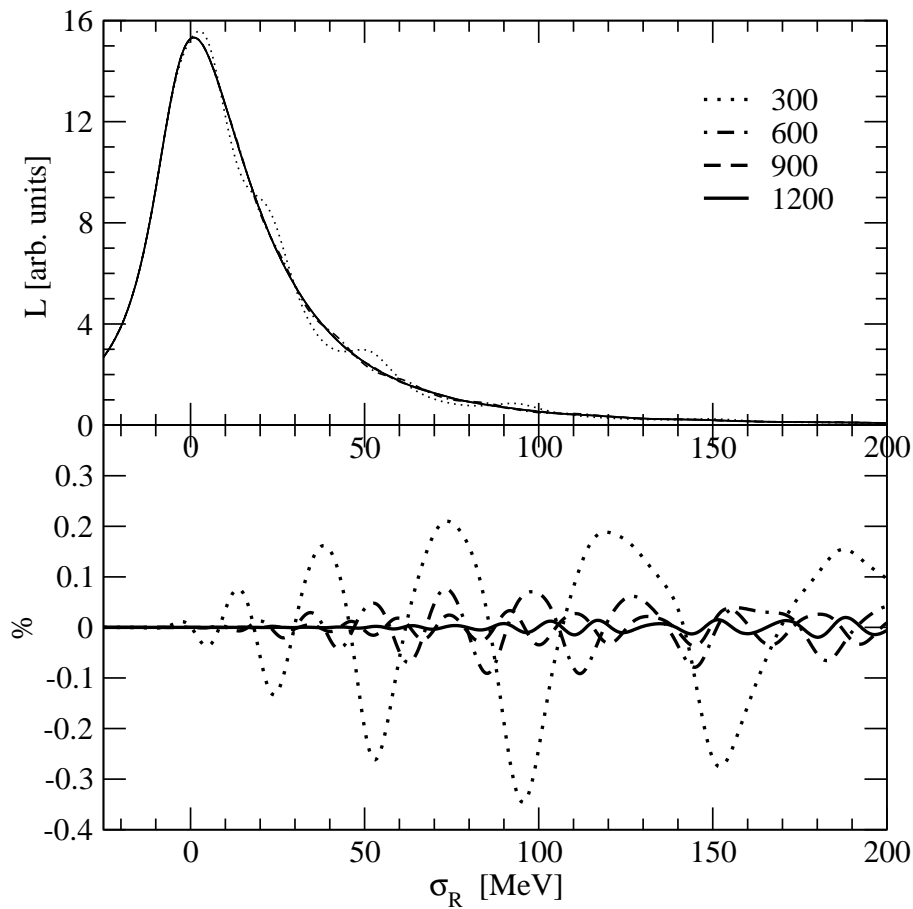}}}
\caption{Upper panel: convergence of the $^4$He photoabsorption inclusive LITs for
         $\sigma_I=10$~MeV and for different numbers N of Lanczos 
         coefficients. Lower panel: 
         relative errors with respect to the converged Lanczos result (N=1500)} 
\label{Fig3:lanc_conv_e10}
\end{figure}
\noindent ii) The reduction in CPU time results from the fast convergence 
of the lowest (and highest) eigenvalues, so that it is not necessary to perform
all the Lanczos steps. In fact, checking the 
convergence of $L(\sigma_R,\sigma_I)$ with respect to the number of Lanczos coefficients 
used in the continued fraction one finds the following: for
$\sigma_R \rightarrow \infty$ only the first Lanczos step is necessary, 
because the entire continued fraction is dominated just by $\sigma_R$. 
Also in the low $\sigma_R$ region, as expected, the continued 
fraction converges rapidly, since for $\sigma_R \rightarrow 0$  the 
dominating contribution to the term $1/(z-\hat{H})$ is given by the lowest
eigenvalue of $\hat H$, which is well approximated using only the first few 
Lanczos coefficients. It turns out that also in the intermediate $\sigma_R$ 
region a rather small number of Lanczos coefficients are sufficient.

\begin{figure}[htb]
\centerline{\resizebox*{12cm}{12cm}{\includegraphics*[angle=0]{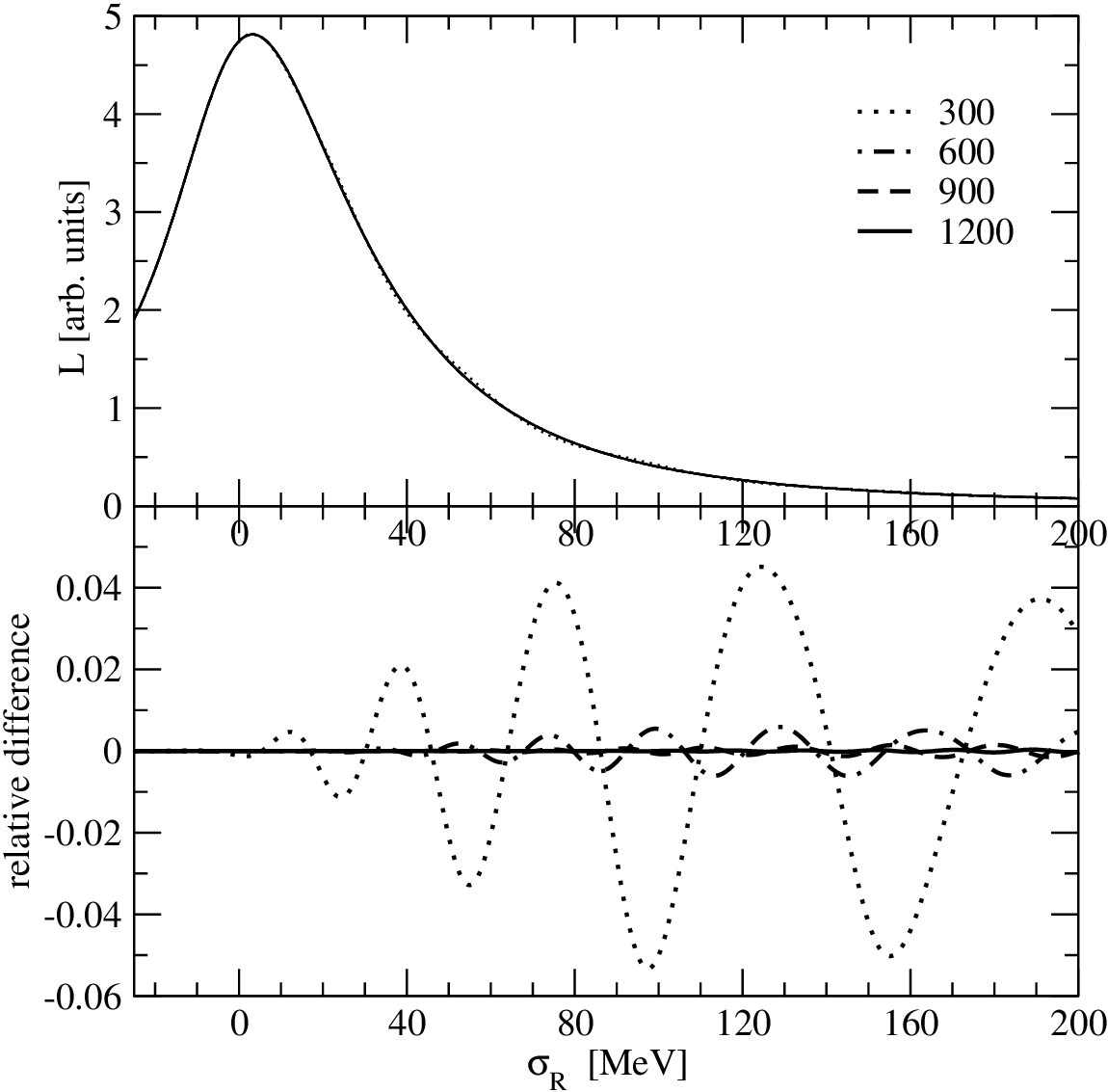}}}
\caption{Same as~\fref{Fig3:lanc_conv_e10}, but for $\sigma_I=20$} 
\label{Fig4:lanc_conv_e20}
\end{figure}

As an example in figures~\ref{Fig3:lanc_conv_e10} and~\ref{Fig4:lanc_conv_e20}  we show  
the convergence patterns with respect to the 
number of Lanczos steps for the LIT of the inclusive photodisintegration of 
$^{4}$He described above. This time, the calculation has been carried out with all
the SYMHH states up to $K_{max}=15$ and with $21$ hyperradial states, i.e.
a total of about $18,000$ basis functions. If written in matrix form the 
Hamiltonian would consume about 2.5 Gb of computer memory, and full 
diagonalization would be a long task.
It can be seen that even for a sizable basis set the transforms converge
very rapidly with only few hundreds Lanczos steps.
As expected the convergence of the LIT is very good near threshold, then the
number of Lanczos steps needed to achieve convergence grows with $\sigma_R$ to
a maximum value at about $\sigma_R\approx 100$~MeV and then, for higher values
of $\sigma_R$ convergence improves again.

To understand the effect of $\sigma_I$ on the convergence pattern it is
instructive to compare figures~\ref{Fig3:lanc_conv_e10} ($\sigma_I=10$~MeV) and
\ref{Fig4:lanc_conv_e20} ($\sigma_I=20$~MeV). Roughly speaking, the convergence
of the LIT improves by almost one order of magnitude when $\sigma_I$ is
increased from 10 to 20~MeV. This follows from the reduction in the sensitivity
of the LIT to the exact position of the eigenvalues of $\hat H$
(see \eref{L_epsnu_diag}), when $\sigma_I$ is large. 
In general for full tridiagonalization the Lanczos algorithm exhibits an
$O(n^3)$ time complexity.  However, 
since the continued fractions converge already with a number of Lanczos 
coefficients much smaller than $n$, the overall time needed to get the Lorentz 
integral transform is proportional to $n^2$. In practice we have found that
about $1000$ Lanczos steps are enough, even for very large bases. 

In summary, reformulating the LIT method with the Lanczos algorithm represents an
extremely powerful way to perform actual LIT calculations. All checks have shown
that it is as accurate as any other method and it is much more efficient from the
computational point of view.

%******************************************
\section{Results}\label{sec:RES}
%******************************************

In this section we present the results of various LIT applications, to few--body systems 
of increasing number of nucleons. Those obtained in the two--body system mainly represent 
important tests of the LIT method. The others are results obtained within different 
potential models, ranging from  semirealistic NN potentials (MT--I/III~\cite{MALFLIET:1969},
TN~\cite{EFROS:1999a}, AV4' and AV8'~\cite{WIRINGA:2002} (solid), MN~\cite{THOMSON:1977}) to realistic ones 
(Paris~\cite{LACOMBE:1980}, AV14~\cite{WIRINGA:1984}, AV18~\cite{WIRINGA:1995}, BonnRA~\cite{MACHLEIDT:1989}). 
Also various versions of the 3N forces have been employed, namely the
Urbana potentials UVIII~\cite{WIRINGA:1991} and UIX~\cite{PUDLINER:1997} and the Tucson--Melbourne 
(TM)~\cite{COON:1979}.

%**********************************************************************
\subsection{Reactions with the two--body system}\label{sec:TWOBODY}
%**********************************************************************

\begin{figure}[htb]
\centerline{\resizebox*{12cm}{12cm}{\includegraphics*[angle=0]{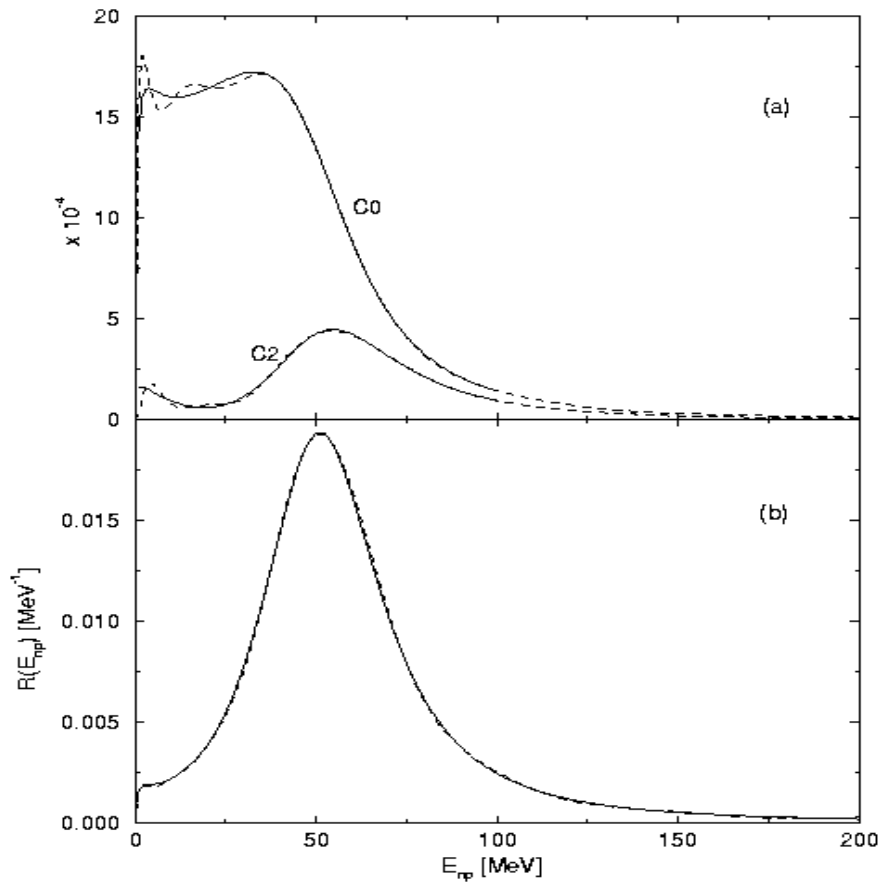}}}
\caption{Longitudinal form factor $R_L(q,E_{np})$ of the $d(e,e'p)n$ reaction
at $q$=440~MeV/c for Paris potential~\cite{LACOMBE:1980} of LIT calculation (dashed curves) 
and calculation with explicit $np$ final state wave functions (full curves): 
contribution of C0 and C2 multipoles (a) and total results (b).}
\label{R1}
\end{figure}

We start the discussion reviewing the very first LIT application, i.e. the calculation of
the deuteron longitudinal form factor $R_L(q,\omega)$ at $q=440$~MeV/c. In ~\cite{EFROS:1994},
where the LIT has been originally proposed, 
\begin{figure}[htb]
\centerline{\resizebox*{12cm}{12cm}{\includegraphics*[angle=0]{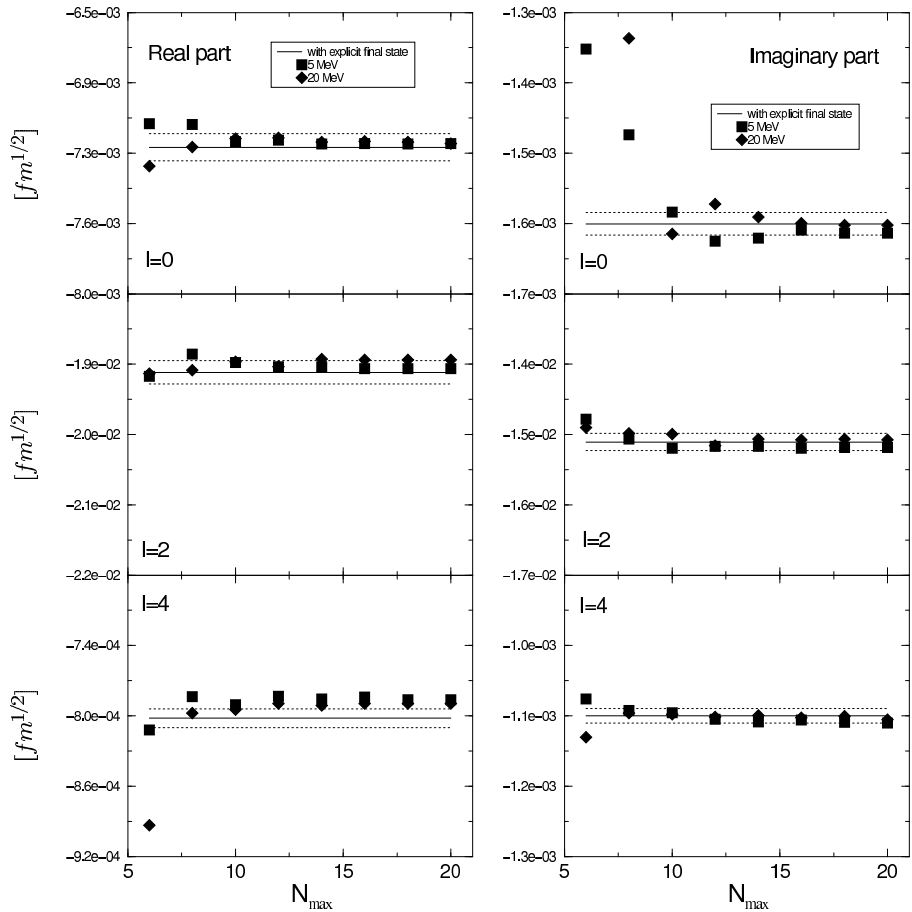}}}
\caption{Deuteron electrodisintegration $d(e,e'p)n$. Real (left) and imaginary (right) parts of Coulomb transitions matrix elements
of multipole order $l$ (Born contribution subtracted) for a LIT calculation at $q$=440~MeV/c and $E_{np}=120$ 
MeV with TN potential as a functions of the number of inversion basis functions
$N_{max}$: diamonds $(\sigma_I=5$~MeV), squares $(\sigma_I=20$~MeV); also shown results for
calculation with explicit $np$ final state wave function (full curves) and deviations of 
$\pm$1\% from these results (dashed curves).}
\label{R2}
\end{figure}
it served as a test case for the applicability of 
the LIT approach. There, in order to have reliable inversion results  
it has been necessary to separate contributions from the Coulomb monopole (C0) and quadrupole 
(C2) transitions from the other multipole transitions (see also \sref{sec:INV}). The former lead to a shoulder of $R_L$ 
close to threshold, while the latter exhibit a peak in the quasi--elastic region.
In~\fref{R1} we show the results of this important LIT test (note: $E_{np}$ is the kinetic energy 
of the outgoing $np$--pair in the centre of mass system). In the upper panel of the figure one sees that the 
C0 and C2 multipoles exhibit a steep rise at the very threshold, leading to somewhat
oscillating inversion results. In~\fref{R1}(b) we present the comparison on the total $R_L$, where also the other 
Coulomb multipoles are taken into account. It is evident that there
is a very good agreement between the LIT result and the result of a calculation with
explicit $np$ continuum state wave functions.

In~\cite{LAPIANA:2000} a similar test study has been carried out for exclusive reactions. 
To this end the longitudinal part of the $d(e,e'N)N$ cross section has been investigated.
In~\fref{R2} we show results for a few selected Coulomb transition matrix elements.
In order to make the comparison more significant the Born contribution (see~\sref{sec:EXCL})
is subtracted. 
One sees that i) the calculation is essentially independent on the used $\sigma_I$--value 
and that ii) there is a very good convergence in $N_{max}$ in the inversion. Also shown are the
results of a conventional calculation. Like for the inclusive test case of~\fref{R1}
they agree very well with the LIT results.

Among the two--body reactions considered within the LIT approach there exist also two interesting test cases
concerning pion photoproduction. In~\cite{REISS:2003} the Kroll--Ruderman term
has been considered and it has been shown that the results are the same
as in a conventional calculation with explicit scattering states. A further study
has been carried out in~\cite{REISS:2005}, where the proper implementation
of $\Delta$ degrees freedom in pion photoproduction has been successfully 
investigated for a LIT calculation.

%**********************************************************************
\subsection{Reactions with three--body systems}\label{sec:THREEBODY}
%**********************************************************************

%********************************************
\subsubsection{Electrons.}\label{sec:ELECTRONS_3}
%********************************************

\begin{figure}[htb]
\centerline{\resizebox*{12cm}{6cm}{\includegraphics*[angle=0]{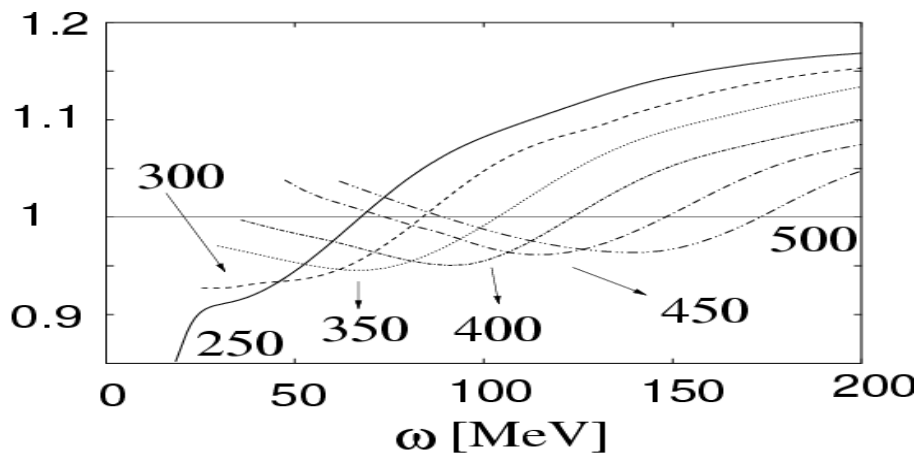}}}
\caption{Ratios $R_L$(with 3N force)/$R_L$(without 3N force) for triton at various
momentum transfers; $q$ values indicated in the figure in units of~MeV/c.}
\label{R3}
\end{figure}
\begin{figure}[htb]
\centerline{\resizebox*{12cm}{11cm}{\includegraphics*[angle=0]{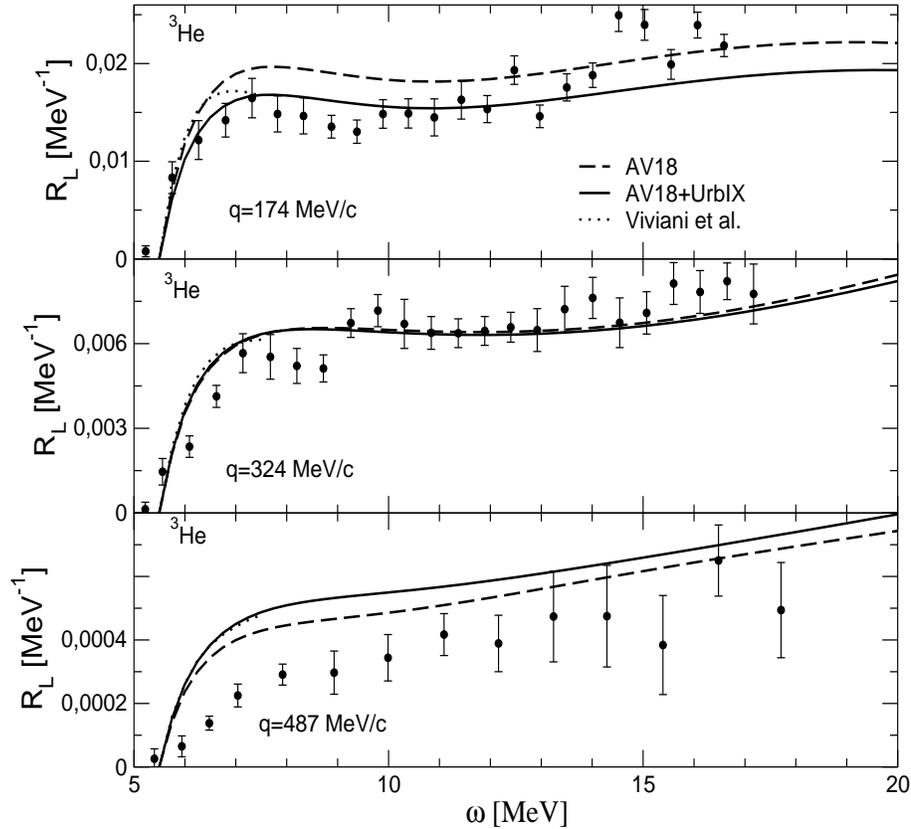}}}
\caption{Comparison of theoretical and experimental $R_L$ of $^3$He: 
AV18+UIX potentials (solid) and AV18 potential (dashed),
experimental data from~\cite{RETZLAFF:1994}, up to three--body break--up threshold.
Also shown theoretical results from~\cite{VIVIANI:2000} with AV18+UIX potentials (dotted).}
\label{R4}
\end{figure}
For the three--nucleon systems first LIT calculations of $R_L(q,\omega)$ ($q$=250, 300
MeV/c) have been made using the Faddeev formalism and taking realistic NN potential 
models~\cite{MARTINELLI:1995}. Like the deuteron cases of the previous section they have 
merely served 
as tests for the applicability of the LIT method. In fact an excellent agreement of the LIT 
results with those of a previous conventional Faddeev calculation has been obtained. 
Later a more systematic study of the effects of various theoretical 
ingredients on $R_L$ has been carried out with the LIT approach for a 
wider momentum transfer range using the CHH expansion technique~\cite{EFROS:2004,EFROS:2005}. 
%where momentum transfers up to $q=700$~MeV/c were considered 
The dependence on the following theoretical ingredients have been investigated: 
NN potential models, 3N potential models, relativistic corrections to the non--relativistic
charge operator (spin--orbit and Darwin--Foldy terms), reference frame dependence. Here 
we give a short summary of the results of~\cite{EFROS:2004}. In general a rather small NN 
potential model dependence has been found, but in some cases there are also larger effects. These 
include the height of the quasi--elastic peak at lower $q$ and the threshold behaviour. In~\fref{R3}
we illustrate the effect of the three--nucleon force (3NF). 
It is typically between 5\% and 10\%, but reaches up to 15\% for the 
low--energy responses, far beyond the quasi--elastic peak. 
The dependence on the 3NF model is very small. Relativistic corrections are unimportant 
at lower $q$, but cannot be neglected for $q > 400$~MeV/c. For momenta up to $q=500$ 
MeV/c the comparison of the theoretical results with experimental data is generally 
rather satisfying. The experimental data, however, are in most cases not precise enough 
to draw conclusions about the 3NF effect. 
A nice exception is the $^3$He low--energy 
response, where a 3NF proves to be necessary to obtain agreement with experiment 
(see upper panel of~\fref{R4}). At higher $q$ and low energy one finds a considerably higher 
$R_L$ response in theory than in  experiment (see lower panel of~\fref{R4}). In~\fref{R4}, 
for energies below the three--body break--up threshold, we also show results from a 
calculation with explicit three--nucleon final state wave functions~\cite{VIVIANI:2000}, where the 
same theoretical ingredients as in our LIT calculation are taken. One sees that there are 
rather good agreements between the results of the two different calculations.

A study of the $R_L$ reference frame dependence has been already initiated in~\cite{EFROS:2004},
but in~\cite{EFROS:2005} has been worked out in more detail and with consideration of higher momentum 
transfers. To this end the $R_L^{\rm fr}(\omega_{\rm fr},q_{\rm fr})$ of different reference 
frames has been calculated and then transformed in a relativistically correct way to the 
laboratory system according to 
\begin{equation}
R_L(\omega,q) = {\frac {q^2} {q_{\rm fr}^2} } {\frac {E_T^{\rm fr}} {M_T} }
R_L^{\rm fr}(\omega_{\rm fr},q_{\rm fr}) \,,
\end{equation}
(note the definition $R_L(\omega,q) \equiv 
R_L^{\rm lab}(\omega_{\rm lab},q_{\rm lab})$) where $E_T^{\rm fr}$ and $M_T$ are energy and mass of the 
target nucleus. 
\begin{figure}[htb]
\centerline{\resizebox*{12cm}{12cm}{\includegraphics*[angle=0]{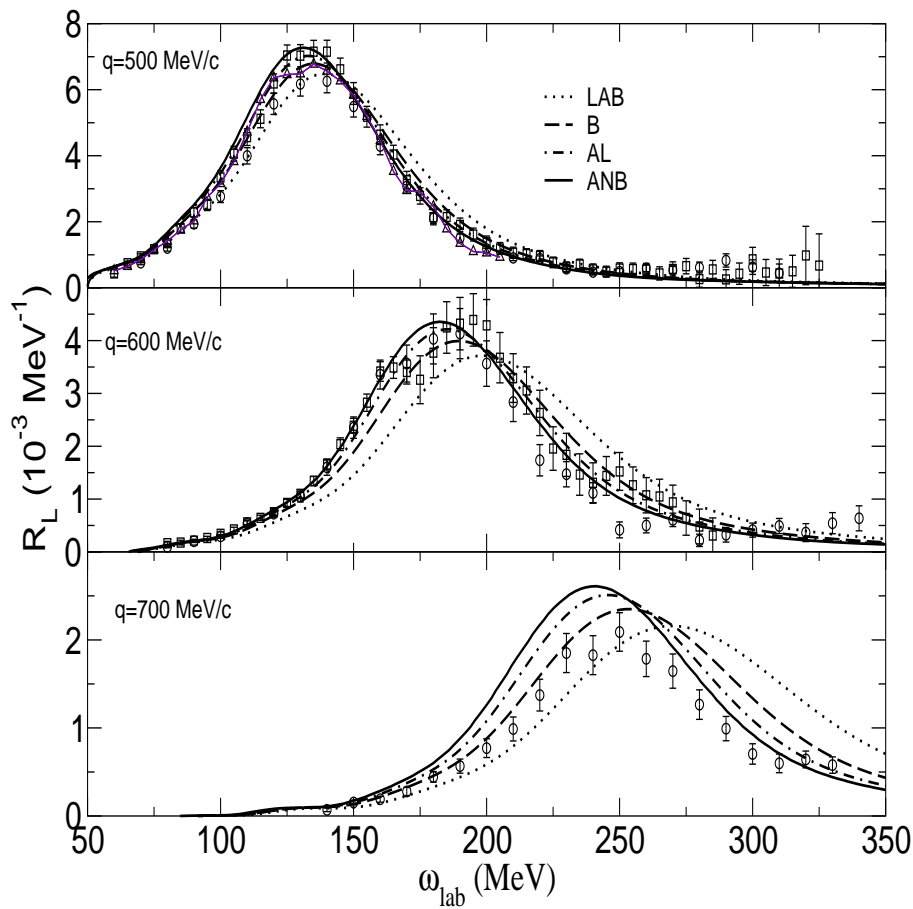}}}
\caption{Frame dependence of $R_L$ of $^3$He: LAB (dotted), B (dashed), 
AL (dash--dotted), ANB (solid); experimental data
from~\cite{MARCHAND:1985} (squares),~\cite{DOW:1988} (triangles),~\cite{CARLSON:2002} (circles).}
\label{R5}
\end{figure}
In~\fref{R5} we show results for four different frames~\cite{EFROS:2005}, which are defined 
via the target nucleus initial momentum
${\bf P}_T^{\rm fr}$: ${\bf P}_T^{\rm lab}$=0,
${\bf P}_T^{\rm B}=-{\bf q}/2$, ${\bf P}_T^{\rm AL}=-{\bf q}$,
${\bf P}_T^{\rm ANB}=-3{\bf q}/2$. It is readily seen 
that one obtains more and more frame dependent results with growing $q$.
As shown in~\cite{EFROS:2005} the frame dependence can be drastically 
reduced if one imposes quasi--elastic kinematics and takes the relativistic relative
momentum between knocked--out nucleon and residual two--body system
as input in the calculation. As seen in~\fref{R6}, in this way one
finds almost frame independent results, which show a very good agreement with
experimental data at $q$ equal to 500 and 600~MeV/c, while at 700~MeV/c the experimental 
peak height is considerably lower than the theoretical one. We would like to point out
that the ANB result of~\fref{R5} lies within the band of almost frame independent
results of~\fref{R6}. Thus one has to consider the ANB frame as the optimal frame for
a non--relativistic calculation. In fact it leads to a proper description of
the quasi--elastic peak region, with no need to assume quasi--elastic kinematics
and thus its validity is not restricted to the peak region
(further explanations why the ANB system is preferable are given in~\cite{EFROS:2005}). 
\begin{figure}[htb]
\centerline{\resizebox*{12cm}{12cm}{\includegraphics*[angle=0]{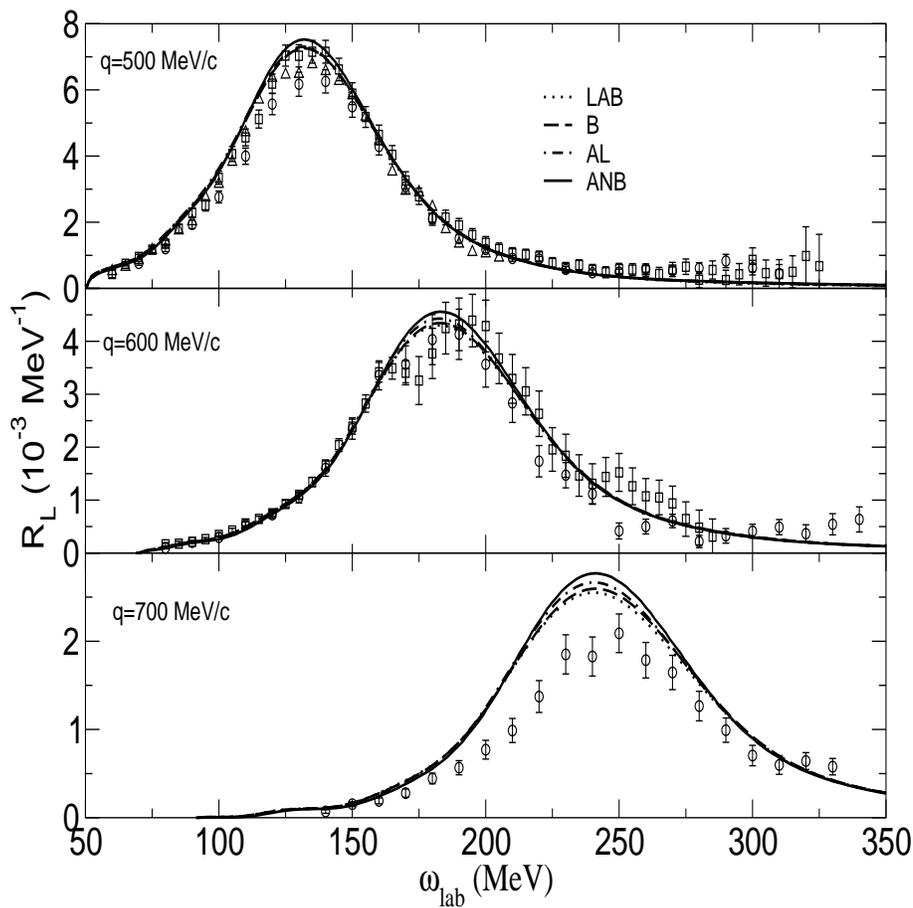}}}
\caption{As in~\fref{R5}, but considering two--body relativistic kinematics for the final state
energy.}
\label{R6}
\end{figure}
%In this section we consider the total photoabsorption cross sections of nuclei.
%This observable, the energy dependent total response of a nucleus to real photons,
%is of great relevance, since it exhibits details of the basic nuclear degrees of
%freedom and thus also of the underlying nuclear dynamics.
%Total photoabsorption cross sections have been calculated with the LIT method up to seven-body
%nuclei. All these calculations, even in the case of the three-nucleon systems, constitute
%the first fully microscopic calculations with inclusion of all relevant break-up channels
%for this observable. Also the in conventional theory well established two-nucleon case
%(see e.g.~\cite{ArS}) was calculated in the LIT approach, but only as a test
%of the JISP potential models~\cite{jisp}. However, also in the present work
%deuteron photodisintegration is considered, here it serves as interesting case study for
%the inversion and for the Lanczos response as well (see sect.?).
\begin{figure}[htb]
\centerline{\resizebox*{12cm}{12cm}{\includegraphics*[angle=0]{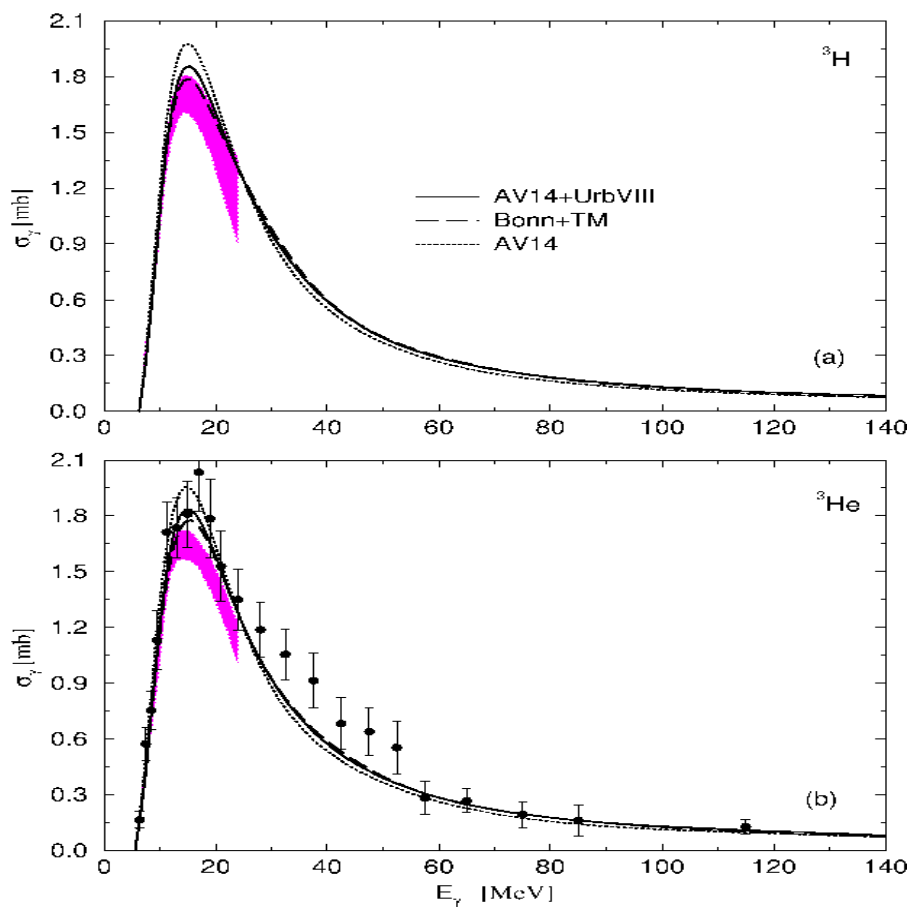}}}
\caption{Total photoabsorption cross section of $^3$H (upper panel) and $^3$He 
(lower panel) with AV14 (dotted), AV14+UVIII (solid) 
and BonnRA+TM (dashed) potentials; 
the shaded area represents the data from~\cite{FAUL:1981}
while the black dots are the data from~\cite{FETISOV:1965}}
\label{R7}
\end{figure}

%***************************************
\subsubsection{Photons.}\label{sec:PHOTONS_3}
%***************************************

Now we turn to the total three--nucleon photoabsorption cross sections. They have been 
studied with semirealistic NN potential models~\cite{EFROS:1997} first, and subsequently also with
realistic nuclear forces (realistic NN potential + 3NF)~\cite{EFROS:2000}.
In both cases CHH expansions have been used.
In~\fref{R7} we report the results of the first calculation of the total
photoabsorption cross section with realistic forces~\cite{EFROS:2000}.
The cross section is rather independent from a complete
nuclear force model with NN and 3N forces. The role of 3NF is not negligible, they lead 
to a reduction of the low--energy peak by about 7\% and to similarly strong increases at 
higher energies.
One finds a rather good agreement of the theoretical results with experimental data.
The experimental error bars, however, are quite large.
In~\cite{GOLAK:2002} the AV18+UIX LIT results, 
which are very similar to the AV14+UVIII and BonnRA+TM results of~\fref{R7}, have been
successfully benchmarked with the Faddeev results of the Bochum--Cracow collaboration. 
In~\cite{GOLAK:2002} also a check of the validity of the unretarded dipole approximation 
has been made. It has been found that retardation and higher multipole contributions are quite small
up to moderate photon energies (below 2\% for $\omega\le 50$~MeV).
 
A high--precision comparison of CHH-- and EIHH--LIT results with the AV18+UIX
nuclear force model has been made in~\cite{BARNEA:2006a}. Finally we would like to mention that various
photonuclear sum rules (polarizability, bremsstrahlungs, and TRK sum rules) have been studied
in~\cite{EFROS:1997,EFROS:2000}.

%************************************************ 
\subsubsection{Neutrinos.}\label{sec:NEUTRINOS_3}
%************************************************ 

Next we consider weak reactions with three--body nuclei.
The calculation of the cross section for neutral current neutrino 
scattering of $^3\rm{H}$ and $^3\rm{He}$ in the supernova
scenario with realistic nuclear Hamiltonian and MEC currents has been presented
in~\cite{OCONNOR:2007}. 
In the supernova scenario it is customary to assume 
that the neutrinos are in thermal
equilibrium. Consequently, their spectra can be approximated by the Fermi--Dirac
distribution with a characteristic temperature $T_{\nu}$ and with zero
chemical potential
\begin{equation}
f({T_{\nu}},k) = \frac{N}{{T_{\nu}}^3} \frac{k^2}{e^{k/{T_{\nu}}}+1} \,,
\end{equation}
where $k$ is the neutrino momentum and 
$N^{-1}=2 \sum_{n=1}^{\infty}(-1)^{n+1}/n^3$  is a normalization 
factor. 
In order to estimate the mean neutrino interaction with 
the nuclear soap
the calculated cross sections must be averaged over
incident energy and angle.
The quantities of interest are the temperature--averaged 
cross sections and energy transfer cross sections:
\begin{eqnarray}
\langle \sigma \rangle_{T_{\nu}} &= \int_{\omega_{\rm th}}^{\infty} d\omega 
\int dk_i \: f({T_{\nu}},k_i) \, \frac{d\sigma}{dk_f} \,,
\label{eq:cross} \\[2mm]
\langle \omega \sigma \rangle_{T_{\nu}} &= \int_{\omega_{\rm th}}^{\infty}
d\omega \int dk_i \: f({T_{\nu}},k_i) \: \omega \, \frac{d\sigma}{dk_f} \,,
\label{eq:etcross}
\end{eqnarray}
\begin{table}
%\begin{ruledtabular}
\begin{tabular}{c|cc|cc}
\hline \hline
${T_{\nu}}$ [MeV] & \multicolumn{2}{c|}{$^3$H} &
\multicolumn{2}{c}{$^3$He} \\[1mm] \hline
1  & 1.97$\times 10^{-6}$ & 1.68$\times 10^{-5}$ 
& 3.49$\times 10^{-6}$ & 2.76$\times 10^{-5}$ \\
2  & 4.62$\times 10^{-4}$ & 4.73$\times 10^{-3}$ 
& 6.15$\times 10^{-4}$ & 5.94$\times 10^{-3}$ \\
3  & 5.53$\times 10^{-3}$ & 6.38$\times 10^{-2}$ 
& 6.77$\times 10^{-3}$ & 7.41$\times 10^{-2}$ \\
4  & 2.68$\times 10^{-2}$ & 3.37$\times 10^{-1}$ 
& 3.14$\times 10^{-2}$ & 3.77$\times 10^{-1}$ \\
5  & 8.48$\times 10^{-2}$ & 1.14                 
& 9.70$\times 10^{-2}$ & 1.25 \\
6  & 2.09$\times 10^{-1}$ & 2.99                 
& 2.35$\times 10^{-1}$ & 3.21 \\
7  & 4.38$\times 10^{-1}$ & 6.61                 
& 4.87$\times 10^{-1}$ & 7.03 \\
8  & 8.20$\times 10^{-1}$ & 13.0                 
& 9.03$\times 10^{-1}$ & 13.7 \\
9  & 1.41                 & 23.4                 
& 1.54                 & 24.6 \\
10 & 2.27                 & 39.3                 
& 2.47                 & 41.2 \\
\hline \hline
\end{tabular}
%\end{ruledtabular}
\caption{Averaged neutrino-- and antineutrino--$^3$H and $^3$He 
neutral--current inclusive inelastic cross sections per nucleon ($A=3$) 
$\langle \sigma \rangle_{T_{\nu}} = \frac{1}{2A} \, \langle \, \sigma_\nu + 
\sigma_{\overline{\nu}} \, \rangle_{T_{\nu}}$ (left columns), and energy 
transfer cross sections, $\langle \omega \sigma \rangle_{T_{\nu}} =
\frac{1}{2A} \, \langle \, \omega \sigma_\nu + \omega
\sigma_{\overline{\nu}} \, \rangle_{T_{\nu}}$ (right columns), 
as a function of the neutrino
temperature ${T_{\nu}}$, in units of $10^{-42} \, \rm{cm}^2$ and 
$10^{-42} \, \rm{MeV}\rm{cm}^2$ respectively.}
\label{tab:nu_A3}
\end{table}
where $k_{i,f}$ are the initial and final neutrino energy,
$\omega = k_i - k_f$ is the energy transfer, and 
$\omega_{\rm th}$ denotes the threshold energy of the break--up reaction.
In Table~\ref{tab:nu_A3}, we present the results of~\cite{OCONNOR:2007}
for the averaged 
neutrino and anti--neutrino neutral--current inclusive inelastic
cross sections and energy transfer cross sections as a function of the neutrino
temperature. The difference between the $^3$H and the $^3$He 
cross sections reflects the difference in thresholds between 
the two nuclei. The mirror symmetry between both nuclei is restored 
with higher neutrino energy. The leading contributions to the 
cross section are the axial $E_1^A$, $M_1^A$ and $E_2^A$ multipoles. The 
relative importance of these multipoles varies as a function
of the momentum transfer, and thus as a function of the neutrino 
temperature. In comparison to inelastic excitations of
$^4$He studied in~\cite{GAZIT:2007}, the cross
sections (the mean values of $^3$H and $^3$He) are about a factor $20$ and $10$ 
times larger at temperatures of $4$~MeV and $6$~MeV, respectively and the energy transfer 
cross sections are $8$ and $2$ times larger.

At low momentum transfer, the Gamow--Teller operator dominates 
for the cross section, consequently the MEC have a large effect 
of about $16\%$ at a temperature of $1$~MeV. At higher momentum 
transfer, higher--order multipoles start to play an important 
role. Due to spatial symmetry, the MEC contribution to these 
multipoles is small and the overall effect of MEC decreases 
rapidly to less than 2\% for temperatures above $4$~MeV. While not directly 
important here, the asymmetry between the scattering of neutrinos and
anti--neutrinos increases with temperature: the difference in the 
energy transfer cross sections grows gradually from 3\% for a neutrino 
temperature of $3$~MeV to more than 50\% for a temperature of $10$~MeV.
Finally, the cutoff dependence of these observables is less than 2\% 
for $1$~MeV and less than 1\% for higher temperatures. 
The estimated precision of these predicted
cross sections is a few percent, which also includes estimates of
the numerical accuracy.
 
%***********************************************************************
\subsection{Reactions with the four--body system}\label{sec:FOURBODY}
%***********************************************************************
%***********************************************************************
\subsubsection{Electrons.}\label{sec:ELECTRONS_4}
%***********************************************************************

\begin{figure}[htb]
\centerline{\resizebox*{12cm}{13cm}{\includegraphics*[angle=0]{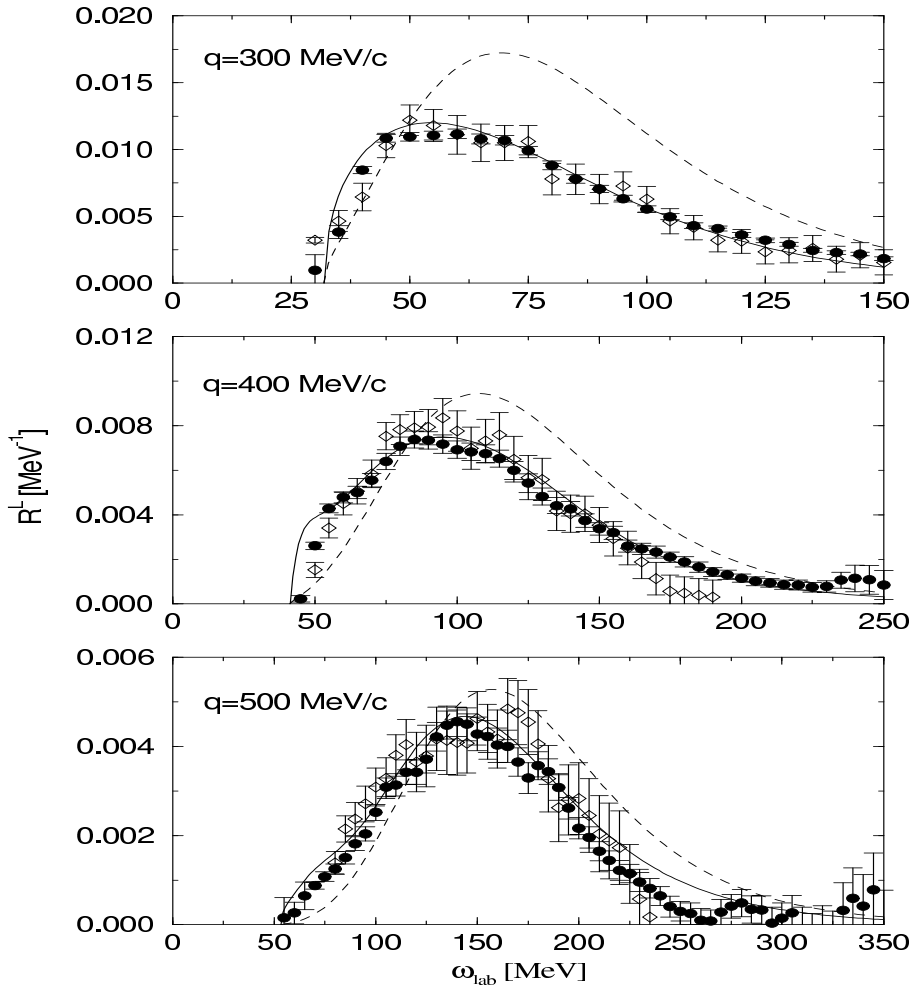}}}
\caption{$R_L$ of $^4$He with the TN potential: plane wave impulse approximation
result via use of spectral function (dashed curves) and full results (solid curves);
experimental data from~\cite{DYTMAN:1988} (diamonds) and~\cite{ZGHICHE:1994} (circles).}
\label{R8}
\end{figure}
\begin{figure}[htb]
\centerline{\resizebox*{12cm}{16cm}{\includegraphics*[angle=0]{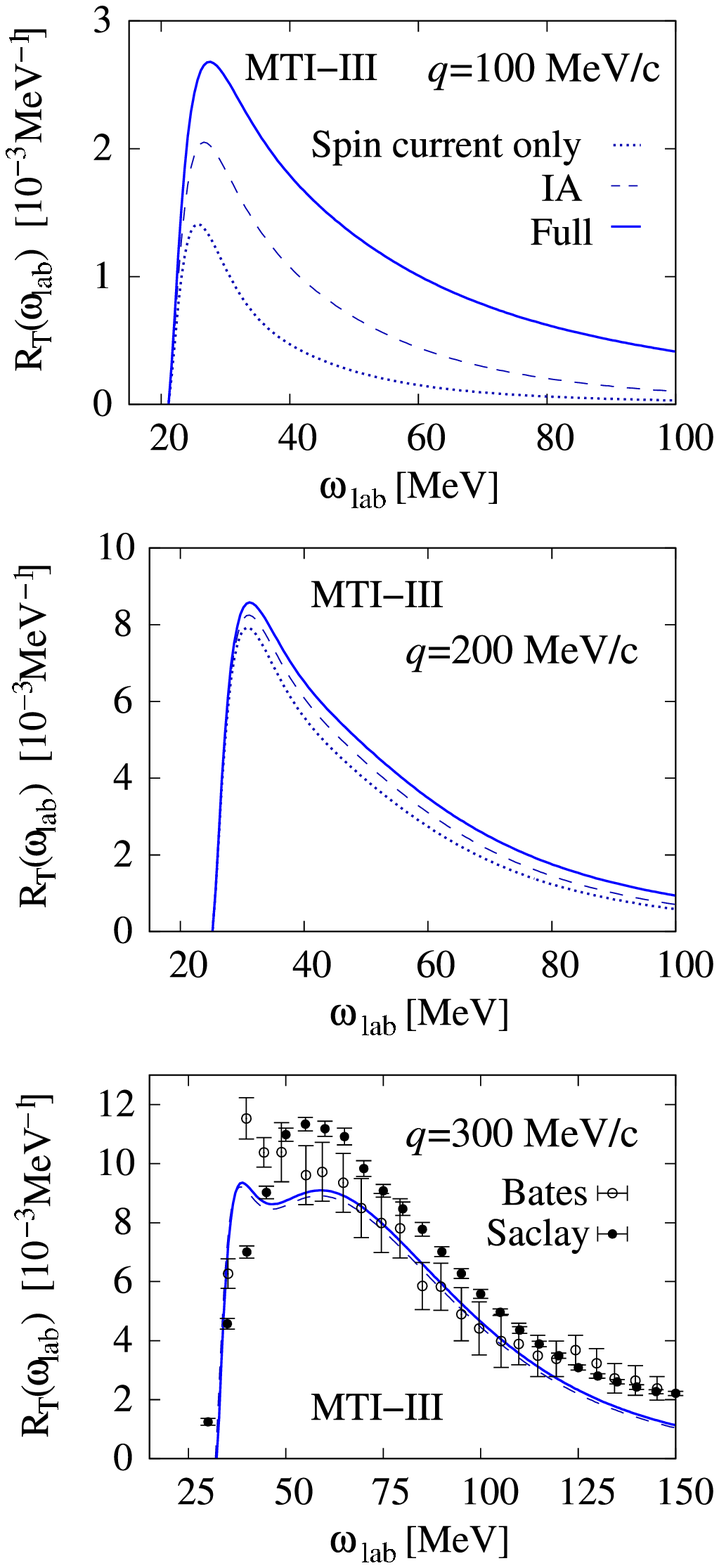}}}
\caption{$R_T$ of $^4$He with the MT--I/III potential:
spin current only (dotted), IA with spin and convection currents (dashed), full 
calculation with the inclusion of a consistent two--body current (solid); experimental data 
from~\cite{DYTMAN:1988} (open circles) and~\cite{ZGHICHE:1994} (solid circles).}
\label{R9}
\end{figure}
The longitudinal form factor $R_L(q,\omega)$ of $^4$He has been 
calculated in~\cite{EFROS:1997a} at 300~MeV/c $\le q \le $ 500~MeV/c with the semirealistic 
TN potential  via the LIT approach and using the CHH expansion
technique. Already before, calculations with a realistic nuclear force had been
carried out using the Green function Monte Carlo approach for the Laplace transform
of the response~\cite{CARLSON:1992}. However, the inversion of this integral transform
is problematic (see discussion in~\sref{sec:OTHERKERNELS}). In~\cite{EFROS:1997a} a 
rather good agreement of 
the LIT results with experimental data has been found, as can be seen in~\fref{R8}. The 
figure also contains LIT results from~\cite{EFROS:1998}, where $R_L$ has been determined 
calculating the spectral function $S(E,k)$ (see~\sref{sec:OTHER}) of the residual three--nucleon 
system in the $^4$He$(e,e'p)X$ reaction. In the
 calculation with a spectral function the FSI of the outgoing
proton with the residual three--nucleon system is not taken into account. The figure
illustrates that the proton FSI is quite important. It leads to shifts of the
quasi--elastic peak positions by about 15~MeV towards lower energy and reduces
the peak heights quite considerably (30\%, 20\% and 10\% at $q=300$, 400, and 500~MeV/c
respectively). Rather strong FSI effects are also found close to the break--up threshold.

Very recently also $R_T(q,\omega)$ has been studied ~\cite{BACCA:2007} with the
MT--I/III potential, using EIHH expansions. 
Besides the non--relativistic 
one--body currents (IA), a consistent meson exchange current (MEC) is taken into account 
explicitly, (with an approximation which, however, is shown to be very good at 
lower momentum transfer). In~\fref{R9} we illustrate $R_T$ for different momentum transfers 
in comparison to the available experimental data. 
The spin current strongly dominates the response function at higher momentum transfer
(here not shown for $q=300$~MeV/c, for more details see~\cite{BACCA:2007}), 
while at $q=100$~MeV/c the convection current gives a large contribution.
One also notes quite an important MEC contribution:
at energies of about 100~MeV MEC lead to an enhancement by a factor of four with respect 
to the IA. With increasing $q$ the enhancing effect of the MEC with respect to the IA is
reduced dramatically. 
Unfortunately, there are no experimental data for the lower two 
$q$--values. At $q=300$~MeV/c one finds a fair agreement with the data at lower and higher
energies, but the quasi--elastic peak height is somewhat underestimated.
\begin{figure}[htb]
\centerline{\resizebox*{12cm}{7cm}{\includegraphics*[angle=0]{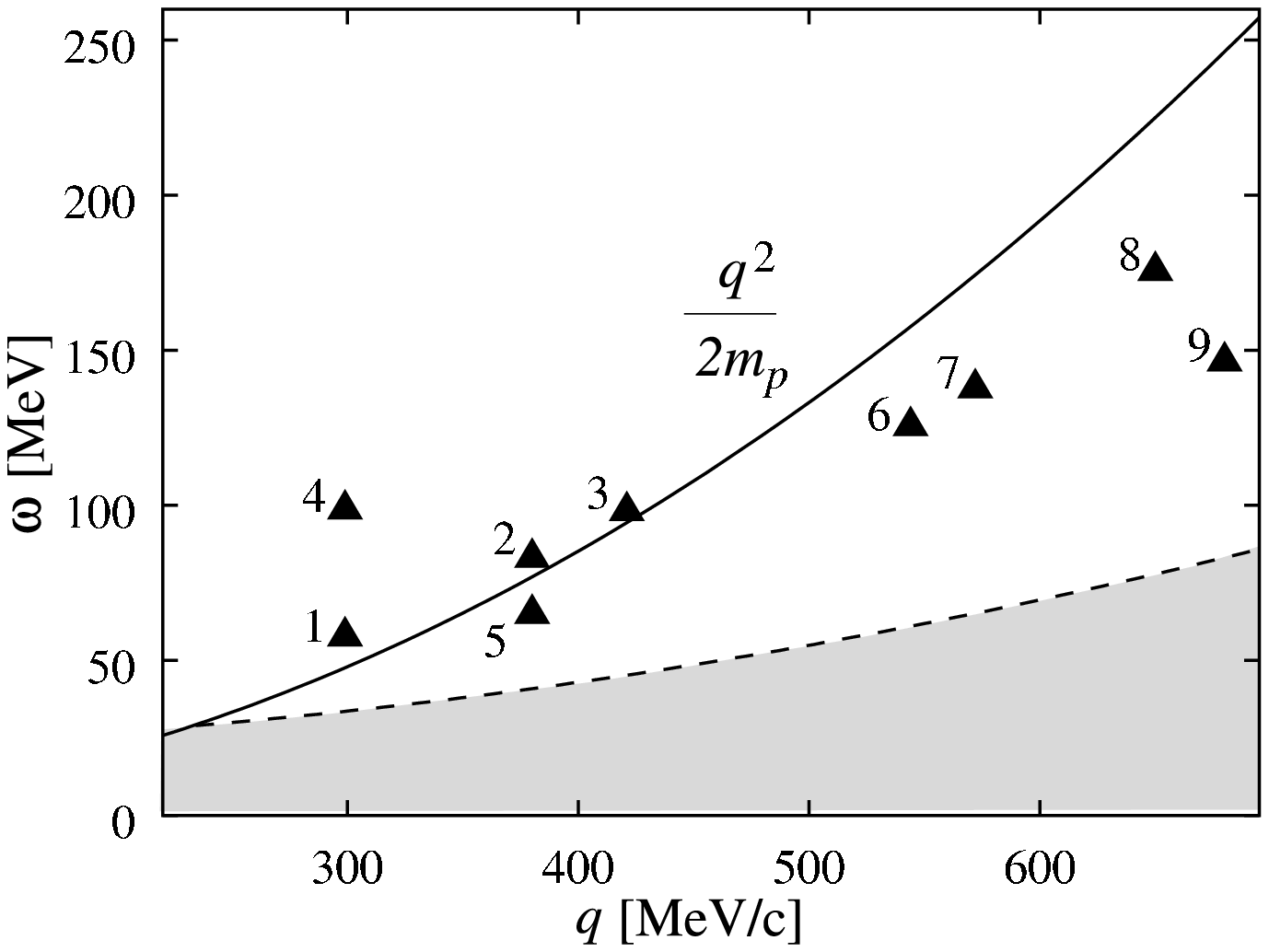}}}
\caption{$^4$He(e,e'p)$^3$H reaction: position of the various kinematics of~\fref{R11} 
with respect to the $\omega=q^2/(2m_p)$ ridge. Shaded area represents the region below the 
break--up threshold.}
\label{R10}
\end{figure}
\begin{figure}[htb]
\centerline{\resizebox*{12cm}{7cm}{\includegraphics*[angle=0]{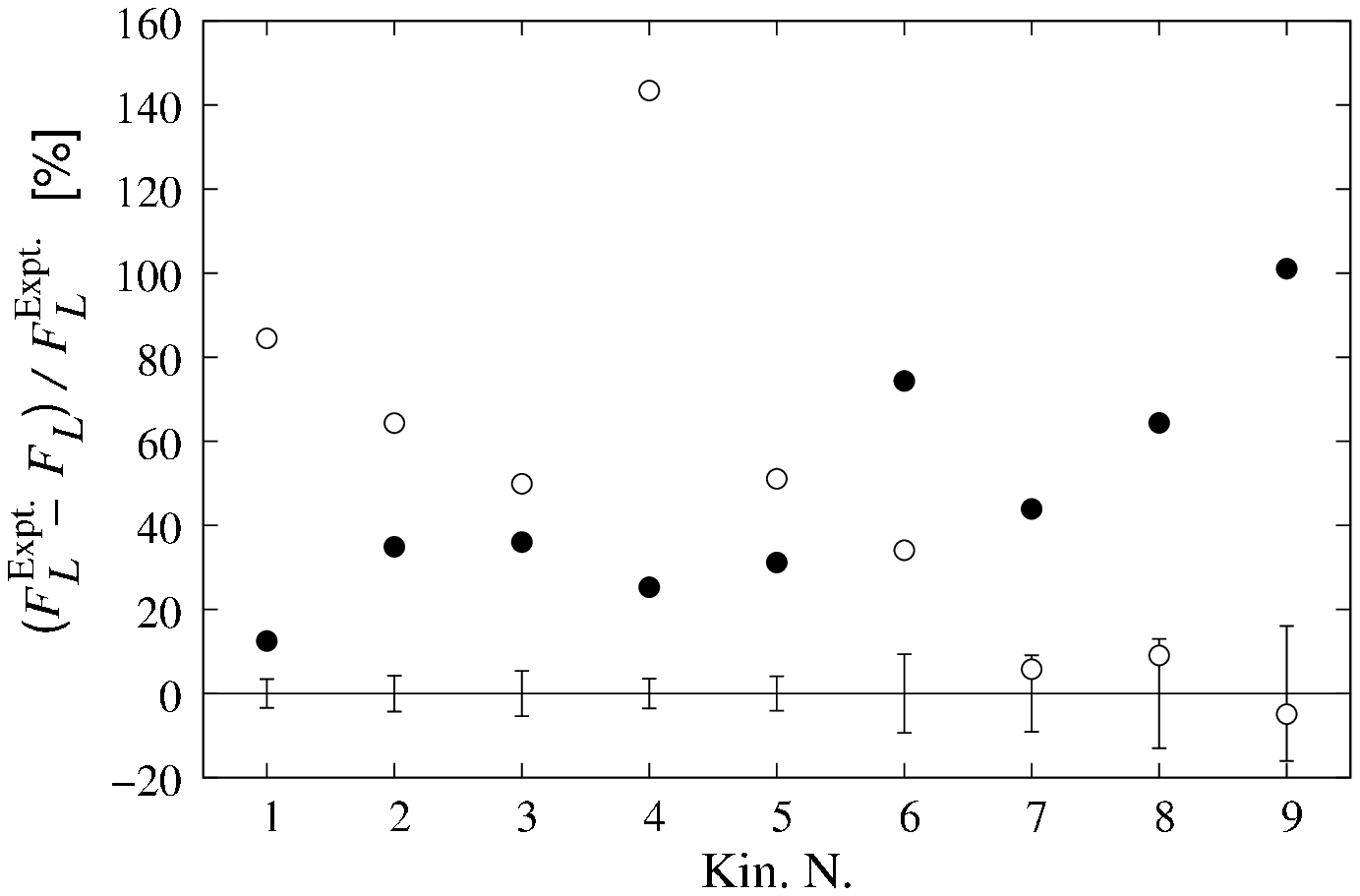}}}
\caption{$^4$He$(e,e'p)^3$H reaction: Percentage deviation of theoretical differential cross 
section from the experimental values of~\cite{DUCRET:1993}: without FSI (open circles) with 
inclusion of full FSI (solid circles).}
\label{R11}
\end{figure}

For the four--body system two electron induced exclusive reactions have been 
considered with the LIT method (CHH expansion technique) and using semirealistic NN potentials: 
the longitudinal parts of the cross sections of $^4$He$(e,e'p)^3$H, close to quasi--elastic
kinematics~\cite{QUAGLIONI:2005}, and of $^4$He$(e,e'd)^2$H, at rather low $\omega$ and
250~MeV/c $\le q \le 450$~MeV/c~\cite{ANDREASI:2006}. For $^4$He$(e,e'p)^3$H 
results on the FSI effects and a comparison with 
experimental data are shown in figures~\ref{R10} and~\ref{R11}. In \fref{R10} the considered
kinematics, labelled with numbers from 1 to 9, are illustrated in the $q$--$\omega$ plane.
In~\fref{R11} the effects of the FSI for the nine cases are shown.  
One sees that FSI reduces/increases 
the cross section below/above $q=500$~MeV/c leading to an improved agreement
with experiment only for $q$ below 500~MeV/c. Presumably for higher $q$ an inclusion
of relativistic effects might be important and possibly lead to a better agreement
with experiment. 

\begin{figure}[htb]
\centerline{\resizebox*{12cm}{7cm}{\includegraphics*[angle=0]{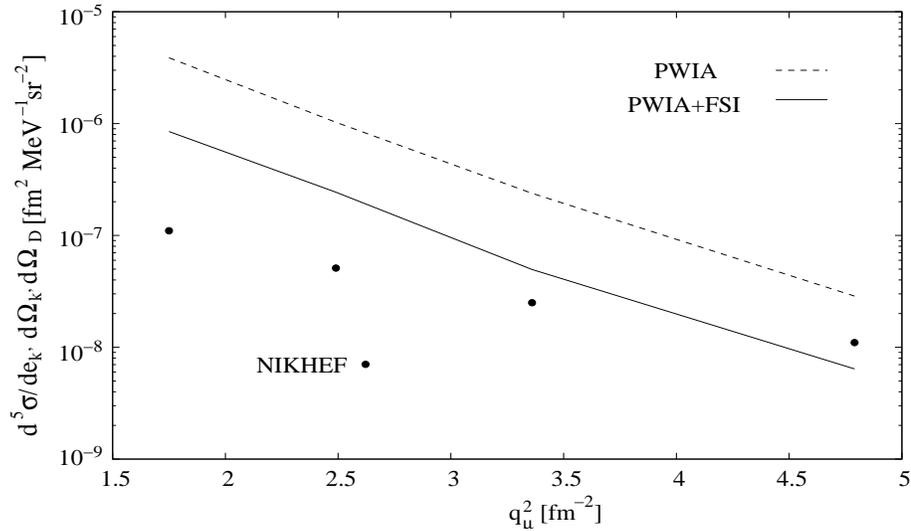}}}
\caption{$^4$He$(e,e'd)d$ reaction, differential cross section at a kinetic energy of the 
outgoing $dd$--pair of 35~MeV and averaged over a missing momentum range between 100 and 150 
MeV as a function of $q_{\mu}^2$ (four--momentum transfer squared): without FSI (dashed),
with FSI (solid). Experimental data from~\cite{ENT:1991}.}
\label{R12}
\end{figure}

For the $(e,e'd)$ reaction of~\fref{R12} one finds an enormous reduction
of the cross section due to FSI, which, however, is not sufficient to describe
the rather small experimental cross section at lower $q$. As discussed in~\cite{ANDREASI:2006}
a consideration of the tensor force could be particularly important for the FSI and 
possibly lead to an improved agreement with experiment.
\begin{figure}[htb]
\centerline{\resizebox*{12cm}{12cm}{\includegraphics*[angle=0]{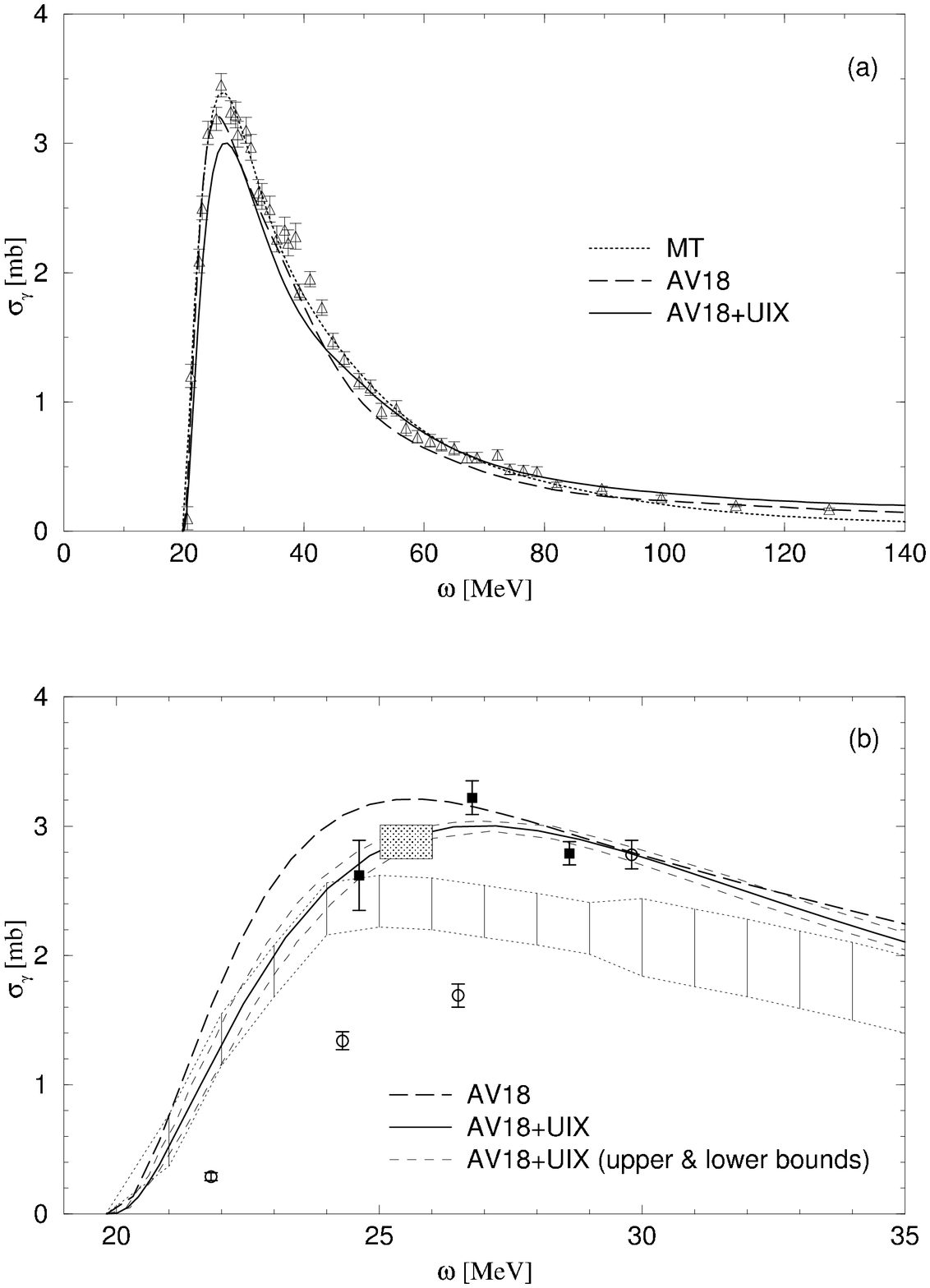}}}
\caption{Total $^4$He photoabsorption cross section: (a) 
MT--I/III (dotted), AV18  (dashed), 
AV18+UIX (solid). Experimental data from~\cite{ARKATOV:1979}.
(b) as (a) but with upper/lower bounds included (estimate of theoretical error of EIHH
convergence); experimental cross sections: dotted box~\cite{WELLS:1992}, squares~\cite{NILSSON:2005}, 
open circles~\cite{SHIMA:2005}. The area between dotted lines has been obtained combining
data from~\cite{BERMAN:1980} and~\cite{FELDMAN:1990}. }
\label{R13}
\end{figure}

%********************************************
\subsubsection{Photons.}\label{sec:PHOTONS_4}
%********************************************

The $^4$He photodisintegration has a rather long history and many different experiments
have been carried out over the years (see e.g.~\cite{QUAGLIONI:2004}). Unfortunately the results of the 
various 
experiments on the total photoabsorption cross section differ quite substantially in the size of 
the giant resonance peak. About ten years ago theory~\cite{ELLERKMANN:1996} and experiment~\cite{BERMAN:1980,FELDMAN:1990}
seemed to converge to a rather low peak cross section.
However, in the LIT calculation of~\cite{EFROS:1997b} 
a much more pronounced giant resonance peak was found. The calculation had been carried out
in the unretarded dipole approximation with 
semirealistic NN potentials models (TN and MT--I/III potentials) using CHH expansions. 
We would like to point out that the unretarded dipole approximation should work even for higher energies
than for the total photoabsorption cross sections of two-- and three--body nuclei, since the
$^4$He root mean square radius is smaller than the radii of $^2$H, $^3$H and $^3$He.
An improved LIT calculation has been later pursued using EIHH expansions~\cite{BARNEA:2001}. Somewhat 
modified results have been found, but the pronounced giant dipole peak
cross section has been confirmed. 
\begin{figure}[htb]
\centerline{\resizebox*{12cm}{7cm}{\includegraphics*[angle=0]{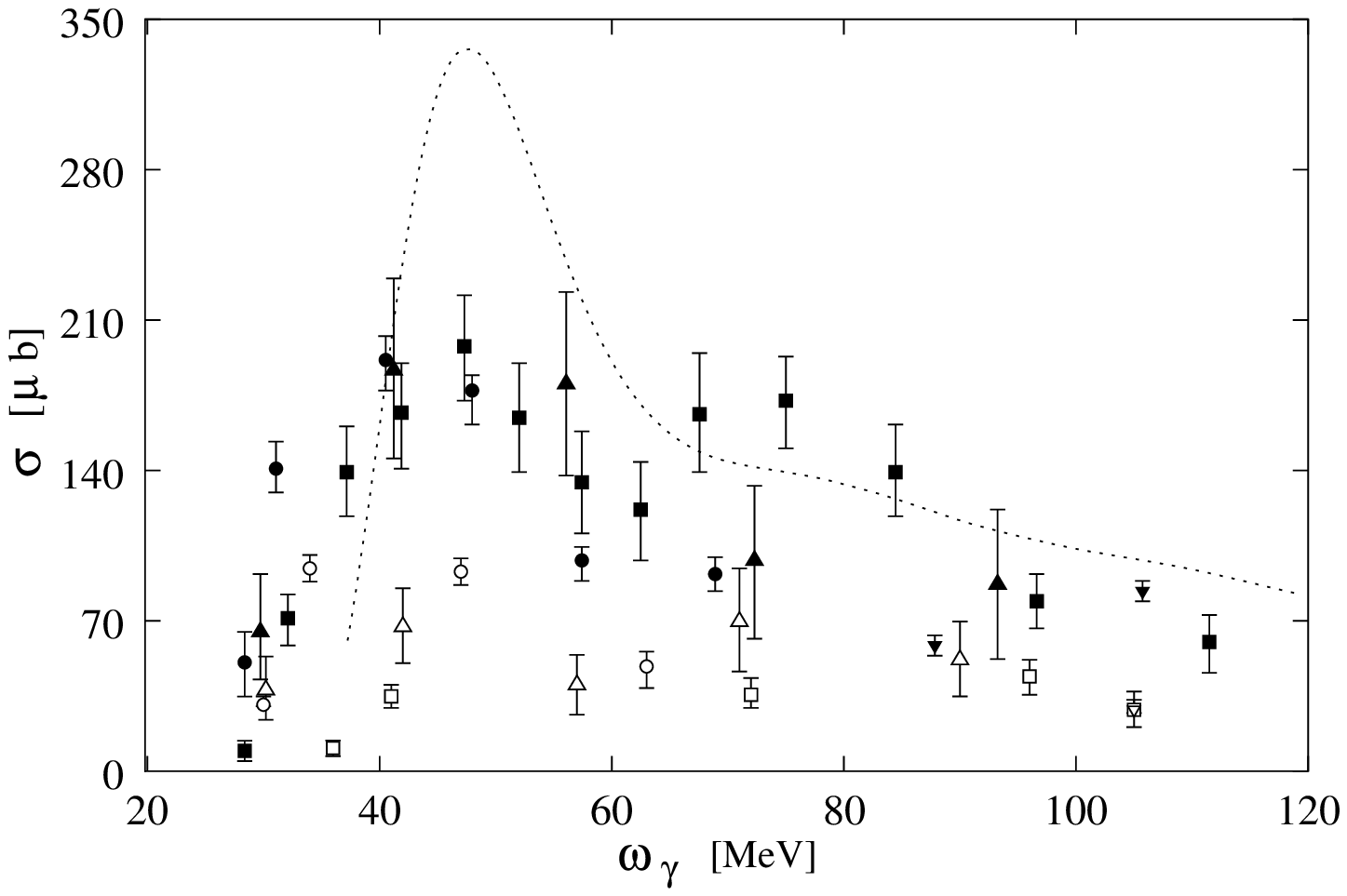}}}
\caption{More--body break--up cross section in $^4$He photodisintegration: total theoretical
result of three-- and four--body break--up (dotted); experimental results for 
$^4$He($\gamma,pn)d$ (full symbols) and $^4$He($\gamma,2p2n)$ (open symbols),~\cite{GORBUNOV:1958} 
(upward triangles),~\cite{ARKATOV:1969} (squares),~\cite{BALESTRA:1979} (circles),
\cite{DORAN:1993} (downward triangles).}
\label{R14}
\end{figure}
The relative importance of the considered reaction and the 
contradiction of theoretical and experimental results have led to a renewed interest in 
the low--energy $^4$He photodisintegration. New experimental data~\cite{NILSSON:2005,SHIMA:2005} as 
well as LIT calculations for the exclusive $^4$He$(\gamma,p)^3$H and $^4$He$(\gamma,n)^3$He
reactions (CHH expansions)~\cite{QUAGLIONI:2004} and a first calculation for the total $^4$He 
photoabsorption cross section with a realistic nuclear force (EIHH expansions)~\cite{GAZIT:2006} 
has become available thanks to the LIT method. The present situation for the total cross section is 
summarized in~\fref{R13}. 
One observes a rather good agreement of the theoretical calculation with the data 
of~\cite{NILSSON:2005}, while there is a large discrepancy with the data of~\cite{SHIMA:2005}. 
One further notes that the effect of the three--nucleon force is quite large, it reduces the giant 
dipole peak height by about 10\% and increases the high--energy cross section quite considerably 
(35\% at pion threshold). The comparison of the full theoretical result with the data 
of~\cite{ARKATOV:1979} is quite satisfying, although the data shows an even higher peak cross 
section. 

Very recently a LIT calculation using  chiral nuclear forces (NN+3N)~\cite{QUAGLIONI:2007} 
has been carried out. The total $^4$He cross section results are very similar
to the AV18+UIX results of~\fref{R13}. Calculations with non--local nuclear force models, where a 
3NF is eliminated by construction  (JISP6 and UCOM~\cite{ROTH:2004} NN potentials) have 
also been performed with the LIT method~\cite{BARNEA:2006,BACCA:2007a}. They confirm the pronounced 
giant dipole cross section, but with peak positions at somewhat higher energies.

In~\cite{GAZIT:2006a} the integrated $^4$He total photoabsorption strength has been considered.
In particular the polarization, bremsstrahlung and TRK sum rules with the AV18+UIX nuclear force
have been investigated.
With their help proton--proton and neutron--neutron distances have been determined and 
a small deviation from the tetrahedral simmetry of the spatial configuration of $^4$He has been found. 

The exclusive reactions $^4$He$(\gamma,N)^3$X have been calculated with the semirealistic MT--I/III 
potentials. The sum of the two channel cross sections can be compared with the 
total $^4$He photoabsorption cross section. Below the three--body break--up threshold both 
results should be identical, since other channels are not open ($^4$He$(\gamma,d)d$ cross 
section is zero in unretarded dipole approximation). On the other hand at higher energies 
the difference between the total $^4$He$(\gamma)$ and the $^4$He$(\gamma,p)^3$H plus $^4$He$(\gamma,n)^3$He
cross sections leads to an estimate for the more--body break--up cross section. The following 
results have been found in~\cite{QUAGLIONI:2004}. Below the three--body break--up threshold the agreement 
between the cross sections of the two kinds of calculations is rather good, but not yet completely 
satisfying. Therefore in~\sref{sec:ALPHA} we have reconsidered a similar consistency check of the LIT 
calculation. There it is shown that the HH expansion has to be carried out to a somewhat higher  
value of the grand angular HH quantum number $K$ than has been done in~\cite{QUAGLIONI:2004} in order to 
find a satisfying agreement between both calculations. Concerning the more--body break--up
cross section, in~\fref{R14} we show the sum of three-- and four--body channel cross sections 
obtained as described above. One sees that there is a fair agreement between experimental 
and theoretical results.

%*************************************************
\subsubsection{Neutrinos.}\label{sec:NEUTRINOS_4}
%*************************************************

The inelastic neutral current neutrino scattering from $^4$He has been calculated
in~\cite{GAZIT:2004}
in the impulse approximation using the AV8' NN potential.
Later in ~\cite{GAZIT:2007} the full nuclear Hamiltonian
with the AV18+UIX potential has been used and MEC have been included.
 
\begin{figure}[htb]
\centerline{\resizebox*{12cm}{7cm}{\includegraphics*[angle=0]{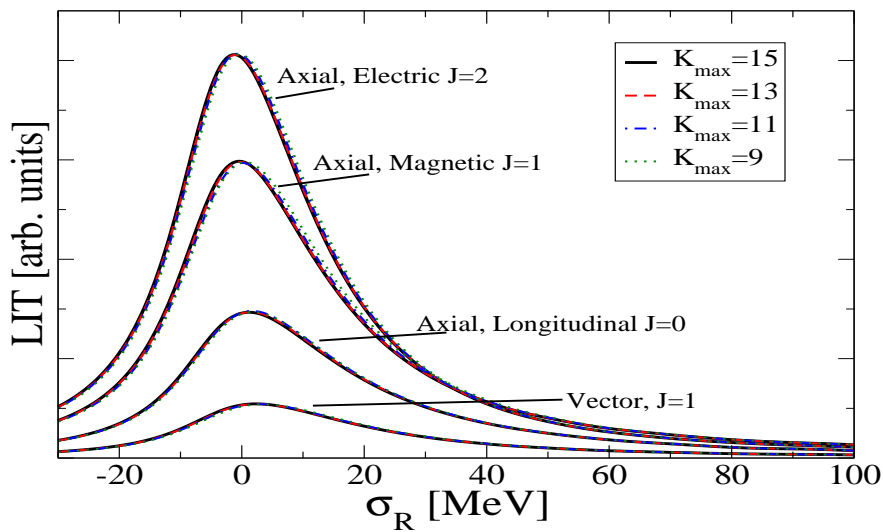}}}
\caption{\label{fig:conv}  LIT convergence in $K_{max}$ for the
leading multipoles.
}
\end{figure}

In the supernova scenario one has to consider neutrinos with up to
about $60$ MeV. Usually, the leading contributions in weak nuclear
processes are the Gamow--Teller and the Fermi operators. Due to the
total angular momentum and spin structure of the $^{4}\mathrm{He}$
nucleus, they are both strongly suppressed. In fact, the
Gamow--Teller operator contributes only due to the small $P$-- and
$D$--wave components of the ground state wave function. 
In addition,
$^{4}\mathrm{He}$ is an almost pure zero--isospin state~\cite{NOGGA:2002,VIVIANI:2005}, 
hence the Fermi operator vanishes. Therefore, the
leading contributions to the inelastic cross section are due to the
axial vector operators $E^A_2, M^A_1, L^A_0, L^A_2 $ and the vector
operators $C^V_1, E^V_1, L^V_1$ (the latter are all proportional to
each other due to the Siegert theorem). For the neutrino energies
considered here it is sufficient to retain contributions up to
$O(q^2)$ in the multipole expansion~\cite{GAZIT:2004}. In 
\fref{fig:conv} we present the convergence in $K_{max}$ for these 
multipoles. It can be seen that the EIHH method results in a
rapid convergence of the LIT calculation to a sub--percentage
accuracy level. 

\begin{table}
\begin{tabular}{c||c|c|c|c}
\hline \hline $T_{\nu}$ [MeV] &  \multicolumn{4}{c} {$\bra \sigma^0_x \ket_T
=  \frac{1}{2} \frac{1}{A} \bra \sigma_{\nu_x}^0+
\sigma_{\overline{\nu}_x}^0 \ket_T$ [$10^{-42}$ cm$^{2}$] }  \\ \hline
 &  AV8'~\cite{GAZIT:2004} & AV18 & AV18+UIX & AV18+UIX+MEC \\
\hline
 4    &  2.09(-3) & 2.31(-3) & 1.63(-3) & 1.66(-3) \\
 6    &  3.84(-2) & 4.30(-2) & 3.17(-2) & 3.20(-2) \\
 8    &  2.25(-1) & 2.52(-1) & 1.91(-1) & 1.92(-1) \\
 10   &  7.85(-1) & 8.81(-1) & 6.77(-1) & 6.82(-1) \\
 12   &  2.05     & 2.29     & 1.79     & 1.80    \\
 14   &  4.45     & 4.53     & 3.91     & 3.93    \\
\hline \hline
\end{tabular}
\caption{{\label{tab:pots}} Temperature averaged neutral current
inclusive inelastic cross section per nucleon (in $10^{-42}$ cm$^{2}$)
as a function of neutrino temperature (in MeV). }
\end{table}

In Table
\ref{tab:pots} we present the temperature averaged total neutral
current inelastic cross section as a function of the neutrino
temperature for the AV8', AV18, and the AV18+UIX nuclear
Hamiltonians and for the AV18+UIX Hamiltonian adding a MEC derived from a chiral 
EFT Lagrangian~\cite{ANANYAN:2002}. From
the table it can be seen that the low--energy cross section is
rather sensitive to details of the nuclear force model (the effect
of 3NF is about $30\%$). This sensitivity gradually decreases with
growing energy. In contrast the effect of MEC is rather small. 
Similar behaviour has also been
observed   for the {\it hep} process~\cite{MARCUCCI:2000}. The small
contribution of the MEC can be understood in the following way. The
symmetry of the vertices in this low--energy approximation dictate
a symmetry between the two nucleons interacting via the meson
exchange. The leading one--body multipoles have negative parity, as
a result the MEC contributions to them is small. In comparison the
MEC correction to the Gamow--Teller is of the same magnitude as the
one--body current, however both terms become marginal with
increasing momentum transfer. Although presented for the neutral
current, these arguments hold true also for the charged currents
since the response functions are related by isospin rotation.
These results are the first fully microscopic study of
$\nu-\alpha$ reactions, using a state of the art nuclear Hamiltonian
and including MEC.
%The
%combination of AV18 and UIX successfully predicts two-- and three--nucleon
%low--energy scattering processes,
The overall accuracy of the calculation is estimated to be 
of the order of $5\%$.
This is mainly due to the strong sensitivity of the
cross section to the nuclear model. The numerical accuracy of the
calculations is of the order of $1\%$, and the cutoff dependence of
the MEC is of the same order. 

%*********************************************************************
\subsection{Reactions with systems with A $>$ 4}\label{sec:MOREBODY}
%*********************************************************************

Different from the lightest nuclear systems the photodisintegrations of nuclei with $A>4$ 
have not yet been studied in microscopic calculations with realistic nuclear forces,
but LIT calculations with semirealistic NN interactions have been carried 
out~\cite{BACCA:2002,BACCA:2004,BACCA:2004a}. 
with the  MT--I/III, MN, and AV4' NN potentials.
Different from the two former potential models the AV4' interaction includes a nonzero 
force also for the relative $P$--waves and also other odd waves.
For the three-- and four--nucleon systems such simple
interactions lead to rather realistic total photoabsorption cross sections~\cite{EFROS:2000,GAZIT:2006},
thus also for $A>4$ cases one can expect that they lead to similarly good results.
\begin{figure}[htb]
\centerline{\resizebox*{12cm}{12cm}{\includegraphics*[angle=0]{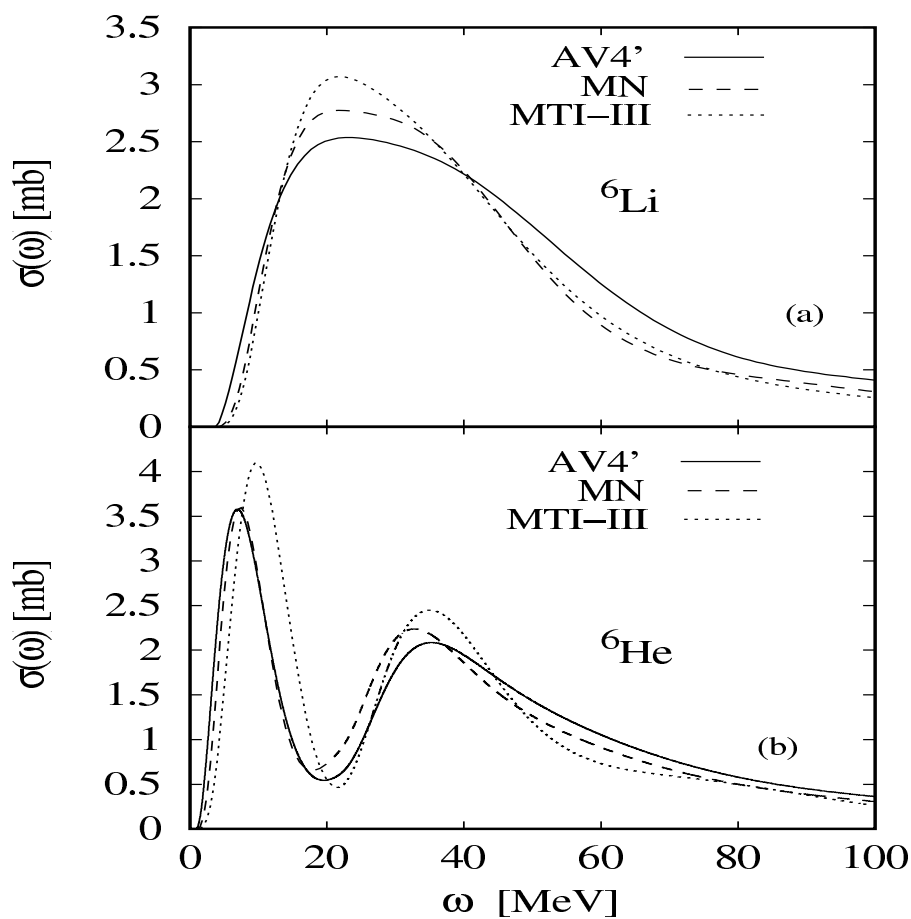}}}
\caption{Total photoabsorption cross section of the six--body nuclei for various 
semirealistic NN potentials: AV4' (solid), MN (dashed), 
MT--I/III (dotted): (a) $^6$Li and (b) $^6$He.}
\label{R15}
\end{figure}
\begin{figure}[htb]
\centerline{\resizebox*{12cm}{12cm}{\includegraphics*[angle=0]{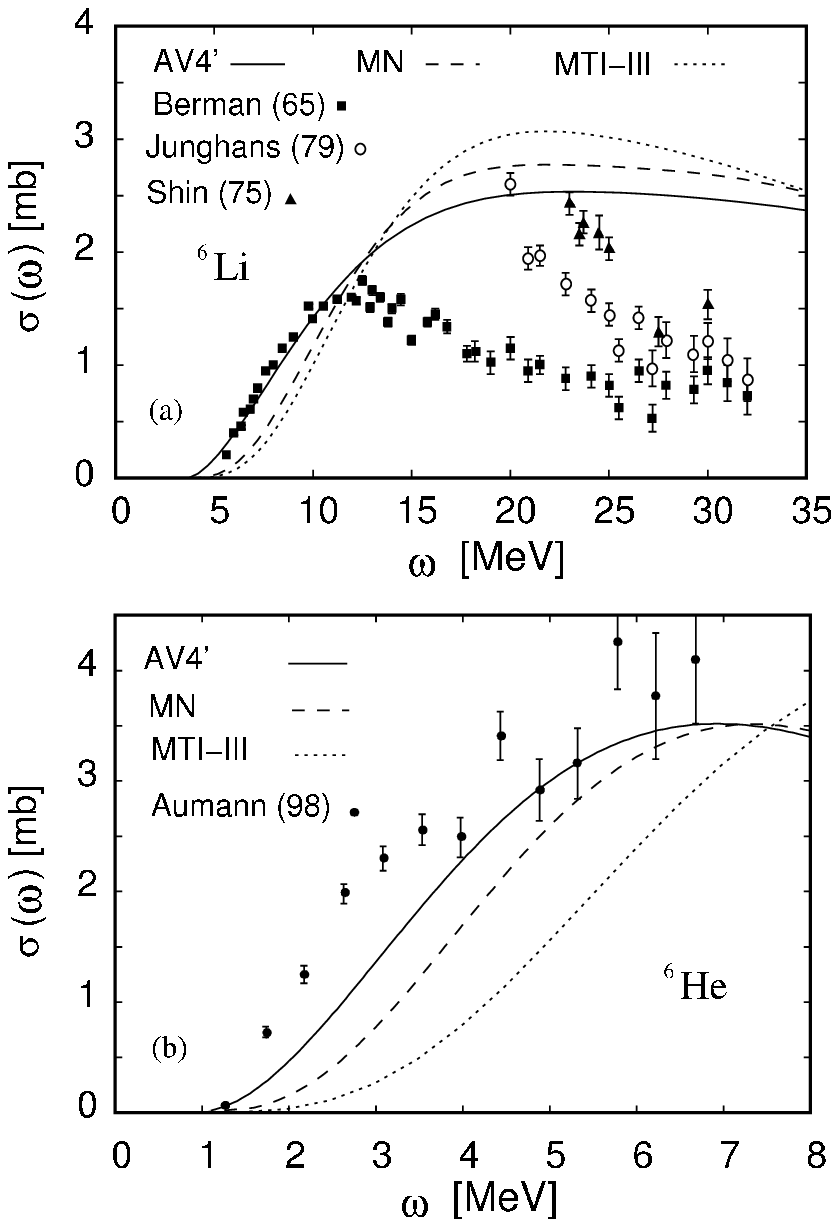}}}
\caption{As in~\fref{R15} but experimental data are also shown: (a) $^6Li(\gamma,n)$ data 
from~\cite{BERMAN:1965} (solid squares) and $^6Li(\gamma,^3$H)$^3$He data from~\cite{JUNGHANS:1979} 
(open circles) and~\cite{SHIN:1975} (solid triangles) added to the ($\gamma$,n) data, (b) $^6$He data 
from~\cite{AUMANN:1999} (solid circles) (theoretical results convoluted with instrumental response
function).}
\label{R16}
\end{figure}

In~\fref{R15} we show the $^6$He and $^6$Li photoabsorption cross sections calculated
with the above mentioned semirealistic potential models. Particularly
interesting is the $^6$He case, which exhibits two separate cross section peaks. In a cluster 
model picture the low--energy peak can be explained by excitations of the surface neutrons 
(soft mode or `pigmy resonance') leading to a final state with an $\alpha$--particle and two 
neutrons, while the second peak corresponds to the  classical electric dipole resonance 
(relative motion of neutrons against protons) with a break--up of the $\alpha$--core. A 
similar double--peak picture is not found for $^6$Li, even though a disintegration
to a final state with an $\alpha$--particle and a $^1S_0$ $np$ pair is possible.
However, such a transition is not energetically as well separated as the $NN+\alpha$ channel
and the $\alpha$ break--up channel in $^6$He. Note that a similar two--body break--up in case of
$^6$He into the $^3$H--$^3$H channel is not induced by the electric dipole operator. 
In~\fref{R16} we show the comparison with experimental data. Compared with the other two
interaction models the AV4' potential leads to a better agreement with experiment at lower
energy both for $^6$He and $^6$Li. In fact for $^6$Li  one finds a rather good description 
of the data up to 10~MeV, while at higher energies more precise data 
are needed in order to draw more definite conclusions. The agreement of theoretical and 
experimental low--energy results is less satisfying for $^6$He. On the other hand one has to take
into account that the $^6$He photodisintegration cross section has not been determined from
a direct measurement, but extracted from the Coulomb excitation in the field of a heavy nucleus.
\begin{figure}[htb]
\centerline{\resizebox*{12cm}{7cm}{\includegraphics*[angle=0]{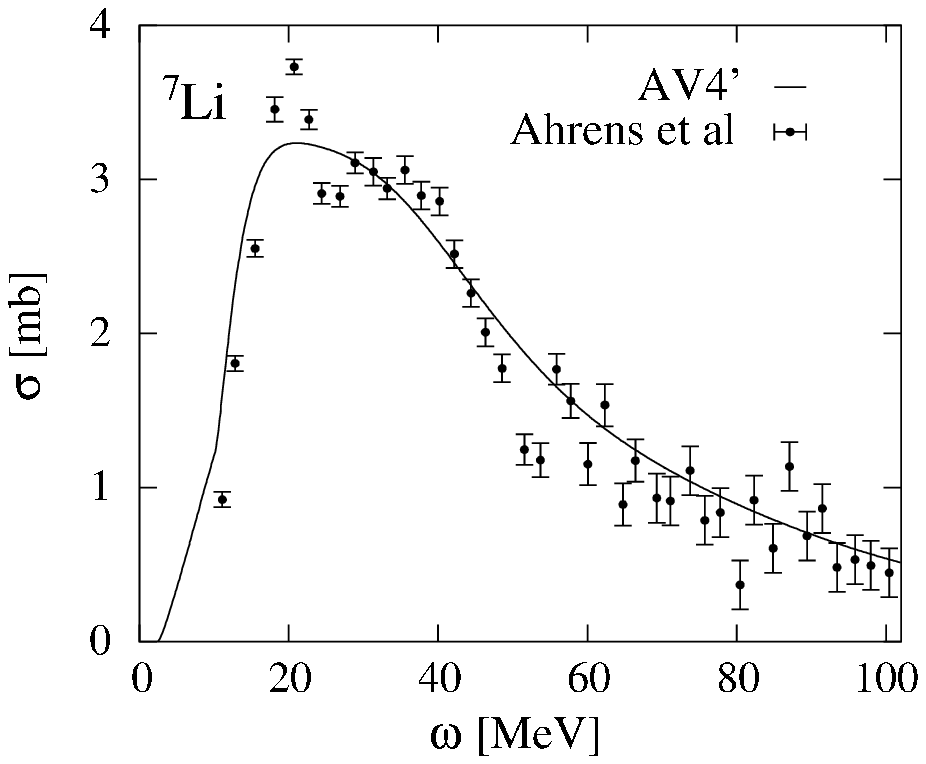}}}
\caption{Total photoabsorption cross section of $^7$Li: theoretical result with
AV4' potential (solid) and experimental data from~\cite{AHRENS:1975} (circles).}
\label{R17}
\end{figure}

Now we turn to the $^7$Li photodisintegration. In~\fref{R17} we show the cross section
calculated with the AV4' potential in comparison with experimental data. One finds quite a good
agreement in the whole considered energy range. The data exhibits a few fluctuations,
but one would need more precise data in order to decide whether such
structures in the cross section really exist. We would like to point out that the experimental
$^7$Li total photoabsorption cross section had been determined via flux attenuation measurements
in forward direction and not by summing up the cross sections of the various break--up
channels. In the near future similar measurements for $^4$He and $^{6,7}$Li are planned 
at MAXLab (Lund) and will hopefully also be performed at the HI$\gamma$S facility
(Duke).

%************************************
\section{Summary}\label{sec:SUMMARY}
%**************************************

We have reviewed the theory, tests and applications of the Lorentz integral 
transform method for perturbation induced reactions into the many--body 
continuum. The method is a novel approach, which allows the microscopic 
calculation of inclusive and exclusive cross sections without using 
complicated many--body continuum wave functions. Since its proposal in 1994 
the method has been applied to quite a number of electroweak processess with 
few--body nuclei, among them are the first calculations with a realistic nuclear 
interaction (NN and 3N forces) for the total $^3$H, $^3$He and $^4$He 
photoabsorption cross sections~\cite{EFROS:2000,GAZIT:2006} 
and the inelastic neutral current neutrino scattering off $^4$He~\cite{GAZIT:2007}. 
Even reactions into the six-- and seven--body continuum have been 
considered~\cite{BACCA:2002,BACCA:2004,BACCA:2004a}
treating the many--body continuum interaction in a rigorous way. We would like to 
point out that at present there are no alternative approaches allowing similar 
ab initio calculations for $A\ge4$.

\section{Acknowledgments}
W.L. and G.O. thanks the Department of Physics of the George Washington University
for the warm hospitality.
The work of N.B. was supported by the ISRAEL SCIENCE FOUNDATION (grant no. 361/05). 
V.D.E. acknowledges support from the RFBR, Grant No. 07--02--01222a,
and the Russian Ministry of Education and Science, Grant No. NS--8756.2006.2.
%%%%%%%%%%%%%%%%%%%%%%%%%%%%%%%%%%%%%%%%%%%%%%%%%%%%%%%%%%%%%%%%%%%%%%
% Section   : CHH
%%%%%%%%%%%%%%%%%%%%%%%%%%%%%%%%%%%%%%%%%%%%%%%%%%%%%%%%%%%%%%%%%%%%%%

\appendix
\section{}\label{sec:CHH}

\noindent{\it The correlated HH method (CHH)} 
\bigskip

In this appendix the CHH method is described. Not only the general idea of the method
is summarized, but we also describe the details of its application, which has led
to the results presented in~\sref{sec:RES}.

The basis functions we use are antisymmetric with respect to particle permutations. Such
a basis is combined in a simple way from hyperspherical harmonics ${\cal{Y}}$
 belonging to a given type of permutational 
symmetry and from spin--isospin functions $\theta$ belonging to a conjugated type of permutational 
symmetry. The set of ${\cal{Y}}$ differs from the unsymmetrized set (5.7) and we denote them  
${\cal{Y}}_{KLM_L,\alpha}^{[f]\mu}$. They are components $\mu$ of a given irreducible 
representation [f] of the permutation group and they have a given grandangular momentum $K$, 
orbital momentum $L$ with projection $M_L$. The label $\alpha$ enumerates the various HH with
the same $K$, $L$, $M_L$, $[f]$ and $\mu$.  The ${\cal{Y}}_{KLM_L,\alpha}^{[f]\mu}$ may be
constructed applying the Young operators to the simple ${\cal Y}_{[K]}(\hat\Omega)$ of (5.7). This
gives them in the form of linear combinations of the ${\cal Y}_{[K]}(P\hat\Omega)$, where
%%%%%%%%%%%%%%%%%%%%%%%%%%%%%%%%%%%%%%%%%%%%%%%%%%%%%%%%%%%%%%%%%%%%%%%%%%%%%%%%%%%%%%%%%%
%We start from the HH of the form ${\cal{Y}}_{K_nL_nM_n\alpha}^{[f]\mu}$. With respect 
%to the notation in~(\ref{HH})
%one notices the addition of two indexes: $\mu$ represents the components 
%of a given irreducible representation $[f]$ of the permutation group.
%The label $\alpha$ enumerates the various HH with the same $K_n,L_n,M_n,[f]$ and $\mu$.
%The ${\cal{Y}}_{K_nL_nM_n\alpha}^{[f]\mu}$ may be constructed applying the Young 
%operators to the simple ${\cal Y}_{[K_n]}(\Omega_N)$ of~(\ref{HH}).
%This gives them in the form of linear
%combinations of the ${\cal Y}_{[K_n]}(P\Omega_N)$, where 
%%%%%%%%%%%%%%%%%%%%%%%%%%%%%%%%%%%%%%%%%%%%%%%%%%%%%%%%%%%%%%%%%%%%%%%%%%%%%%%%%%%%%%%%%%%
$P$ are particle permutations.
These HH must be  combined 
with spin--isospin functions $\theta_{SM_STM_T}^{[{\bar f}]{\bar\mu}}$ as follows
\be
\sum_\mu\{{\cal{Y}}_{KL,\alpha}^{[f]\mu}\otimes
\theta_{STM_T}^{[{\bar f}]{\bar\mu}}\}_{JM}.\la{yt}
\ee
The spin--isospin functions have given spin $S,M_S$ and  isospin $T,M_T$ quantum numbers
 and belong 
to the conjugated representation $[\bar f]$ of the permutation group.
The spin--isospin function with a $\bar \mu$  value is conjugated to 
the spatial function
with  given $\mu$ so that
the functions (\re{yt}) 
are antisymmetric
with respect to particle permutations. They possess a given total momentum $J$
and its projection $M$. When solving the 
dynamic equations, products of the functions (\re{yt}) and
hyperradial functions $R_\beta(\rho)$ can be taken as spatial basis functions $\chi$ to
expand $|\tilde\Psi\rangle$ in coordinate representation.
For brevity we shall write the total basis functions as
\be
\phi_i=\sum_\gamma \chi_{i,\gamma}\theta_\gamma,\la{bf}
\ee    
where $\theta_\gamma$ are the spin--isospin functions of the above mentioned form, and
$\chi_{i\gamma}$ are  the conjugated spatial functions.
 
The short range repulsion 
at small distances, present in the central components of conventional 
NN interactions, slows
down the convergence of the HH expansion~(\ref{bf}) considerably. To speed up the convergence
one may include correlation factors in the basis functions.
%When HH are taken in the first mentioned form inclusion of the correlation factors is natural 
%since this does not complicate noticeably the calculation.  
In~\cite{FENIN:1972} it has been  shown that, in the case 
of central NN potentials, the inclusion of a Jastrow correlation factor
$\omega=\prod_{i<j}f(r_{ij})$ 
in the HH expansion  accelerates the convergence drastically. 
To compensate for the strong short range repulsion potential that is present in the A--body
dynamic equation the pair correlation function $f(r_{ij})$ should approximately
be a solution of the two--body equation with this potential. When the mentioned simple
Jastrow correlation factor is employed this may be fulfilled
only for $s$--states of the NN pair and only when the triplet and singlet 
potentials $\hat{V}_{31}(r_{ij})$ and $\hat{V}_{13}(r_{ij})$ are close
to each other in the region of short--distance repulsion. 
To go beyond these frames a more general ansatz has been proposed~\cite{EFROS:1999b,EFROS:2000}:
\be
|\Psi\rangle=\hat{\omega}\sum_ic_i\phi_i,\la{ac}
\ee
where $\phi_i$ are the HH--spin--isospin--hyperradial basis functions (\re{bf}) 
and $\hat{\omega}$ is the correlation operator of the form 
\be
\hat{\omega}={\cal S}\prod_{i<j}\sum_{S,T}f_{ST}(r_{ij})\hat{P}_{ST}(ij).\la{ocr}
\ee
Here $\hat{P}_{ST}(ij)$ are the projection operators onto nucleon pair $(ij)$
states of spin $S$ and isospin $T$ and $\cal{S}$ is the particle symmetrization operator.
The inclusion of  $\cal{S}$ is necessary since $\hat{P}_{ST}(ij)$ for different pairs $(ij)$
do not commute with each other. Operators of a similar type  have been used
in direct variational calculations of light nuclei, e.g. in~\cite{WIRINGA:1991}. In~(\re{ac})
$|\Psi\rangle$ stands for the initial state $|\Psi_0\rangle$, or for the LIT 
function $|\tilde\Psi\rangle$.

This form of the  correlation operator is easily incorporated into the calculations by first 
constructing the set of coefficients $\langle\theta_{\gamma'}|\hat{\omega}|\theta_{\gamma}\rangle$
and then writing
\be
|\Psi\rangle=\sum_ic_i\tilde{\phi}_i,\la{ac1}
\ee
where
\be
\tilde{\phi}_i=\sum_{\gamma'} \tilde{\chi}_{i,\gamma'}\theta_{\gamma'},\qquad 
\tilde{\chi}_{i,\gamma'}=\sum_\gamma\langle\theta_{\gamma'}|\hat{\omega}|\theta_{\gamma}\rangle
\chi_{i,\gamma}.
\ee
The pair correlation functions entering (\re{ocr}) are constructed as follows. At
 $r\le r_0$ ($r\equiv r_{ij}$) they are solutions of the two--body equation
\be
-\frac{\hbar^2}{M}\left[\frac{d^2f}{dr^2}+\frac{2}{r}\frac{df}{dr}-\frac{l(l+1)}{r^2}f\right]
+[\hat{V}_{ST}^c(r)+\hat{V}_{ST}^m(r)]f=0 \la{tbe}
\ee 
where $f=f_{ST}$. Here  $l=0$ for even states and $l=1$ for odd states of a nucleon pair.  
The potential $\hat{V}_{ST}^c$
is the central component of the NN interaction. The potential $\hat{V}_{ST}^m$ takes approximately into 
account the other components of the NN interaction and also the interaction produced by other 
nucleons. (In the case $S=T=1$ the corresponding $p$--wave NN interaction is $j$--dependent which
is disregarded in (\re{tbe}).) The potentials in (\re{tbe}) should ensure that $f'(r_0)=0$ at some
point $r_0$. Equation (\re{tbe}) is solved in the range $0\le r\le r_0$ and at $r>r_0$ one sets 
$f(r)=f(r_0)=1$.

In the $A=3$ calculations with realistic NN+3N forces presented in this review 
the procedure described above has been  applied.
For even NN states the potential $\hat{V}_{ST}^m$ was disregarded in (\re{tbe}). In the $A=4$ calculations
with central NN forces the simple Jastrow correlation factor has been used with an average 
potential equal to $1/2[\hat{V}_{31}(r)+\hat{V}_{13}(r)]$. 

In order to diminish the number of basis states  
in the correlated HH calculations one may optimize
the potentials $\hat{V}_{ST}^m$
in (\re{tbe}). In addition, one may apply different 
correlation operators (\re{ocr}) to different parts of $|\tilde\Psi\rangle$ or $|\Psi_0\rangle$. 
The correlation functions may also be obtained from equations
of (\re{tbe}) form for finite energies. These energies may be fixed from the condition 
that $f_{ST}''$ is 
continuous, i.e. $f_{ST}''(r_0)=0$~\cite{FENIN:1972}, or in another way. 
In the ground  state case the parameters determining $f_{ST}(r)$
may be optimized from the minimum energy requirement or from the condition
\be
{\rm min}\,\,\{\langle({\hat H}-E_0)\Psi_0|({\hat H}-E_0)\Psi_0\rangle\}
.\la{m1}
\ee
In the integral transform case these parameters may be chosen from the condition
\be
{\rm min}\,\,\{\langle({\hat H}-{\tilde E}){\tilde\Psi}-Q|({\hat H}-{\tilde E}){\tilde\Psi}-Q\rangle\}
\la{m2}\ee
%\be
%{\rm min}\,\,\{\,[\langle({\hat H}-{\tilde E}){\tilde\Psi}|-\langle Q|\,]\,\,
%[\,|({\hat H}-{\tilde E}){\tilde\Psi}\rangle-|Q\rangle]\,\}
%\la{m2}\ee
for some typical $\tilde E$ values, where $|Q\rangle$ indicates the source terms 
in the inhomogeneous equations (see \sref{sec:INCL}).
In general, conditions of (\re{m1}) and (\re{m2}) form 
may be used to fix non--linear parameters in variational few--body calculations
that are not based on the minimum energy principle. In particular in variational methods
for calculating reactions the condition
\be
\langle({\hat H}-E)\Psi^{\pm}|({\hat H}-E)\Psi^{\pm}\rangle={\rm min}
\ee
may be used for this purpose.

Besides correlations caused by the strong short distance repulsion in NN potentials
long range tensor correlations also require to reach high $K$ values in the HH expansion.
In the $A=4$ CHH calculations presented here, as well as in some of the $A=3$ ones
a  selection of correlated HH has been used.
In general only HH  that have small $l_n$ and $K_{n-1}$ values before applying 
the Young symmetrization 
operators are to be retained~\cite{EFROS:1972,EFROS:1978}. This prescription is justified for 
non--correlated HH and in the case of the ground state, but seems to work
in our case as well. A more detailed selection within these frames 
has been done (see also~\cite{VIVIANI:2005}.)
%In some cases it occurs that the inclusion of terms representing two--body correlations is 
%also most important to take into account long range correlations. 
For example, in the case of $A=3$ bound state it has been found that for
high $K$ values only those HH are to be retained that in the notation of~\sref{sec:HHCOORD}
had  quantum numbers 
$l_2=2$ and  $l_1=0$ before  applying the Young symmetrization operators.

Convergent results have been obtained in the calculations with the 
CHH method (for judging the accuracy reached see~\cite{BARNEA:2006a} in $A=3$ calculations). 
Also in~\cite{VIVIANI:1995,VIVIANI:1998,VIVIANI:2001} 
accurate results have been obtained in the $A=4$ case with realistic 
nuclear forces in a CHH basis for the ground state and low--energy
scattering. There, different Jastrow
type correlation factors for different parts of wave functions have been applied. 
It is ineteresting to notice that, due to flexibility of the basis, accurate results 
e.g. for the $\alpha$--particle ground state with realistic NN+3N interactions, have been obtained 
retaining only low $K$ values (up to $K=8$).

A final comment regards the practical way to calculate matrix elements of the Hamiltonian
between CHH states. In the $A=3$ case an analytic
integration over the Euler angles that determine the orientation of the system as
a whole has been performed~\cite{EFROS:2002}. The remaining three--dimensional
integration has been carried out with regular quadratures. 
In the $A>3$ case the Monte Carlo integration has been applied.

%%%%%%%%%%%%%%%%%%%%%%%%%%%%%%%%%%%%%%%%%%%%%%%%%%%%%%%%%%%%%%%%%%%%%%
% Section   : Effective Interaction
%%%%%%%%%%%%%%%%%%%%%%%%%%%%%%%%%%%%%%%%%%%%%%%%%%%%%%%%%%%%%%%%%%%%%%

\section{}\label{sec:EIHH}

\noindent{\it The effective interaction HH method (EIHH)}
\bigskip

In the effective interaction approach
~\cite{ZHENG:1994,NAVRATIL:2000,HJORTHJENSEN:1994,BARNEA:2000,BARNEA:2001a}
the lowest eigenvalues of an $A$--body Hamiltonian 
\begin{equation}\label{1}
  \hat{H}^{[A]} = \hat{H}_0+\hat{V}\;,
\end{equation}
is treated in the following way. The Hilbert space of 
$\hat{H}^{[A]}$ is divided into a model space and a residual space,
through the use of the  
eigenprojectors $\hat{P}$ and $\hat{Q}$ of $\hat{H}_0$, which satisfy the 
relations 
\begin{equation}\label{2}
  [\hat{H}_0,\hat{P}]=[\hat{H}_0,\hat{Q}]=0\,;\,\,\,\,\hat{Q}\hat{H}_0\hat{P}=\hat{P}\hat{H}_0Q=0\,;\,\,\,\,\hat{P}+\hat{Q}=1\,.
\end{equation}
The Hamiltonian $\hat{H}^{[A]}$ is then replaced by the 
effective model space Hamiltonian
\begin{equation}\label{3}
  \hat{H}^{[A]eff} = \hat{P}\hat{H}_0 \hat{P} + \hat{P} \hat{V}^{[A]eff} \hat{P} \;
\end{equation}
that by construction has the same energy levels as the low 
lying states of $\hat{H}^{[A]}$. 
In general the effective interaction appearing 
in~(\ref{3}) is an $A$--body interaction, and its 
construction is usually as difficult as finding the full--space 
solutions of the $A$--body problem.
Therefore, one has to approximate 
$\hat{V}^{[A]eff}$. 
However, one must build 
the approximate effective potential in such a way that it coincides with the 
bare one for $P\to 1$, so that 
an enlargement of the model space 
leads to a convergence of the eigenenergies to the {\it true} values.
The EIHH method is developed along these lines.

In the HH formalism the internal $A$--body Hamiltonian $\hat{H}^{[A]}$
is written as 
\begin{equation}\label{HA}
\hat{H}^{[A]} = \hat{T}_\rho+\hat{T}_K(\rho) + \hat{V}^{[A]}(\rho,\hat\Omega)\,,
\end{equation}
where 
\begin{equation}\label{6}
  \hat{V}^{[A]}(\rho,\hat\Omega)\equiv\sum_{i<j}^{A} v_{ij}
\end{equation}
denotes the bare two--body potential and 
\begin{equation}\label{7}
\hat{T}_\rho=- \frac{1}{2m}\Delta_{\rho}\,,\,\,\,\,\,
       \hat{T}_K(\rho)=\frac{1}{2 m} \frac{\hat{K}_{A-1}^2}{\rho^2} 
\end{equation} 
are the hyperradial and hypercentrifugal kinetic energies, 
respectively. In the previous equation $\Delta_{\rho}$ is the 
Laplace operator with respect to the hyperradial coordinate $\rho$,
while  $\hat{K}_{A-1}$ is the hyperspherical grand 
angular momentum (see equations~(\ref{Laplacen})-(\ref{Kn})). 
The hyperradial kinetic energy $\hat{T}_\rho$ and the residual Hamiltonian
\begin{equation}\label{9}  
  \hat{{\cal H}}^{[A]}(\rho)\equiv \hat{T}_K(\rho)  +  \hat{V}^{[A]}(\rho,\hat\Omega)
\end{equation} 
are often considered separately. The 
Hamiltonian $\hat{{\mathcal H}}^{[A]}(\rho)$, often used as a starting point
in atomic and molecular calculations, is called 
adiabatic~\cite{MACEK:1968} as the hyperradial coordinate 
$\rho$ is a {\it  slow} coordinate with respect to the 
hyperangles. 
In the EIHH method~\cite{BARNEA:2000,BARNEA:2001a} the effective interaction is calculated
starting with this adiabatic Hamiltonian.
The unperturbed Hamiltonian $\hat{H}_0$ is 
chosen to be $\hat{T}_K(\rho)$ with the hyperspherical harmonics ${\cal Y}_{[K_{A-1}]}$
as eigenfunctions.
The model space $P$ is defined as the complete
set of HH basis functions with generalized angular momentum 
quantum number $K \leq K_P$, and the Q--space as
the complete set of HH basis functions with $K > K_P$. The 
states are denoted by $\{|p\ket\,;\,\, p= 1,2,...,
n_P\}$ for the P--space and $\{|q\ket\,;\,\, 
q=n_{p+1},n_{p+2},...,n_Q \}$ for the Q--space. 
Of course, in principle one has $n_Q \to \infty$. 
In actual calculations one considers finite Q--spaces.
However, $n_Q$ must be 
sufficiently large. 

For each value of the hyperradius $\rho$ an effective 
adiabatic Hamiltonian is constructed
\begin{equation}\label{10}
\hat{{\mathcal H}}^{[A]eff}(\rho,\hat\Omega)= \hat{P} \hat{T}_K(\rho) \hat{P}  +  
                 \hat{P}\,\hat{V}^{[A]eff}(\rho,\hat\Omega)\,\hat{P}\,.
\end{equation}
However, as already pointed out, the effective potential 
would be a complicated $A$--body interaction, therefore
$ \hat{V}^{[A]eff}$ is approximated by a sum of {\it quasi--two--body} 
terms 
\begin{equation}\label{11}
  \hat{V}^{[A]eff}\simeq \,\,\sum_{i<j}^A \hat{v}^{[2]eff}_{i,j}\,.
\end{equation}
%In the following it will be shown how $v^{[2]eff}_{i,j}$
%is derived ensuring that the effective 
%potential satisfies the above mentioned condition, i.e. 
%$\,\,\,v^{[2]eff}_{i,j}$ tends to the bare $v_{ij}$ for 
%$\hat{P} \to 1$ . 
Due to the use of antisymmetric wave functions
one only needs to calculate the effective interaction operator 
relative to one pair, since
\begin{equation}\label{12}   
  \bra\,\hat{V}^{[A]eff}\, \ket \simeq \bra\, \sum_{i<j}^A 
  \hat{v}^{[2]eff}_{i,j}\,\ket=\frac{A(A-1)}{2}
  \bra\, \hat{v}^{[2]eff}_{A,(A-1)}\,\ket \;.
\end{equation}
%Below it will become clear why the choice to express 
%$\bra\,V^{[A]eff}\, \ket$ in terms
%of the pair potential between the last two particles $A$ and $(A-1)$
%is particularly convenient.
\noindent
The {\it quasi--two--body} effective potential $\hat{v}^{[2]eff}_{A,(A-1)}$ 
is determined as follows.
First for each value $\rho$ of the hyperradial coordinate
one defines a {\it quasi--two--body} adiabatic Hamiltonian containing the 
hypercentrifugal kinetic energy and the bare potential between 
the last two particles (the prefix {\it quasi} is justified, because this Hamiltonian 
depends on the $A$--body coordinate $\rho$)
\begin{equation}\label{H2}
  \hat{{\mathcal H}}^{[2]}(\rho\, ; \theta_{A-1},\hat{\eta}_{A-1}) = 
  \hat{T}_K(\rho) + \hat{v}_{A,(A-1)}
  (\sqrt{2}\rho \sin \theta_{A-1}\cdot \hat{\eta}_{A-1}) \;,
\end{equation}
%where $\hat {\eta}_N$ is the unit vector associated with 
%the last Jacobi vector
%\begin{equation}\label{eta_N}
%   \vec\eta_N = \sqrt{\frac{1}{2}}(\bi{r}_{A-1}-\vec r_A)
%\end{equation}
%and $\theta_N$ is the hyperangle defined through the 
%relation
%\begin{equation}\label{15}
%  \eta_N = \rho \sin \theta_N \;.
%\end{equation}
%We emphasize here that we are using reversed order $A$--body Jacobi 
%coordinates 
%\begin{equation}\label{15bis}
%  \bfeta_i = \sqrt{\frac{A-i}{A+1-i}}
%  \left(\bi{r}_i - \frac{1}{A-i}\sum_{j=i+1}^A \bi{r}_j \right) \;.
%\end{equation}
The Hamiltonian of~\eref{H2} is then diagonalized on the 
$A$--body HH basis. Such a diagonalization is
easily performed since $\rho$ is only a parameter in $\hat{{\mathcal H}}^{[2]}$, 
(there are no derivatives with respect to $\rho$) 
and for each value of $\rho$ the Hamiltonian 
$\hat{{\mathcal H}}^{[2]}(\rho \,; \theta_{A-1}, \hat{\eta}_{A-1})$ depends only 
on three variables. This is just due to our choice of the $A$--$(A-1)$ 
pair in~(\ref{12}).
%Equation~(\ref{12}). 
The obtained eigenstates are denoted by 
$|j(\rho)\ket$ as they are continuous functions of $\rho$. 

One proceeds applying the Lee--Suzuki~\cite{SUZUKI:1980,SUZUKI:1982,SUZUKI:1983}
similarity transformation to $\hat{{\mathcal H}}^{[2]}(\rho\,;\theta_{A-1},
\hat{\eta}_{A-1})$ in order to get the corresponding hermitian
effective Hamiltonian
\begin{equation}\label{16}
\hat{{\mathcal H}}^{[2]eff}(\rho\,;\theta_{A-1},\hat{\eta}_{A-1})
             = \hat{U}^\dagger(\rho)\,\hat{{\mathcal H}}^{[2]}(\rho\,;
             \theta_{A-1},\hat{\eta}_{A-1})\, 
             \hat{U}(\rho)\,,
\end{equation}
where
\begin{equation}\label{17}  
\hat{U}(\rho)= (\hat{P}+\hat{\omega }(\rho))
         \frac{1}{\sqrt{\hat{P}(1+\hat{\omega }(\rho)^{\dagger}\hat{\omega }(\rho))\hat{P}}}\,.
\end{equation}
The operator $\hat{\omega }(\rho)$ is obtained using the following
property~\cite{SUZUKI:1980} 
\begin{equation}\label{18}
 \hat{\omega }(\rho)=\hat{Q}\, \hat{\omega }(\rho)\, \hat{P}\,.
\end{equation}
The matrix $\hat{\omega }(\rho)$ is calculated for each value of $\rho$ taking the 
$n_P$ states $|j(\rho)\ket$ with the lowest eigenvalues. 
Each of these states leads
to the following system of $(n_Q-n_P)$ equations
\begin{equation}\label{19}
   \bra q|j(\rho)\ket=\sum_{p} \bra q|\hat{\omega }(\rho)|p\ket\,\bra 
   p|j(\rho)\ket\,.
\end{equation}
The $n_P(n_Q-n_P)$ matrix elements $\bra q|\hat{\omega }(\rho)|p\ket$ are 
obtained by solving the equation system~(\ref{19}).
Once the effective {\it quasi--two--body} Hamiltonian $\hat{{\mathcal H}}^{[2]eff}$ 
is constructed,
the effective potential is obtained by a
subtraction of the hypercentrifugal kinetic energy
\begin{equation}\label{20}
   \hat{v}^{[2]eff}_{A,(A-1)} = \hat{{\mathcal H}}^{[2]eff}(\rho) - \hat{T}_K(\rho) \;.
\end{equation}
%In the following we will denote $v^{[2]eff}_{A,(A-1)}$ simply by 
%$v^{[2]eff}$.
Using this $\rho$--dependent effective potential 
and taking into account~(\ref{10})--(\ref{12}) one solves 
the $A$--body problem with the effective Hamiltonian 
\begin{equation}\label{HAeff}
\hat{H}^{[A]eff}=\hat{T}_\rho+\hat{{\mathcal H}}^{[A]eff}=\hat{T}_\rho+\hat{T}_K+\sum_{i<j}\hat{v}^{[2]eff}_{ij}
\end{equation}
in the P--space. 
One repeats the procedure enlarging the P--space up to a convergence 
of the low--lying energies of the $A$--body system. 

\begin{figure}[htb]
\centerline{\resizebox*{9cm}{11cm}{\includegraphics*[angle=0]{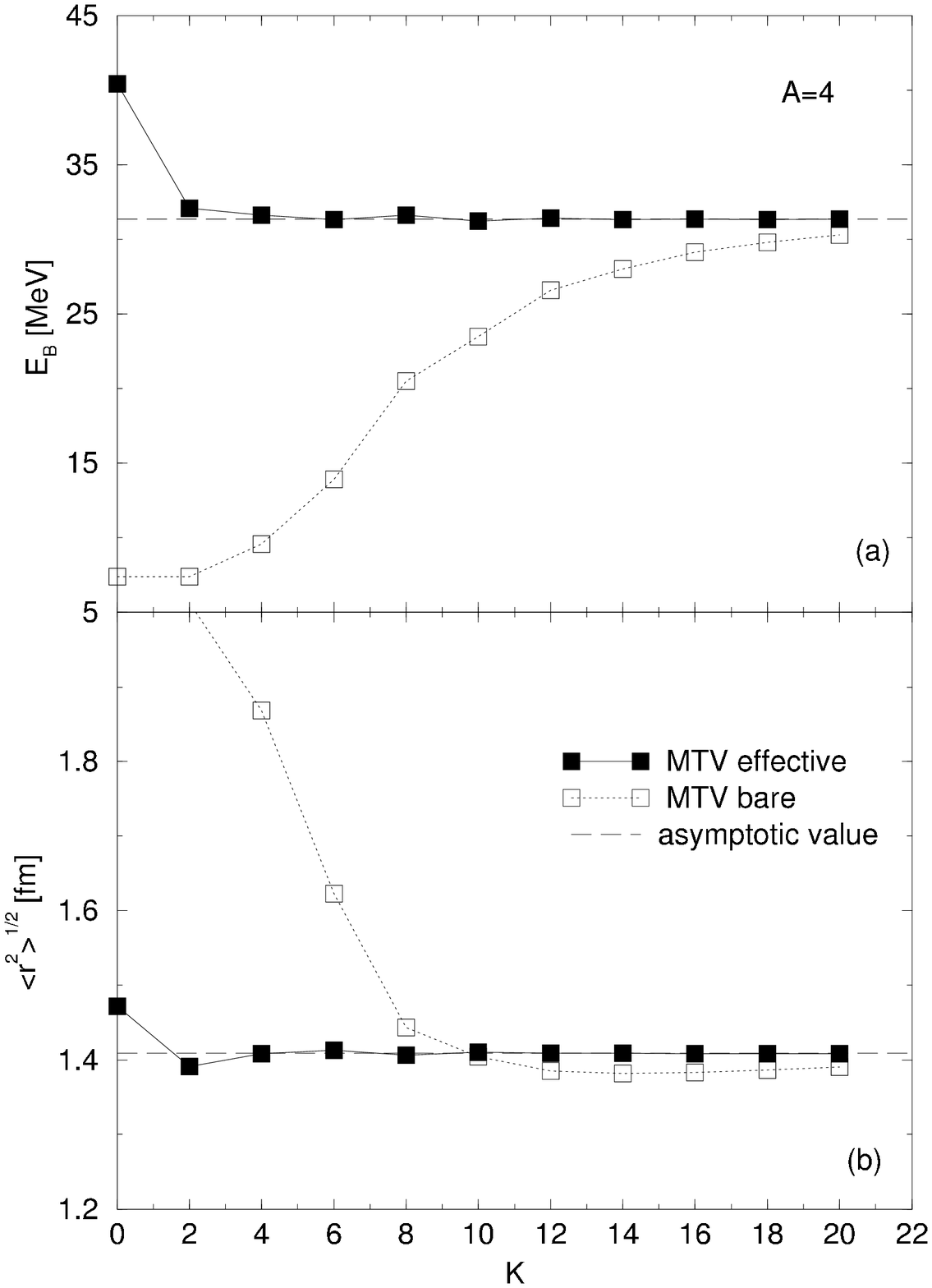}}}\label{MTV_conv}
\caption{Binding energy (a) and root mean square radius (b) of $^4$He for the MTV potential,
as a function of the maximal hyperangular momentum $K_{max}$. 
The asymptotic value has been indicated by a dashed line.} 
\end{figure}
We would like to emphasize the following points: 

\noindent
(i) it is evident that $\hat{U}(\rho)
\to 1$ for $P\to 1$ and thus 
$\hat{v}^{[2]eff}$ converges to the bare $\hat{v}_{A,(A-1)}$; therefore  
the energy spectrum converges to the exact one;

\noindent
(ii) in such a construction of an effective interaction the hyperradius 
is a parameter rather than a coordinate, and $\hat{v}^{[2]eff}_{ij}$ is determined for 
various fixed $\rho$ values; therefore, while being a two--body
interaction, it depends on the whole $A$--body system via 
this collective coordinate; 

\noindent
(iii) there is an additional dependence of $\,\hat{v}^{[2]eff}$
on the quantum number $K_{A-2}$ of the residual system (see 
equations~(19) and (20) of~\cite{BARNEA:2000});

\noindent
(iv) via the operator $\hat{U}(\rho)$ the effective potential 
$\hat{v}^{[2]eff}$ contains information about a large
part of the PQ--space interaction, hence the convergence to 
the exact eigenvalues
of $\hat{H}^{[A]}$ is accelerated with respect to the normal
HH expansion;

\noindent
(v) application of the transformation operator $\hat{U}(\rho)$ to the
hyperradial kinetic energy operator yields corrections to
$\hat{T}_{\rho}$ that go beyond the adiabatic effective interaction~\cite{BARNEA:2003};

\noindent
(vi) the convergence of the EIHH method can be further improved through
the use of a $3$--body effective interaction~\cite{BARNEA:2003}.

\noindent
The effectiveness of the EIHH method can be best demonstrated comparing 
the convergence rate of a normal HH expansion with that of an EIHH one for
some typical observable. As an example in figure B1 such
a comparison is shown for the binding energy and  root mean square matter
radius of a four--particle system interacting through the MTV 
NN potential. From the
figure it can be seen that, whereas the EIHH method gives almost converged 
results for $K_{max}=4$, the  HH calculation with the bare interaction misses about 
1.5~MeV for $K_{max}=20$.

\section{}\label{sec:APMULT}

\noindent{\it The multipole expansions of the response functions}
\bigskip

If the Hamiltonian is rotationally invariant it is useful to expand 
the states $Q$ and $Q'$ in (\re{rr}) over states possessing given values $J$ and $M$ of the
total angular momentum and its projection. Then the whole calculation may be done    
in separate subspaces of states belonging 
to given $J$ and $M$. Furthermore, the calculations are $M$ independent.
Here we shall present the required relationships for the case of longitudinal and transverse
electrodisintegration response functions of a nucleus.

We start from (\ref{long}), (\ref{trans}), omitting the indexes `$\alpha$' and `$\beta$'
\begin{eqnarray}
& & r^L(q,\omega)=\frac{1}{2J_0+1}\sum_{M_0,J,M}
\sum\!\!\!\!\!\!\!\int\,df \langle\Psi_0(J_0,M_0)|\hat{\rho}^\dag({\bf q})|\Psi_f(J,M)\rangle \nonumber\\
& &
\times \langle\Psi_f(J,M)|\hat{\rho}({\bf q})|\Psi_0(J_0,M_0)\rangle\,\delta\left(E_f(J)-E_0+\frac{q^2}{2 M_T}-\omega\right),\la{llong}
\end{eqnarray}
\begin{eqnarray}
& & r^T(q,\omega)=\frac{1}{2J_0+1}\sum_{M_0,J,M}
\sum\!\!\!\!\!\!\!\int\,df \langle\Psi_0(J_0,M_0)|\hat{{\bf j}}^{T\dag}({\bf q})|\Psi_f(J,M)\rangle\nonumber\\ 
& & \times\langle\Psi_f(J,M)|\hat{{\bf j}}^T({\bf q})|\Psi_0(J_0,M_0)\rangle\,
\delta\left(E_f(J)-E_0+\frac{q^2}{2 M_T}-\omega\right)\,,\la{ttrans}
\end{eqnarray}
where  the rotational quantum numbers $J$ and $M$ have been put in evidence.
One then substitutes the expansions
\ber
\hat{\rho}(\bi{q})=4\pi\sum_{jm}i^j\hat{\rho}_{jm}(q)Y_{jm}^*({\hat {\bf q}}),\la{rrhoexp}\\
\hat{{\bf j}}_T(\bi{q})=4\pi\sum_{\lambda={\rm el,mag}}
\sum_{jm}i^{j-\epsilon}\hat{T}_{jm}^{\lambda}(q){\bf Y}_{jm}^{\lambda *}({\hat {\bf q}}) 
\la{jjexp}
\eer
into (\ref{llong}), (\ref{ttrans})
and apply the operation $(4\pi)^{-1}\int d{{\bf\hat q}}$ to both sides of the arising expressions.  
This operation does not affect the 
responses while in their right--hand sides it leads to a 
simplification due to the orthonormality property of spherical harmonics or
vector spherical harmonics. This gives 
\ber
r^L(q,\omega)=\frac{4\pi}{2J_0+1}\sum_{M_0,J,M,j,m}\sum\!\!\!\!\!\!\!\int\,df
\langle\Psi_0(J_0,M_0)|\hat{\rho}^\dag_{jm}|\Psi_{f}(J,M)\rangle\nonumber\\
\times\langle\Psi_{f}(J,M)|\hat{\rho}_{jm}|\Psi_0(J_0,M_0)\rangle\delta\left(E_{f}-E_0+\frac{q^2}{2M_T} 
-\omega\right),
\eer
\ber
r^T(q,\omega)=\frac{4\pi}{2J_0+1}\sum_\lambda\sum_{M_0,J,M,j,m}\sum\!\!\!\!\!\!\!\int\,df
\langle\Psi_0(J_0,M_0)|\hat{T}^{\lambda\dag}_{jm}|\Psi_{f}(J,M)\rangle\nonumber\\
\times\langle\Psi_{f}(J,M)|\hat{T}^\lambda_{jm}|\Psi_0(J_0,M_0)\rangle\delta\left(E_{f}-E_0+\frac{q^2}{2M_T} 
-\omega\right).
\eer
Now we express these formulae in terms of the quantities
\be
Q_{JM}^j=\left(\hat{\rho}_{j}\otimes\Psi_0(J_0)\right)_{JM},\qquad
Q_{JM}^{j\lambda}=\left(\hat{T}_{j}^\lambda\otimes\Psi_0(J_0)\right)_{JM}.\la{q}
\ee 
They play the role of source terms in the LIT calculation. 
The notation of (\re{q}) type  means the Clebsch--Gordan coupling of the tensors of     
ranks $j$ and $J_0$  to the tensor of rank $J$. Using the 
orthogonality property of this transformation
we finally obtain:
\begin{eqnarray}
r^L(q,\omega)&=&\frac{4\pi}{2J_0+1}\sum_{Jj}(2J+1)(r^L)_J^j,\la{rrl}\\
r^T(q,\omega)&=&\frac{4\pi}{2J_0+1}\sum_{\lambda={\rm el,mag}}\sum_{Jj}(2J+1)(r^T)_J^{j\lambda},
\la{rrt}
\end{eqnarray}
where
\begin{eqnarray}
(r^L)_J^j=\sum\!\!\!\!\!\!\!\int\,df \langle Q_{JM}^j|\Psi_{f}(J,M)\rangle & &
\langle\Psi_{f}(J,M)|Q_{JM}^j\rangle \times\nonumber \\ & &  \times\delta(E_{f}-E_i+\frac{q^2}{2M_T} 
-\omega),\la{rlj}
\end{eqnarray}
\begin{eqnarray}
(r^T)_J^{j\lambda}=\sum\!\!\!\!\!\!\!\int\,df \langle Q_{JM}^{j\lambda}|\Psi_{f}(J,M)\rangle & &
\langle\Psi_{f}(J,M)|Q_{JM}^{j\lambda}\rangle\times\nonumber \\ & &  \times\delta(E_{f}-E_i+\frac{q^2}{2M_T} -\omega).\la{rtj}
\end{eqnarray}
The quantities on the right--hand sides of~(\re{rlj}),~(\re{rtj}) do not depend on $M$. 
In fact one may note that
\ber  
\langle\Psi_{f}(J,M)|Q_{JM}^j\rangle=(2J+1)^{-1/2}(\Psi_{f}(J)||\hat{\rho}_j||\Psi_0(J_0)),\la{rme1}\\
\langle\Psi_{f}(J,M)|Q_{JM}^{j\lambda}\rangle=(2J+1)^{-1/2}(\Psi_{f}(J)||\hat{T}_j^\lambda||\Psi_0(J_0)),
\la{rme2}\eer
where the right--hand sides include the conventional reduced matrix elements. 

Similar to (\re{tra}) we consider the partial transforms 
$(L^L)_J^j$ and $(L^T)_J^{j\lambda}$ of the contributions (\re{rlj}), (\re{rtj}). 
Taking into
account the completeness of the set $\Psi_{f}(J,M)$ in the subspace of states
with given $J$ and $M$ one calculates these partial transforms as 
\be
( L^L)_J^{j}=
\langle({\hat H}-E_0-\sigma_R+i\sigma_I)^{-1}{Q}^{j}_{JM}
|({\hat H}-E_0-\sigma_R+i\sigma_I)^{-1}{Q}'^{j}_{JM}\rangle,\la{cal1}
\ee
\be
( L^T)_J^{j\lambda}=
\langle({\hat H}-E_0-\sigma_R+i\sigma_I)^{-1}{Q}^{j\lambda}_{JM}
|({\hat H}-E_0-\sigma_R+i\sigma_I)^{-1}{Q}'^{j\lambda}_{JM}\rangle\,.\la{cal2}
\ee
Similar to~(\ref{rme1}) and~(\ref{rme2}) the quantities~(\ref{cal1}) and~(\ref{cal2})
are $M$--independent. 
The total transforms of the responses~(\re{rrl}) and~(\ref{rrt}) are
\be
L^L(q,\sigma_R)=
\frac{4\pi}{2J_0+1}
\sum_{Jj}(2J+1)(L^L)_J^{j}(q,\sigma_R),
\la{ssuml}\ee
\be
L^T(q,\sigma_R)=
\frac{4\pi}{2J_0+1}\sum_{\lambda={\rm el,mag}}
\sum_{Jj}(2J+1)(L^T)_J^{j\lambda}(q,\sigma_R).
\la{ssumt}\ee
To obtain the responses~(\re{rrl}) and~(\ref{rrt}) from these transforms the integral equations of the
form (\re{tra}) are to be solved. Alternatively,
one may invert the single terms  separately and then sum up
the corresponding partial contributions to the responses~(\re{rrl}) and~(\ref{rrt}).

The source terms (\re{q}) possess definite parities. If $\hat{\cal P}$ is the parity operator then
\begin{eqnarray}
{\hat {\cal P}}\hat{\rho}_{jm}{\hat {\cal P}}&=&(-1)^j\hat{\rho}_{jm},\\
{\hat {\cal P}}\hat{T}_{jm}^{\rm el}{\hat {\cal P}}&=&(-1)^j\hat{T}_{jm}^{\rm el},\\
{\hat {\cal P}}\hat{T}_{jm}^{\rm mag}{\hat {\cal P}}&=&(-1)^{j+1}\hat{T}_{jm}^{\rm mag}.
\end{eqnarray}
Therefore dynamic calculations proceed in subspaces of states with given $J,M$ and given parities
$P=P(j)P_0$ where $P_0$ is the parity of the ground state and $P(j)$ is $(-1)^j$ or $(-1)^{j+1}$. 
\noappendix

\end{document}